\documentclass[12pt,preprint]{aastex}
\begin{document}
\title{A Tight Connection between Gamma-Ray Outbursts and Parsec-Scale Jet Activity in the Quasar 3C~454.3}
\author{Svetlana G. Jorstad\altaffilmark{1,2}, Alan P. Marscher\altaffilmark{1}, Paul S. Smith\altaffilmark{3}, Valeri M. Larionov\altaffilmark{2,9,17}, Iv\'an Agudo \altaffilmark{4,1}, Mark Gurwell\altaffilmark{5}, Ann E. Wehrle\altaffilmark{6}, Anne L\"{a}hteenm\"{a}ki\altaffilmark{7},
Maria G. Nikolashvili\altaffilmark{8}, Gary D. Schmidt\altaffilmark{10}, Arkady A. Arkharov\altaffilmark{9}, Dmitry A. Blinov\altaffilmark{2},  Kelly Blumenthal\altaffilmark{1}, Carolina Casadio\altaffilmark{4}, Revaz A. Chigladze\altaffilmark{8}, Natatia V. Efimova\altaffilmark{2,9}, Joseph R. Eggen\altaffilmark{11}, 
Jos\'e L. G\'omez\altaffilmark{4}, Dirk Grupe\altaffilmark{12}, Vladimir A. Hagen-Thorn\altaffilmark{2,17}, Manasvita Joshi\altaffilmark{1},
Givi N. Kimeridze\altaffilmark{8}, Tatiana S. Konstantinova\altaffilmark{2}, Evgenia N. Kopatskaya\altaffilmark{2}, Omar M. Kurtanidze\altaffilmark{8}, Sofia O. Kurtanidze\altaffilmark{8},
Elena G. Larionova\altaffilmark{2}, Liudmilla V. Larionova\altaffilmark{2}, Sigua A. Lorand\altaffilmark{8}, 
Nicholas R. MacDonald\altaffilmark{1}, Jeremy D. Maune\altaffilmark{11}, Ian M. McHardy\altaffilmark{13}, 
H. Richard Miller\altaffilmark{11}, Sol N. Molina\altaffilmark{4}, Daria A. Morozova\altaffilmark{2}, Terri Scott\altaffilmark{1}, Brian Taylor\altaffilmark{1,14}, Merja Tornikoski\altaffilmark{7}, Ivan S. Troitsky\altaffilmark{2}, Clemens Thum\altaffilmark{15}, Gary Walker\altaffilmark{16}, Karen E. Williamson\altaffilmark{1}, Stephanie Sallum\altaffilmark{16,18}, Santina Consiglio\altaffilmark{16,19}, and Vladimir Strelnitski\altaffilmark{16}} 

\altaffiltext{1}{Institute for Astrophysical Research, Boston University, 725 Commonwealth Avenue, Boston, MA 02215}
\email{jorstad@bu.edu}
\altaffiltext{2}{Astronomical Institute, St. Petersburg State University, Universitetskij Pr. 28, Petrodvorets, 
198504 St. Petersburg, Russia}
\altaffiltext{3}{Steward Observatory, University of Arizona, Tucson, AZ 85721-0065}
\altaffiltext{4}{Instituto de Astrof\'{\i}sica de Andaluc\'{\i}a, CSIC, Apartado 3004, 18080,
Granada, Spain}
\altaffiltext{5}{Harvard-Smithsonian Center for Astrophysics, 60 Garden St., Cambridge, MA 02138}
\altaffiltext{6}{Space Science Institute, 4750 Walnut St., Boulder, CO 80301}
\altaffiltext{7}{Aalto University Mets\"ahovi Radio Observatory Mets\"ahovintie 114,
FIN-02540 Kylm\"al\"a, Finland }
\altaffiltext{8}{Abastumani Astrophysical Observatory, Mt. Kanobili, Abastumani, Georgia}
\altaffiltext{9}{Main (Pulkovo) Astronomical Observatory of RAS, Pulkovskoye shosse, 60, 196140,
St. Petersburg, Russia}
\altaffiltext{10}{National Science Foundation, 4201 Wilson Ave., Arlington, VA, 22230 USA}
\altaffiltext{11}{Department of Physics and Astronomy, Georgia State University, Atlanta, GA 30303-3083}
\altaffiltext{12}{Department of Astronomy and Astrophysics, Pennsylvania State University, 525 Davey Lab, University Park, PA 16802}
\altaffiltext{13}{Department of Physics and Astronomy, University of Southampton, Southampton, SO17 1BJ,
United Kingdom}
\altaffiltext{14}{Lowell Observatory, Flagstaff, AZ 86001}
\altaffiltext{15}{Instituto de Radio Astronom\'{i}a Milim\'{e}trica, Avenida Divina Pastora, 7, Local 20, E--18012 Granada, Spain}
\altaffiltext{16}{Maria Mitchell Observatory, 4 Vestal St., Nantucket, MA 02554}
\altaffiltext{17}{Isaac Newton Institute of Chile, St. Petersburg Branch, St. Petersburg, Russia}
\altaffiltext{18}{EAPS, MIT, 77 Massachusetts Ave., Cambridge, MA 02139}
\altaffiltext{19}{Department of Physics and Astronomy, University of California, Los Angeles, CA}

\shorttitle{Connection of $\gamma$-rays to the jet in 3C~454.3}
\shortauthors{Jorstad et al.}
\begin{abstract}
We analyze the multifrequency behavior of the quasar 3C 454.3 during three prominent $\gamma$-ray outbursts: 2009 Autumn, 2010 Spring, and 2010 Autumn. The data reveal a repeating pattern, including a triple flare structure, in the properties of each $\gamma$-ray outburst, which implies similar mechanism(s) and location for all three events. The multi-frequency behavior indicates that the lower frequency events are co-spatial with the $\gamma$-ray outbursts, although the $\gamma$-ray emission varies on the shortest timescales. We determine that the variability from UV to IR wavelengths during an outburst results from a single synchrotron component whose properties do not change significantly over the different outbursts. Despite a general increase in the degree of optical linear polarization during an outburst, the polarization drops significantly at the peak of the $\gamma$-ray event, which suggests that both shocks and turbulent processes are involved. We detect two disturbances (knots) with superluminal apparent speeds in the parsec-scale jet associated with the outbursts in 2009 Autumn and 2010 Autumn. The kinematic properties of the knots can explain the difference in amplitudes of the $\gamma$-ray events, while their millimeter-wave polarization is related to the optical polarization during the outbursts. We interpret the multi-frequency behavior within models involving either a system of standing conical shocks or magnetic reconnection events located in the parsec-scale millimeter-wave core of the jet. We argue that $\gamma$-ray outbursts with variability timescales 
as short as $\sim$3 hr can occur on parsec scales if flares take place in localized regions such as turbulent cells.
\end{abstract}
\section{Introduction}
The basic cause of the extremely high nonthermal luminosity and pronounced variability
of flux and polarization in the blazar class of active galactic nuclei (AGN) can be explained
through the paradigm of a relativistic jet of high-energy plasma \citep[e.g.,][]{bk79,MG85,sik09}.
However, our understanding remains limited about the physical processes, such as the compression
and heating of the plasma and production of relativistic electrons that generate the emission,
as well as the driver behind the rapid fluctuations in the flow speed (and possibly direction), 
magnetic field, and number of radiating electrons in the jet. Studies of
large samples of blazars are valuable for defining the statistics of the observed properties,
such as the probability function of the flow velocity and the correlation between apparent velocities
of knots in the jet and the observed level of $\gamma$-ray emission
\citep[e.g.,][]{Lister11}. More detailed observations of individual 
objects can provide a wealth of information as well, ranging from time profiles of major events
(e.g., flares), physical properties of emission features such as knots displaying apparent
superluminal motion, and the evolution of well-sampled spectral energy distributions (SEDs) at different times \citep[e. g.,][]{MAR10,J10,IVAN11A,IVAN11B}
This is especially true when a blazar undergoes a singular event that can be readily identified at
different wavebands \citep{ANN12}.

The quasar 3C~454.3 (redshift $z = 0.859$) is a prime example of a blazar that exhibits
such singular events. \citet{vil07}, \citet{RAI08}, \citet{HT09}, and \citet{J10} have analyzed comprehensive  multi-waveband observations of an extraordinary radio to X-ray outburst in 2005, as well as
major, but less pronounced, flares over the following two years.
After the launch of the {\it Astro-rivelatore Gamma a Immagini LEggero} ({\it AGILE})
and {\it Fermi Gamma-ray Space Telescope} orbiting
observatories, 3C~454.3 displayed unprecedentedly bright $\gamma$-ray outbursts in
late 2009, April 2010, and late 2010 \citep{ACKER10,ABDO11,BON11,RAI11,verc11,ANN12}. During the 2010
event, 3C~454.3 reached the highest $\gamma$-ray flux ever detected from a single non-transient
cosmic source. A variety of telescopes observed contemporaneous outbursts from millimeter
wavelengths to $\gamma$-rays. Analysis of the resulting rich dataset serves as a valuable
probe into the structure and physical conditions of the jet at distances within dozen's parsecs
from the central engine, as well as the changes in those conditions that cause such an outburst.

Here we perform an analysis of the trio of outbursts from late 2009 to
early 2011. We combine observations from millimeter wavelengths (mm-wave)
to $\gamma$-ray energies and compare the timing of features in the light curves, polarization variations vs.\ 
time curves, and mm-wave images (from the Very Long Baseline Array --- VLBA --- in both total
and polarized intensity) to provide a comprehensive
description of the variations in emission and structure of the jet during the outbursts. The
data reveal repeated patterns of variability during the outbursts,
implying that the location in the jet and physical conditions are similar for the different events.
We are able to infer the location of the emission sites relative to a bright, essentially
stationary feature found on the upstream end of the mm-wave images, referred to as the ``core.''
The location constrains the source of seed photons that are scattered to $\gamma$-ray
energies.

We present the observations in \S{2}, followed by analyses of the data in \S{3}--\S{6}. In \S{7}
we discuss the implications of the data and offer a physical interpretation of the
outbursts. We draw conclusions in \S{8}.

\section{Observations and Data Reduction}
We have used data obtained for the quasar 3C~454.3 from 2009 April 15 to 2011 August 1 from $\gamma$-ray to millimeter (mm) wavelengths at: 
1) 0.1-300~GeV,  2) 0.3-10 keV, 3) 2030-3501 {\AA}, 4) optical {\it BVRIJHK} bands, 
5) 4-21~$\mu$m, 6) 70-500~$\mu$m, 7) 350~GHz (0.85~mm), 230~GHz (1.3~mm),  86.24~GHz (3.5~mm),  43~GHz (7~mm), and  37~GHz (8~mm). 
The observations from optical
to mm-wavelengths as well as the data reduction at all wavelengths were performed by the authors. 
Throughout the paper we use {\it Reduced Julian Date}, RJD, which is RJD=JD-2450000.0;
the analyzed period in RJD dates is from RJD: 4937.5 to RJD: 5774.5. Current standard cosmological 
constants with $\Omega_{\rm m}=0.3$, $\Omega_\Lambda=0.7$, and Hubble constant 
$H_\circ$=71~km~s$^{-1}$~Mpc$^{-1}$ are used in calculations; this gives a scale of 7.7~pc per 
milliarcsecond (mas) at the quasar redshift.

\subsection {Multi-Frequency Light Curves} 
The $\gamma$-ray data are collected with the Large Area Telescope (LAT) of the {\it Fermi Gamma-ray Space Telescope}.
We construct a daily $\gamma$-ray light curve of the quasar
using {\it Pass 7} photon and spacecraft data, version {\it V9r23p1} of the Fermi Science Tools, and 
the instrument responses for the {\it gal\_2yearp7v6\_v0} and {\it iso\_p7v6clean.txt}
diffuse source models. We model the $\gamma$-ray emission from 3C~454.3
and other point sources within 15 degrees radius of the quasar with spectral models, as found in the 2FGL catalog 
of sources detected by the LAT. We have fixed the catalog's spectral parameters of sources within 
the area and searched for values of flux normalization parameters with 1~day integration intervals of 
photons between 0.1 and 200~GeV using the standard unbinned likelihood analysis. This produces a $\gamma$-ray light curve 
with 823 measurements, with 59 values representing only upper limits to the $\gamma$-ray emission. 
The flux is considered detected if the test statistic, $TS$, provided by the analysis exceeds 10, 
which corresponds to approximetely a 3$\sigma$ detection level \citep{2FGL}. 

We have acquired X-Ray Telescope (XRT) and Ultraviolet and Optical Telescope (UVOT) data of 3C~454.3 from 2009 April 25 to 2011 August 1
from the {\it Swift} archive and processed them with the {\it HEAsoft} version 6.11 software package.
We have obtained 201 measurements of the flux at 0.3-10~keV. The XRT observations were carried out in a mixture of Photon Counting (PC) and
Windowed Timing (WT) modes \citep{HILL04}. We counted photons with apertures as proposed by \cite{RAI11}, a 30 pixel circular region ($\sim$71 arcsec) and an annular region with inner and outer radii of 110 and 160 pixels for the source and background measurements, respectively. All of the observations collected in PC mode were near or exceeded 0.5 counts s$^{-1}$, indicating possible photon pile-up that was corrected by eliminating 3-5 central pixels. A new exposure map was generated using the Swift XRT task {\it xrtexpomap}. For each observation collected in WT mode, we created a box-shaped extraction region individually sized to exclude the point at which the pixels dropped to less than 2 counts. The Swift XRT task {\it xrtmkarf} was applied on all extracted spectra. The spectra were rebinned with the FTOOLS task {\it grppha} to include a minimum of 20 photons in each channel. We used XSPEC version 12.7.0 to fit the data with a single power law model while the hydrogen column density was fixed at 1.34 $\times 10^{21}$cm$^{-2}$ \citep{VIL06}. 

For the UVOT data, we extracted the magnitude and its error using the tool UVOTSOURCE, specifying a region with a circular aperture of 7 arcseconds, and a background annulus region centered on the object with inner and outer radius of 22 and 25 arcseconds, respectively. The magnitudes were corrected for Galactic extinction following the procedures outlined in \citet{CCM89}, with A(V)=0.355~mag and E(B-V)=0.107~mag \citep{SFD98}. We converted the magnitudes to fluxes using the central wavelengths for each filter as calibrated by \citet{POOLE08}.

The optical photometric data in {\it BVRI} bands were collected at various telescopes:
1) the 1.83~m Perkins telescope of Lowell Observatory (Flagstaff, AZ); 
2) the 1.54~m Kuiper
and 2.3~m Bok telescopes of Steward Observatory (Mt. Bigelow and Kitt Peak, AZ);
3) the 70~cm AZT-8 telescope of the Crimean Astrophysical Observatory (Nauchnij, Ukraine);  
4) the 40-cm LX-200 telescope of St.~Petersburg State University (St. Petersburg, Russia); 
5) the 2.2~m telescope of Calar Alto  Observatory (Almer\'ia, Spain);  
6) the 2~m Liverpool telescope of the Observatorio del Roque de Los Muchachos (Canary Island, Spain); 
7) the 1.25~m telescope of Abastumani Astrophysical Observatory (Mt. Kanobili, Georgia); 
8) the 60~cm telescope of the Maria Mitchell Association (Nantucket, MA); 
9) the 1.3m telescope at the Cerro Tololo Inter-American Observatory (CTIO), and 
10) the UVOT of {\it Swift}.    

Near-infrared, $JHK$, photometric data (near-IR) were collected at the 1.1~m telescope of the Main (Pulkovo) Astronomical Observatory of the Russian Academy of Sciences located at Campo Imperatore, Italy \citep{LAR08}. The optical and near-IR data were supplemented by measurements by the SMARTS consortium, posted at their website \citep{ERIN12}. 
The data have been corrected for Galactic extinction. The conversion factors calculated by \citet{MEAD90} were used to convert magnitudes into flux densities.

Mid-infrared (mid-IR) observations were carried out on 2010 November 3 at the NASA Infrared Telescope Facility (IRTF) with the MIRSI
camera \citep{KASSIS08}. The observations were performed in three bands centered at 4.9, 10.6, and 20.7~$\mu$m with total on-source 
integration times of 480, 720, and 960~s, and frame times of 200, 24, and 4~ms, respectively. The comparison star {\it 63 Peg} from the MIRAC3 Users's Manual was observed before and after each observation of the quasar to provide the flux density calibration. The data were reduced with an IDL script supplied by the IRTF staff. This resulted in flux measurements of 293$\pm$21, 699$\pm$37, and 1293$\pm$197~mJy at 4.9, 10.6, and 20.7~$\mu$m, respectively.  

Far-infrared (far-IR) photometric data were collected from 2010 December 25 to 2011 January 10 at 250, 350, and 500~$\mu$m with the SPIRE
photometer (13 measurements at each wavelength) and with the PACS photometer at 70 and 160 $\mu$m (15 measurements at each wavelength) on board the {\it Herschel} satellite. The details of the observations and data reduction along with a table of flux densities can be found in \citet{ANN12}. 

The 0.85~mm (350~GHz) and 1.3~mm (230~GHz) measurements were obtained at the Submillimeter
Array (SMA), Mauna Kea, Hawaii within a monitoring program of compact extragalactic radio sources that can be used as calibrators at mm and sub-mm wavelengths \citep{GUR07}. The data at 0.85~mm (47 data points) and 1.3~mm (215 data points) are supplemented by measurements at the IRAM telescope at 1.3~mm (22 data points) and 3.5~mm (28 data points). The data reduction procedure of the IRAM data can be found in \citet{IVAN10}. We will refer to the combined light curve at 1.3~mm as ``the 1~mm light curve''.

The 8~mm (37~GHz) observations were performed with the 13.7~m telescope at Mets\"ahovi Radio Observatory of Aalto University, Finland (307 measurements). The flux density calibration is based on observations of DR~21, with 3C~84 and 3C~274 used as secondary calibrators. A detailed description of the data reduction and analysis is given in \citet{FIN98}.

\begin{figure}
\plotone{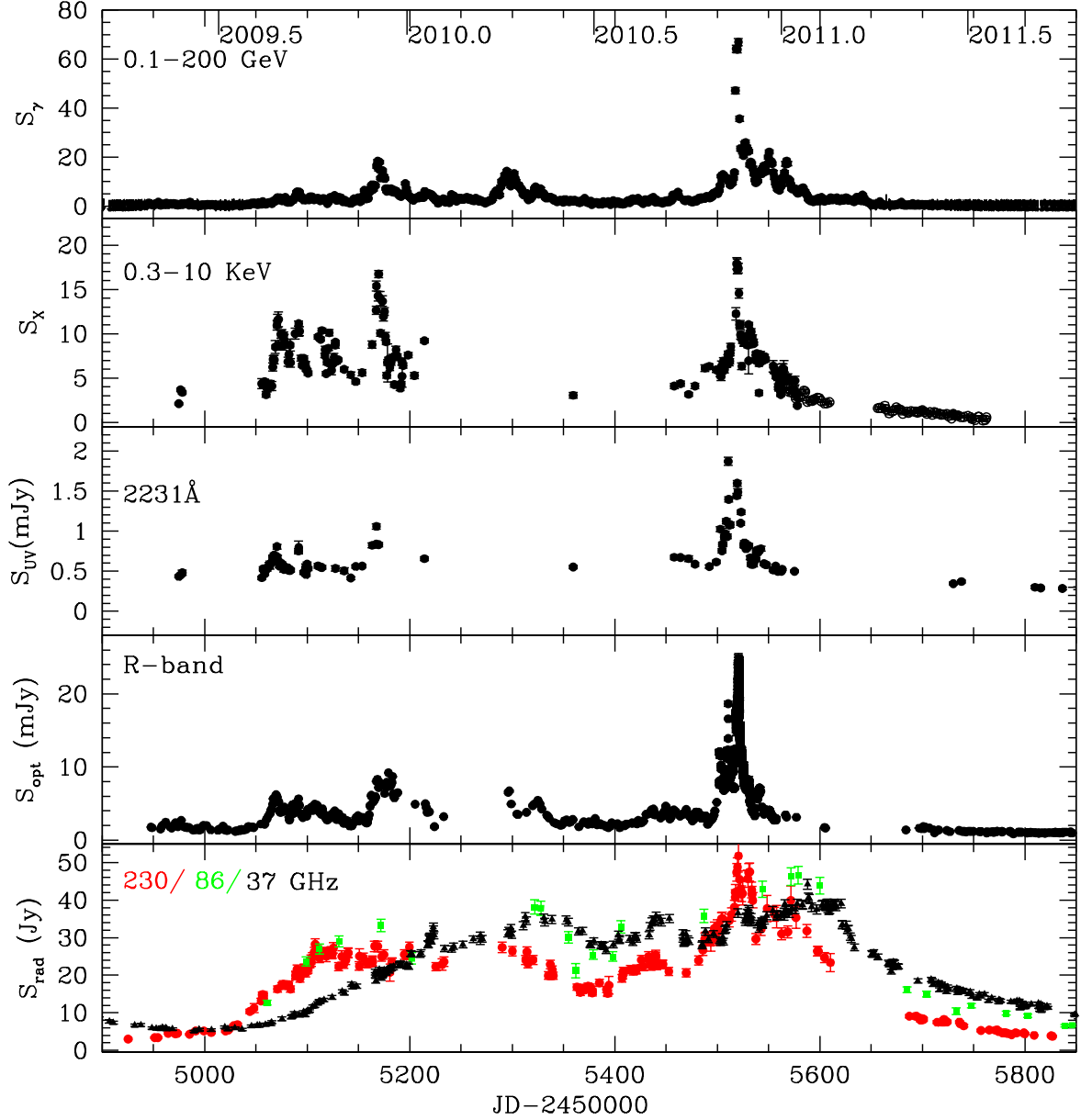}
\caption{Light curves of the quasar 3C~454.3 at different frequencies. From the top: 1) {\it Fermi} LAT $\gamma$-ray flux 
with 1~day binning interval in units of 10$^{-6}$~photon~cm$^{-2}$~$s^{-1}$; 2) {\it Swift} X-ray flux in units
of 10$^{-11}$~erg~cm$^{-2}$~$s^{-1}$; 3) UVOT flux measurements at 2231{\AA}; 4) optical light curve in R band; and 
5) flux densities at 230~GHz (1.3~mm, red circles), 86~GHz (3~mm, green squares), and 37~GHz (8mm, black triangles).} 
\label{mainLC}
\end{figure}

Figure ~\ref{mainLC}
shows the $\gamma$-ray, X-ray, UV, optical $R$ band, and radio light curves 
from 2009 April 25 to  2011 August 1 (RJD: 4947-5775). Simple visual inspection of the light curves reveals a prolonged, $\sim$600~day,
state of high activity at $\gamma$-rays, from 2009 August (RJD$\sim$5050) to 2011 March (RJD$\sim$5650),
that coincides with a high state seen in the 1~mm light curve. Within this active state three major $\gamma$-ray outbursts occurred, 
in 2009 December, 2010 April, and 2010 November. 

\subsection{Observations of Spectrum and Polarization}
The optical polarization measurements were performed at telescopes 1-5 as listed above 
in {\it R} band, except for LX-200 of St.~Petersburg State University, where the observations were carried out without a filter with the central wavelength $\lambda_{\rm eff}\sim 670$~nm, and the spectropolarimetric observations at Steward Observatory (see below). The observations at the Calar Alto Observatory were carried out 
within the MAPCAT (Monitoring AGN with Polarimetry at the Calar Alto
Telescopes)\footnote{http://www.iaa.es/$\sim$iagudo/research/MAPCAT/MAPCAT.html} program.
The details of optical polarization observations and data reduction can be found in \citet{J10}.

\begin{figure}
\plotone{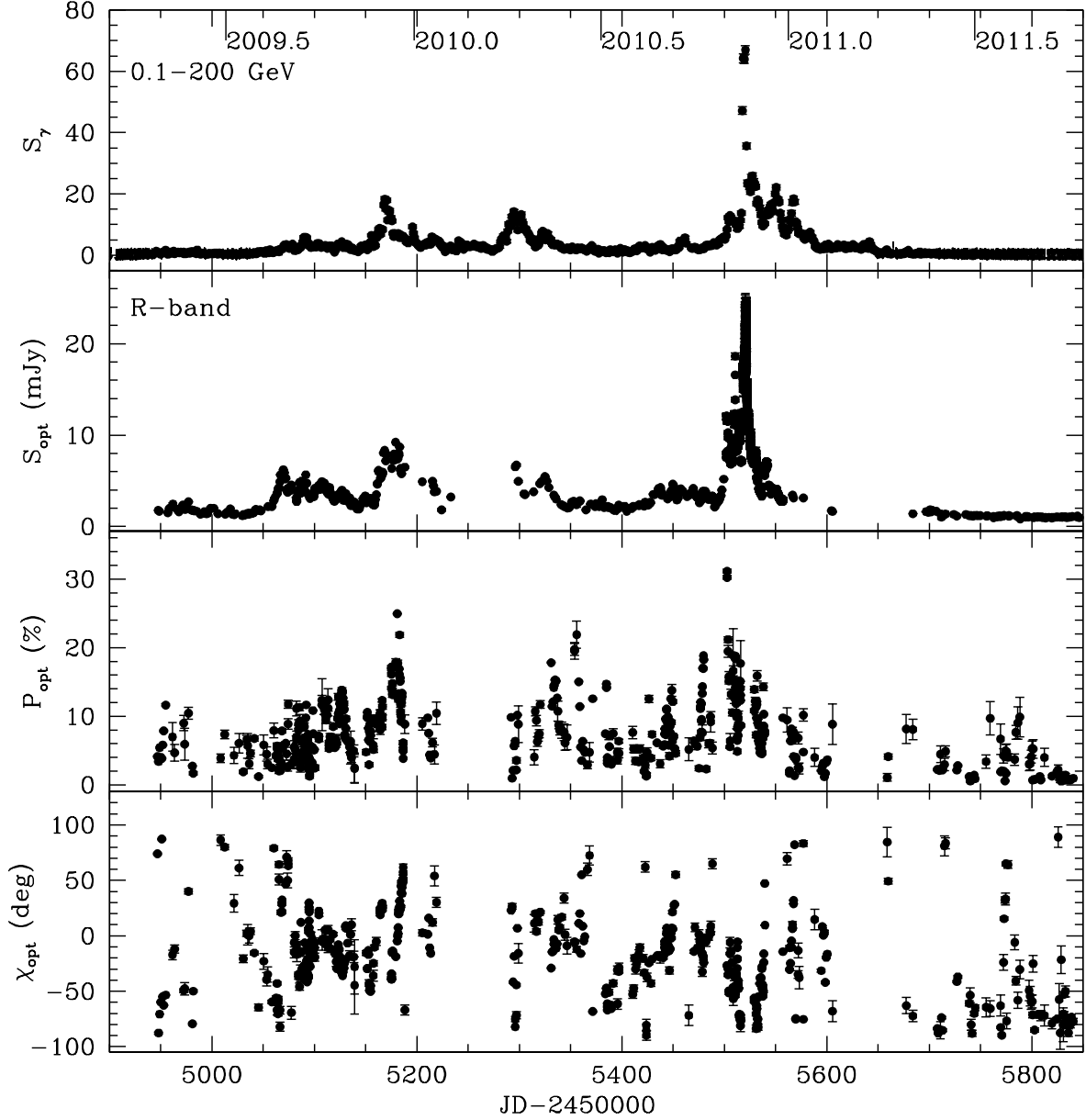}
\caption{Optical polarization curves of the quasar 3C~454.3 along with the $\gamma$-ray and optical light curves. From the top: 1) $\gamma$-ray light curve with 1~day binning interval in units of 10$^{-6}$~photon~cm$^{-2}$~s$^{-1}$; 2) optical light curve in R band;
3) degree of optical linear polarization; 4) position angle of optical polarization.} \label{mainPC}
\end{figure}

The spectropolarimetric observations of 3C~454.3 at Steward Observatory
were part of a currently operating program to monitor bright $\gamma$-ray blazars from the {\it Fermi} LAT-monitored blazar list\footnote{http://james.as.arizona.edu/$\sim$psmith/Fermi}. The observations were performed using the CCD Imaging/Spectro-polarimeter (SPOL;\citealt{SPOL}), yielding spectra that span the range of 4000-7550~{\AA} with a dispersion of 4~{\AA} per pixel. Depending on the width of the slit used for the observation, the resolution was typically between 16 and 24~{\AA}. The flux density averaged over 5400-5600~{\AA} was scaled to agree with that determined from the synthetic {\it V} band photometry performed on the same night. As a result, 181 calibrated spectra of the quasar were obtained during the period from 2009 April 25 to 2011 August 1.
In polarization mode, the full-resolution Stokes spectra were obtained to calculate the linear polarization parameters within 5000-7000~{\AA} (244 spectra). Details of the spectropolarimetric data reduction can be found in \citet{SMITH09}. The combined optical polarization data obtained from the telescopes used for this study consist of 523 measurements of the degree, $P$, and position angle, $\chi_{\rm opt}$, of polarization. The data are displayed in Figure \ref{mainPC}. 

Polarization observations of the quasar at 1.3 and 3~mm were obtained at the IRAM 30~m telescope 
within the MAPI\footnote{MAPI: Monitoring of AGN with Polarimetry at IRAM-30~m} and POLAMI\footnote{POLAMI: Polarimetric AGN Monitoring at the IRAM-30 m-Telescope} polarimetric programs. Each program performs monthly monitoring of a sample of $\gamma$-ray blazars, with both samples including 3C~454.3. The data were reduced in the same manner as described in \citet{IVAN10}. The values of $P$ at all wavelengths were corrected for statistical bias \citep{WK74}.

\subsection{VLBA Observations}
We observed 3C~454.3 with the VLBA in the course of a program
of monthly monitoring of bright $\gamma$-ray blazars  at 43~GHz (7~mm)\footnote{http://www.bu.edu/blazars/VLBAproject.html} and more dense monitoring during campaigns in 2009 October 12-25, 2010 April 7-20, 
and  2010 October 31 - November 13.  Within these campaigns, the quasar was observed three times. During the period from April 2009 to August 2011, we obtained 35 total
and polarized intensity images at a resolution of $\sim$0.3$\times$0.1~milliarcseconds (mas).
We performed the data reduction in the manner of \citet{J05} using the Astronomical Image Processing
System (AIPS) and Difmap \citep{DIF97}. 
The electric vector position angle
(EVPA) was calibrated by different methods. Over the period 2009 April -- 2009 
December we used the NRAO polarization data base\footnote{http://www.vla.nrao.edu/astro/calib/polar/} that
provides EVPAs at 43~GHz for several sources in our sample 
(0420$-$014, 0528+134, OJ287, 1156+295, 3C~279, BL~Lac, and 3C~454.3) obtained with the 
Very Large Array (VLA), which we compared with VLBA integrated EVPAs at simultaneous or nearly simultaneous epochs.
We obtained polarization measurements during the campaigns with the VLA 
on 2009 October 14 (sources: 0235+164, 0528+134, BL~Lac, and 3C~454.3)
and with the EVLA on 2010 April 10 (sources: 1156+295, 3C~279, 1308+326, and OT+081) and
on 2010 November 2 (sources:  0235+164, 0528+134, 0716+710, and OJ287). 
At epochs where VLA/EVLA data were not available, we used the D-terms method \citep{Dterm}.
The calibration was checked for consistency between epochs by comparing EVPAs of polarized jet features located $\ge$ 1~mas
from the core in 0528+134, 3C~273, 3C~345, CTA102, and BL~Lac that had stable EVPAs based on VLBA observations with
simultaneous VLA/EVLA observations. 
The accuracy of the EVPA calibration is within 5-10 degrees. We have corrected the EVPA values of the core 
using the most recent estimate of the Faraday rotation measure in the core region of 3C~454.3, 
$RM=1320\pm170$~rad~m$^{-2}$~\citep{AGS11}.
The accuracy of the flux density calibration as revealed by comparison
between the VLBA integrated flux and VLA/EVLA flux obtained at simultaneous epochs is within 5\%. 

\section{Structure and Timescales of the Outbursts} 
We define $\gamma$-ray outbursts using the criterion that the $\gamma$-ray flux, $S_\gamma$, calculated with an integration interval of 1~day within the energy range from 0.1 to 200~GeV,
must exceed 2$\times$10$^{-6}$~phot~cm$^{-2}$~s$^{-1}$ and never drop below this level during the event. Therefore,  
the duration of a $\gamma$-ray outburst is determined by a period when $S_\gamma>$~2$\times$10$^{-6}$~phot~cm$^{-2}$~s$^{-1}$. This criterion is arbitrary, however, it agrees with visual inspection of the light curve of 3C~454.3. It defines the three brightest $\gamma$-ray states of the quasar as follows: outburst I from 2009 November 9 to 2010 January 29 (RJD: 5145-5226), 
outburst II from  2010 March 21 to 2010 May 25 (RJD: 5277-5342), and outburst III from  2010 October 10 to 2011 January 30 (RJD: 5480-5592). 
We employ the same periods to analyze X-ray and optical outbursts. Unfortunately, during the main part of outburst II the quasar was too close to the Sun, resulting in very limited X-ray and optical observations for this event.

\subsection{Gamma-Ray Outbursts}
\begin{figure}
\plotone{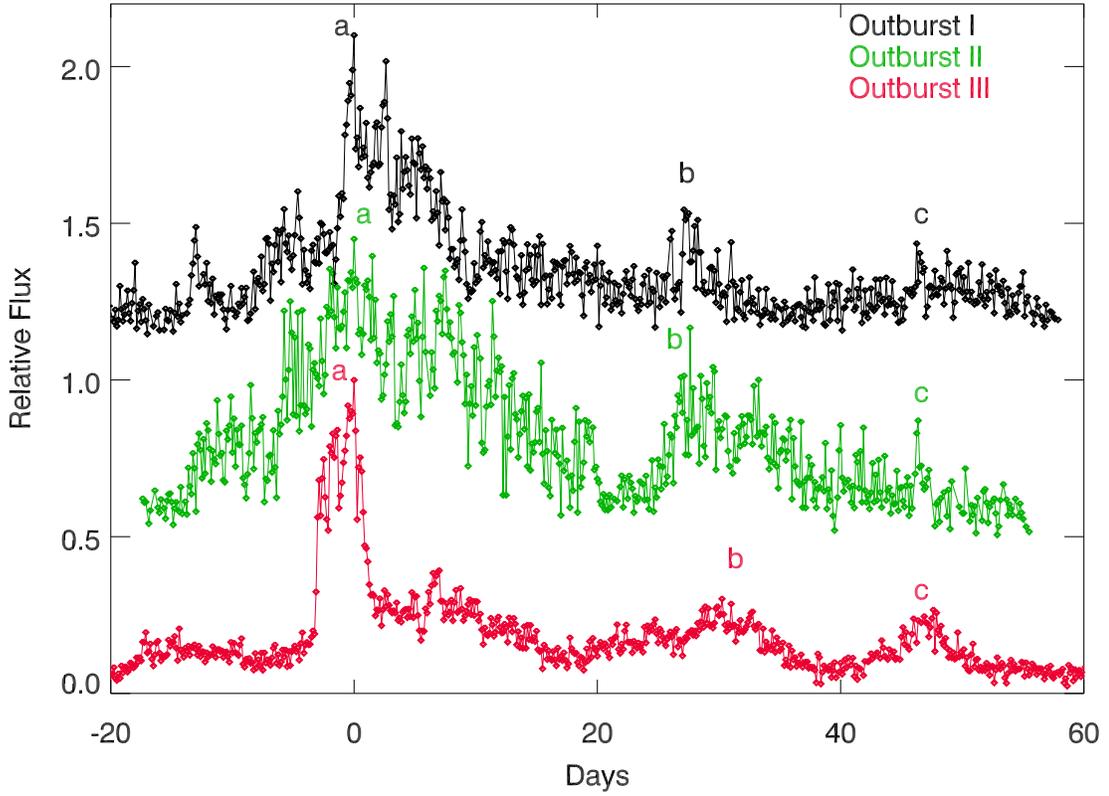}
\caption{Gamma-ray light curves during outbursts I (black), II (green), and III (red) relative to $T_\gamma^{\rm max}$ of each outburst and normalized to corresponding $S_\gamma^{\rm max}$; light curves I and II are shifted by 1.1 and 0.45, respectively, for clarity. The three main flares during each outburst are designated as $a$, $b$, and $c$ (see Table~\ref{Gparm}).} \label{3G}
\end{figure}

We have calculated $\gamma$-ray light curves with a 3~hr integration interval during outbursts I, II, and III using the same
approach as described in \S~2.1. This results in 646 (38), 526 (24), and 908 (39) measurements for outbursts I, II, and III, respectively, with the numbers in parentheses being non-detections. We have ignored the non-detections in our analysis since they represent a small fraction of the data. The light curves, presented in Figure~\ref{3G}, are normalized to the maximum flux density of each outburst, with time $t=0$ set to the date of the maximum. 
Figure~\ref{3G} shows that the structure of the outbursts is similar, although values of the maximum flux differ. We identify three flares, $a$, $b$, and $c$, within each outburst, separated by troughs with durations comparable to those of the flares. Peaks $b$ and $c$ have similar delays (within 0.5--3~days) with respect to the main peak of flare $a$. The primary difference in the profiles of the three outbursts is connected with the shape of flare $a$. We define the duration of the main flare, $a$, as the full width at half maximum (FWHM) of a Gaussian that fits the flare profile near maximum flux, $\Delta T_\gamma^{\rm a}\sim$11, 20, and 5 days for oubursts I, II, and III, respectively. All three outbursts have pre-flare and post-flare ``plateaus'' of enhanced $\gamma$-ray emission, as discussed by \citet{ABDO11}. We determine the duration of a plateau as
the time interval within which the standard deviation of the average $\gamma$-ray flux does not exceed 2$\sigma{S_{\rm ave}}$, where $\sigma{S_{\rm ave}}$ is the average uncertainty of individual measurements. The duration of the plateaus differs from outburst to outburst, although the duration of the pre-flare plateau, $\Delta T_\gamma^{\rm pre}$, is almost equal to the duration of the post-flare plateau, $\Delta T_\gamma^{\rm post}$, for each outburst. In addition, the flux levels of the pre- and post-flare plateaus are comparable, except for outburst III. The entire duration of flare $a$, which includes $\Delta T_\gamma^{\rm pre}$, $\Delta T_\gamma^{\rm a}$, and  $\Delta T_\gamma^{\rm post}$, is comparable for all outbursts (28, 33, and 31~days for outbursts I, II, and III, respectively), as is the period between the peaks of flares $a$ and $c$, equal to 46, 46, and 48~days for outbursts I, II, and III, respectively. 
{\it The similarity in structure of the $\gamma$-ray outbursts argues in favor of the same mechanism(s) and location of $\gamma$-ray production for all three events.} \citet{J10} have previously reported a triple flare structure of optical outbursts in 3C~454.3 that coincide with the time of passage of superluminal knots through the mm-wave core of the jet. The measured time interval between the first and third peaks of these earlier events was $\sim 50$~days, which is only slightly longer than the interval between the peaks of flares $a$ and $c$ observed for $\gamma$-ray outbursts I, II, and III. Parameters of the $\gamma$-ray outbursts studied here are given in Table~\ref{Gparm}.

We have determined timescales of $\gamma$-ray flux variability, $\tau_\gamma$,
using the formalism suggested by \citet{BJO74}:  $\tau \equiv \Delta t/\ln\;(S_2/S_1)$, where $S_{\rm i}$ is the flux density at epoch $t_{\rm i}$, with $S_2 > S_1$, and $\Delta t = |t_2-t_1|$. We have calculated the timescale of variability for all possible pairs of flux measurements within 3~days of each other if, for a given pair, $S_2-S_1>3({\sigma}S_1+{\sigma}S_2)/2$, where ${\sigma}S_{\rm i}$ is the uncertainty of an individual measurement, and if the test statistic of the $\gamma$-ray measurement $TS>25$ for both measurements.
Using the derived values of $\tau_\gamma$, we have searched for a minimum timescale of variability among pairs.  Table~\ref{Gparm} gives minimum timescales of $\sim$3-4~hr with the $\gamma$-ray flux changing by a
a factor of $\ge$2. Table~\ref{Gparm} shows that a short timescale of variability can occur at different stages of an outburst, although the occurrence of $\tau_\gamma^{\rm min}$ takes place during the pre-flare plateau for all three outbursts. The range is consistent with the minimum timescales of variability reported by \citet{ACKER10}, and \citet{ABDO11}, as well as \citet{FOS11}, who apply different methods for estimation of $\tau_\gamma$. Note that \citet{FOS11} have calculated $\gamma$-ray light curves  using a time bin equal to the GTI ({\it good time interval}). Such a method should produce the most accurate flux estimates at short timescales for observations performed in scanning mode, since it allows one to find shortest intervals for binning with a sufficient number of photons for a good statistic. Agreement between
our results suggests that  $\tau_\gamma^{\rm min}\sim$3-5~hrs might be an upper limit for the minimum timescale of variability defined by the GTI. In addition, \citet{FOS11} report the time of the global peak of outburst III to be RJD: 5520.573 - 5520.627, which matches $T_\gamma^{\rm max}$ very well. We have also analyzed the distribution of $\tau_\gamma$ values that fall within the range 0$-$72~hr to determine
a typical timescale, $\tau_{\gamma,2}$, for the $\gamma$-ray emission to change by a factor of $\ge$ 2. Table~\ref{Gparm} shows that this timescale of variability is similar for all three outbursts, $\tau_{\gamma,2}$=20$\pm$1~hr. 

\subsection{X-Ray Outbursts}
\begin{figure}
\plotone{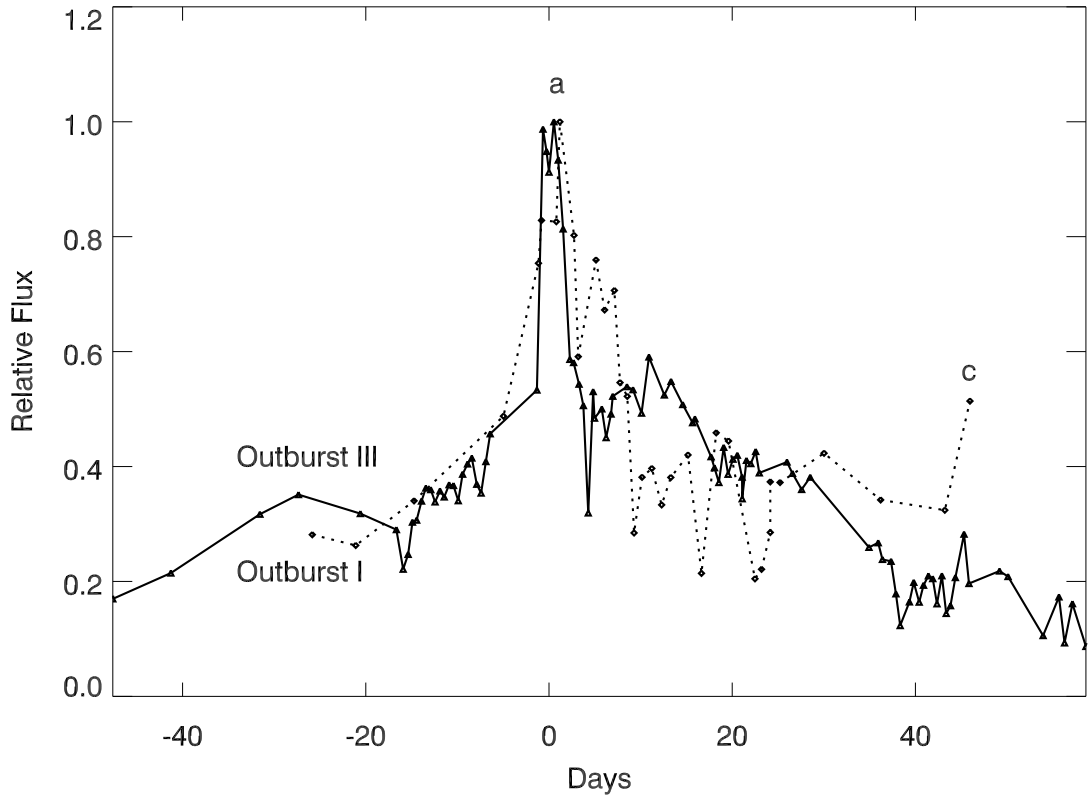}
\caption{ X-ray light curves during outbursts I (diamonds, dotted line) and III (triangles, solid line) normalized to the corresponding maximum and centered with respect to the corresponding peak of $\gamma$-ray outburst.}
\label{2X}
\end{figure}

Figure~\ref{2X} displays the X-ray light curves for outbursts I and III normalized to the maximum flux density of each X-ray outburst and centered with respect to the date of the maximum of corresponding $\gamma$-ray outburst. Table \ref{Xparm} lists the parameters of the X-ray outbursts. Although the X-ray data are much more sparsely sampled than the $\gamma$-ray light curves, the global X-ray and
$\gamma$-ray peaks of outbursts I and III coincide within $\sim$1~day, with the $\gamma$-ray peak of outburst I leading by $\sim$1.2~day while the $\gamma$-ray peak of outburst III is delayed by $\sim$1.0~day. 
 In addition, the duration of the main X-ray event
in flare $a$ is similar to $\Delta T_\gamma^{\rm a}$ for both outbursts, and flare $a$ of outburst III has pre-flare and post-flare plateaus contemporaneous with their $\gamma$-ray counterparts. There are also indications of the presence of a post-flare plateau and flare $c$ in outburst I (the durations of the plateaus were determined in the same manner as for the $\gamma$-ray events). The main difference between the X-ray and $\gamma$-ray outbursts is the timescale of variability $\tau_{\rm X}$, calculated in the same manner as $\tau_\gamma$, except for the condition for the test statistic. The fastest events were observed when the X-ray flux changed by a factor of 1.8 in 27~hrs, which corresponds to $\tau_{\rm X}^{\rm min}\simeq 6\tau_{\gamma}^{\rm min}$, and  the typical flux doubling timescale is $\sim$2~days,  which gives $\tau_{\rm X,2}\simeq 2\tau_{\gamma,2}$

\subsection{Optical Outbursts}
\begin{figure}
\plotone{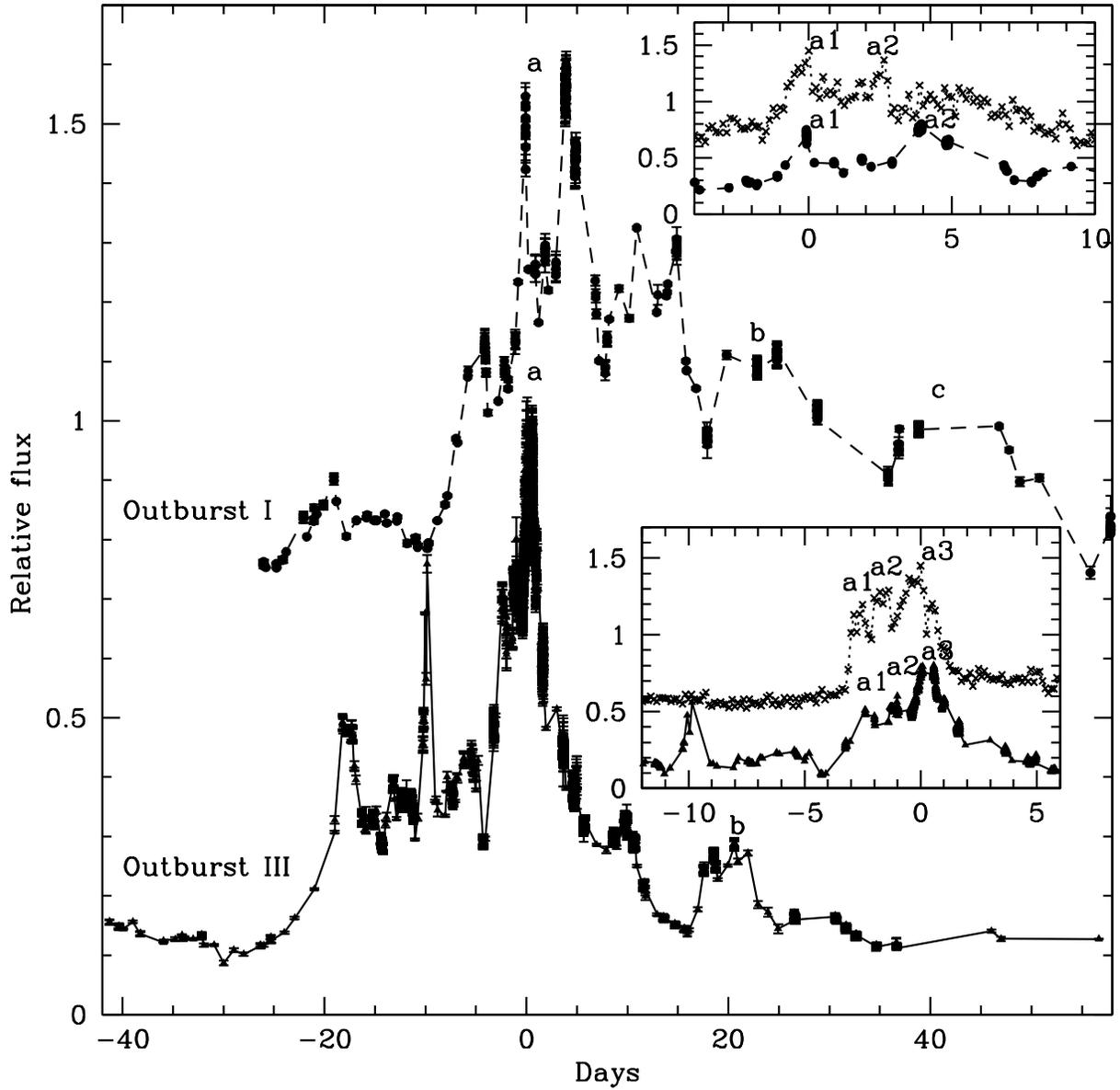}
\caption{ Optical $R$-band light curves during outbursts I (circles, dash line) and III (triangles, solid line) normalized to the corresponding maximum and centered with respect to the corresponding peak of $\gamma$-ray outbursts.
The three main flares during each outburst are designated as $a$, $b$, and $c$ (see Table \ref{Optparm});
optical light curves I and III are shifted by 0.6 and -0.2, respectively, for clarity.
The top insert shows the structure of flare $a$ at optical (circles, dash line) and $\gamma$-ray (crosses, dotted line) wavelengths for
outburst I; the bottom insert shows the structure of flare $a$ at optical (triangles, solid line) and $\gamma$-ray (crosses, dotted line) wavelengths for outburst III; the $\gamma$-ray fluxes are calculated with a 3~hr binning interval.} \label{2GO}
\end{figure}

We perform the same analysis of the structure of the $R$ band optical light curves (Fig.~\ref{2GO} and Table~\ref{Optparm}) during outbursts I, II, and III as for the $\gamma$-ray and X-ray light curves. Figure~\ref{2GO} shows that the two well-sampled optical outbursts, I and III, have a complex structure 
of the main flare, $a$: flare $a$ of outburst I has 2 peaks, $a1$ and $a2$ (Fig.~\ref{2GO}, the top insert)
and flare $a$ of outburst III has at least three peaks, $a1$, $a2$, and $a3$ (the bottom insert in Fig.~\ref{2GO}). Remarkably, flare $a$ at $\gamma$-ray energies possesses similar structure for both outbursts, with the global $\gamma$-ray peak ($a1$ for outburst I and $a3$ for outburst III) coinciding with a prominent optical counterpart within the 3~hr $\gamma$-ray sampling, although the ratio of the fluxes of the $\gamma$-ray peaks can be different from those at optical wavelengths. The similarity in structure of flare $a$ at optical and $\gamma$-ray frequencies implies that the flaring emission at the two wavelengths originates in the same region. The difference between the relative amplitudes of $\gamma$-ray vs. optical peaks can be explained as the result of differences in relativistic boosting of $\gamma$-ray and optical emission, as proposed by \citet{RAI11}, or by variations in the density of seed photons available for scattering to $\gamma$-ray energies, as suggested by \citet{verc11}. The latter is additionally supported by existence of orphan optical outbursts, for example, a very sharp spike at $\sim$10~days before the maximum  when the fastest optical variability during outburst III was observed (Table~\ref{Optparm}) without an obvious counterpart in the $\gamma$-ray light curve (Figure~\ref{2GO}, the bottom insert).  

Figure~\ref{2GO} and Table~\ref{Optparm} show that optical outbursts I and III have pre-flare and post-flare plateaus 
that are contemporaneous with the corresponding $\gamma$-ray plateaus, although the relative flux level of the pre-flare plateau is higher with respect to the global maximum at optical wavelengths than at $\gamma$-ray energies. The durations of the optical plateaus are similar to the $\gamma$-ray values. The durations of the plateaus in $R$ band were determined by the criterion that the flux variations within a plateau should not exceed 30\% of the average flux value. This criterion is similar to that used for the $\gamma$-ray data analysis, since the average 1$\sigma$ uncertainty of a $\gamma$-ray measurement is $\sim$17\% (Table~\ref{Gparm}) while the 1$\sigma$ uncertainty of an optical flux is $\sim$2\% (Table~\ref{Optparm}). Note that the duration of the pre-flare plateau of outburst III is a factor of 2 shorter than $\Delta T_{\gamma}^{\rm pre}$. However, $\Delta T_{\rm opt}^{\rm pre}$ would match $\Delta T_{\gamma}^{\rm pre}$ if the
pre-flare plateau were not interrupted by the orphan optical flare mentioned above. For outbursts I and III the entire optical flare $a$ ($\Delta T_{\rm opt}^{\rm a}$+$\Delta T_{\rm opt}^{\rm pre}$+$\Delta T_{\rm opt}^{\rm post}$) has a similar duration as its $\gamma$-ray counterpart. The post-flare optical variability does not correspond as closely to the $\gamma$-ray variations as during the pre-flare and main flare stages. Nevertheless, optical outbursts I and II have counterparts to $\gamma$-ray flares $b$ and $c$ (see Table \ref{Optparm} and Fig.~\ref{2GO}) that peak within 0.5-5~days of the corresponding $\gamma$-ray flares. Flare $b$ is distinct during outburst III as well, although it precedes the $\gamma$-ray flare $b$ by $\sim$9~days.

Table~\ref{Optparm} shows that the minimum timescale of variability is $\sim$18-24~hr with the optical 
flux changing by a factor of 1.5-2.5 (see also \citealt{RAI11,verc11}). Comparison of Tables~\ref{Gparm},~\ref{Xparm}, and \ref{Optparm} reveals that the timescale of optical variability is different from 
$\tau_\gamma$ and similar to $\tau_{\rm X}$, which is longer by a factor of 5 than the minimum timescale of the $\gamma$-ray flux. Note that during flare $a$ of outburst III (from RJD: 5500 to 5540) we obtained $\sim$2400 measurements in $R$ band, which is suitable for revealing a timescale of variability as short as $<$1~hr. The typical doubling timescale is also different for $\gamma$-ray and optical variations, with $\tau_{\rm opt,2}\approx (2-3)\tau_{\gamma,2}$. 

\subsection{Millimeter-Wave Outbursts}
\begin{figure}
\plotone{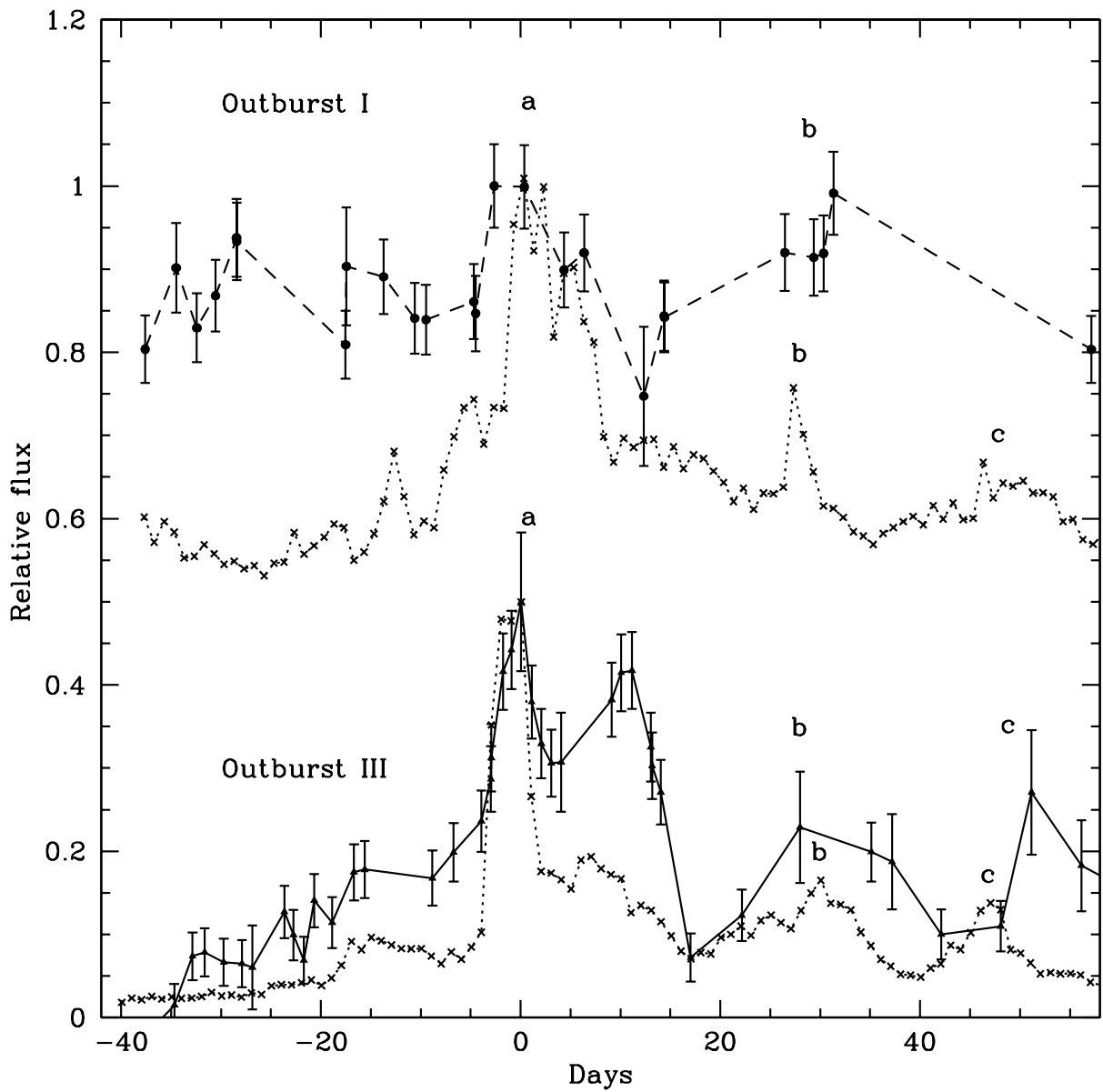}
\caption{Light curves at 1~mm during outbursts I (circles, dash line)
and III (triangles, solid line), normalized to the corresponding maximum
and superposed with the corresponding $\gamma$-ray light curves (crosses,
dotted lines). The $\gamma$-ray light curves
are normalized to twice the value of the corresponding maximum. All
light curves are centered with respect to the corresponding peak of the
$\gamma$-ray outbursts. The three main flares during each outburst are
designated as $a$, $b$, and $c$. For clarity, the $\gamma$-ray light
curve during outburst I is shifted by $+$0.5, while during outburst III
the 1mm light curve is shifted by $-$0.5.
The $\gamma$-ray fluxes are calculated with a 1~day binning interval.}
\label{2mm}
\end{figure}

Parameters of outbursts I and III at 1~mm are presented in
Table~\ref{mmparm}.  Figure~\ref{2mm} shows the structure of outbursts I
and III with  respect to $T_\gamma^{\rm max}$ of the corresponding
$\gamma$-ray outburst. (Unfortunately, as in the case of
the X-ray and optical light curves, observations at 1~mm miss outburst
II.) Strikingly, the global peaks of the mm-wave and $\gamma$-ray
outbursts coincide within hours for both events,
while during the dramatic outburst in 2005 the global peak at 1~mm
was delayed with respect to that at optical wavelengths by
$\sim$2~months \citep{RAI08,J10}. Moreover, the duration of
flare $a$ is similar at mm-waves and $\gamma$-rays for both outbursts,
and pre-flare and post-flare plateaus are apparent for outburst III. In
addition, both outbursts at 1~mm contain flare $b$, which coincides with
the corresponding flare $b$ at $\gamma$-rays within 2~days, and flare $c$
is seen in the 1~mm light curve of outburst III only 3.5~day later than
$\gamma$-ray flare $c$. Taking into account the dramatic difference in
the opacity at $\gamma$ and mm wavelengths, such similarity requires the
$\gamma$-ray events to take place in a region that is optically thin at 1~mm.
The main differences between the $\gamma$-ray and mm-wave events are
connected with the amplitude and timescale of variability. According to
Tables~\ref{Gparm} and \ref{mmparm}, the size of the emission region at
1~mm is $\sim$100-300 times larger than that at $\gamma$-rays.
  
\subsection{Correlation Analysis \label{sDCF}}
\begin{figure}
\epsscale{1.0}
\plottwo{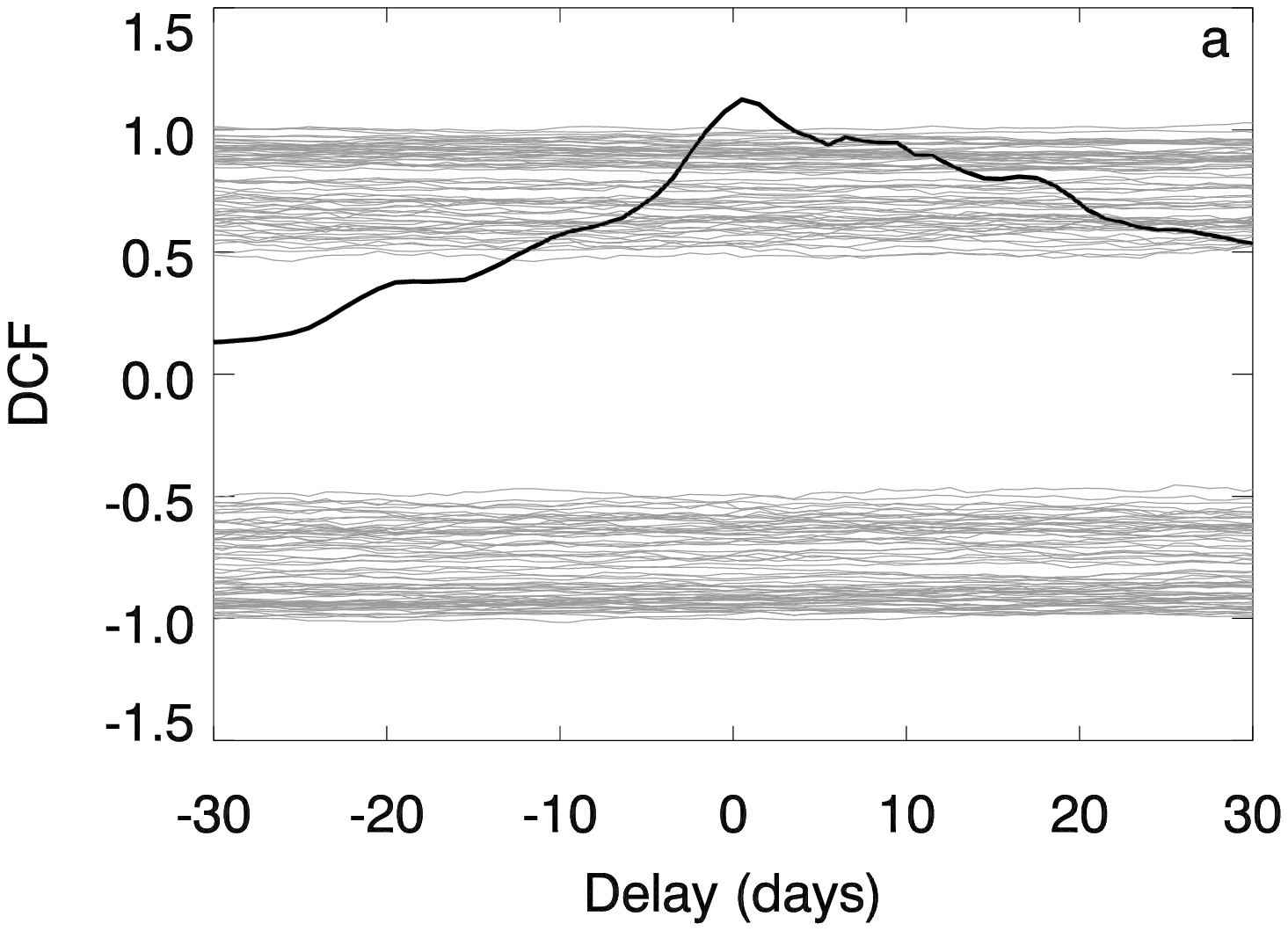}{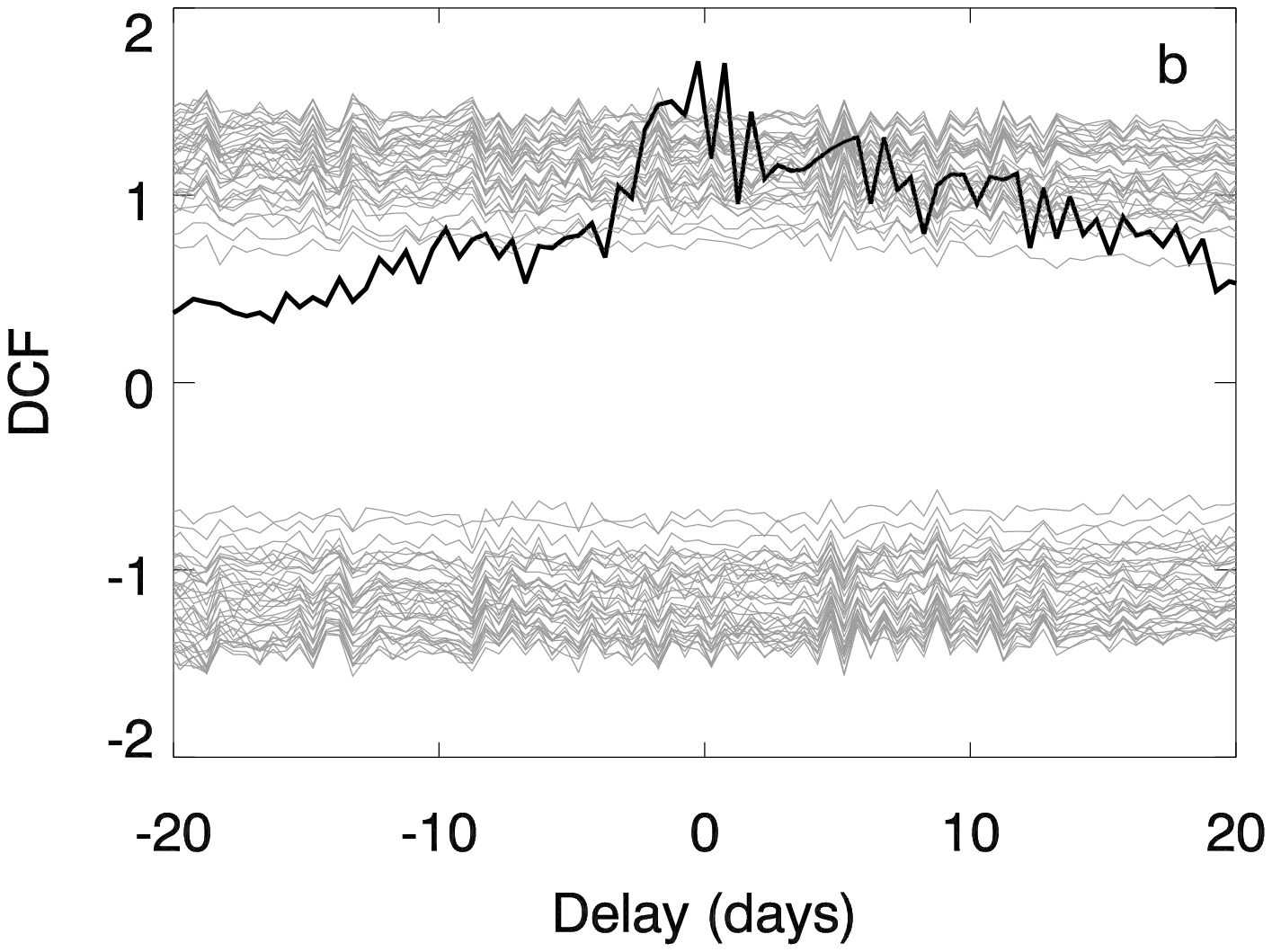}
\caption{{\it Left:} Discrete cross-correlation function (DCF) between the $\gamma$-ray and optical light curves (black); the gray curves at positive (negative) DCF values denote 99.7\% confidence limits relative to stochastic variability for different combinations of the PSD slope $b$ (see \S 3.3). {\it Right:} DCF between the X-ray and optical light curves (black). Negative delay means that high-energy flux variations lead those at optical wavelengths.} \label{DCF}
\end{figure}

We perform a discrete cross-correlation analysis between the $\gamma$-ray and optical light curves and between the X-ray and optical light curves. For the purpose of this analysis, we construct a $\gamma$-ray light curve with an integration time of 12~hr in the same manner as described in \S~2.1. We use the original sampling of the X-ray light curve, which corresponds to a minimum time interval between two measurements of $\sim$12~hr, and bin the optical light curve with a 12~hr minimum interval, although the light curves have gaps ranging from days to months. We calculate the discrete cross-correlation function (DCF) using the algorithm developed by \citet{EK88}, and determine the significance of the correlation with the approach suggested by \citet{RITA08} and \citet{MM10}. We follow \citet{TK95} by simulating 5000 light curves with the same mean and standard deviation as the observed light curves. The statistics of the flux variations are described by a power spectral density, $PSD$, with a power-law shape, $PSD\propto f^{-b}$, where a different value of $b$ from 1 to 2.5, in steps of 0.2, is adopted for each set of simulations \citep{RITA12}. Figure~\ref{DCF} shows the DCF between the $\gamma$-ray and optical light curves ({\it left}) and between the X-ray and  optical light curves ({\it right}). The DCF between the $\gamma$-ray and optical light curves is symmetric within $\pm 2$ days of the peak, with no delay between variations at two wavelengths
$>$12~hr, since such a delay would produce, at least, an asymmetry in the DCF peak. The peak is significant at a level $>99.7$\%. The DCF between the X-ray and optical light curves has a peak near zero delay as well; however, the position of the maximum of the centroid calculated for points exceeding 99.7\% significance gives a delay of the X-ray with respect to optical variations of 0.5$\pm$1~day. This result is consistent with the finding of \citet{RAI11} that the X-rays lag the optical flux variations by 1.0$\pm$1.0~days during the period 2008-2009. 

A correlation analysis between the $\gamma$-ray light curve from 2008 August 5 to 2011 October 21
and 1~mm light curve from 2008 January 12 to 2011 October 27 was performed by \citet{ANN12}. These authors
found a significant correlation between variations at the two wavelengths for delays from $-$1.5 to $+$3.5~day,
which suggests that mm-wave variations are either simultaneous with $\gamma$-ray variations or slightly
precede the latter. 

Our analysis of the multi-frequency light curves therefore reveals a strong similarity in the general structure of contemporaneous $\gamma$-ray, X-ray, optical, and mm-wave outbursts and a statistically significant
correlation between variations at different wavelengths. This suggests that:
\begin{enumerate}
\item{The same ensemble of relativistic electrons that produces variable synchrotron optical emission participates in the production of variable $\gamma$-ray emission. This follows from the lack of significant time delays between the variations at these two wavebands. In fact, for values of the magnetic field typically
inferred in the flaring regions of jets, $\sim$0.1-1 G, essentially the same energies of electrons are involved in both optical and $\gamma$-ray emission.}
\item{Given the above similarity in electron energies, the existence of orphan optical flares implies that either the density of seed photons changes with time, or changes in the strength and/or direction of the magnetic field cause such flares.}
\item{The emission regions at all four wavelengths are at least partially co-spatial, with the $\gamma$-ray emission exhibiting the fastest variability, $\tau_\gamma^{\rm min}\approx 1/5\tau_{\rm opt}^{\rm min}$, while $\tau_{\rm opt}^{\rm min}\approx\tau_{\rm X}^{\rm min}$ and $\tau_{\rm mm}^{\rm min}\approx 30\tau_{\rm opt}^{\rm min}$. Either the $\gamma$-ray flux is more sensitive to changes in the physical parameters than is the optical/X-ray flux, or the $\gamma$-ray emitting plasma fills $\sim 1/5$ of the optical/X-ray emission region. The similar X-ray and optical timescales of variability suggest that the corresponding emission regions are fully co-spatial. Although the mm-wave emission region is $\sim$100 times larger than that at $\gamma$-rays, a strict correspondence between the global maxima
of the $\gamma$-ray and mm-wave outbursts implies that the mm-wave region has sub-structures of different  
sizes, e.g.  $\sim$0.01~pc (the size of a turbulent cell within which the magnetic field is considered to be uniform \citealt{ANN12}), $\le$0.4~pc (the size of the mm-wave core, see \S~\ref{jet}), and $\sim$1~pc (the size of a superluminal knot), with the most compact and variable features co-spatial with the $\gamma$-ray emission region.}
\item{The delay of X-ray with respect to optical flux variations, plus the similarity of the timescales of variability, imply a delay in the arrival of seed photons before they are scattered to X-ray energies. This favors a synchrotron origin of the seed photons from a location near to, but not coincident with, the scattering electrons. Such a situation can occur in the synchrotron self-Compton (SSC) mechanism for X-ray production, since there is a light-travel delay of synchrotron seed photons from the flare as they cross the flaring region if the angle to the line of sight is close to zero \citep{SMM04}, as in 3C~454.3 ($\Theta_\circ\sim 0.3^\circ-1.3^\circ$, see \S~\ref{jet}).}
\item{The dramatic difference in the amplitudes of the $\gamma$-ray and optical outbursts, the presence of orphan flares, and the tight correlation between $\gamma$-ray and optical variations without a significant delay favor the process of scattering of external photons by relativistic electrons that produce synchrotron optical emission up to $\gamma$-rays (the external Compton mechanism, EC) as 
the main mechanism for production of the $\gamma$-ray emission during the outbursts. However, a contribution from SSC cannot be avoided, since synchrotron seed photons are also produced by the relativistic electrons involved in the $\gamma$-ray production.}
\end{enumerate}

\section{Behavior of the Parsec Scale Jet \label{jet}} 
\begin{figure}
\epsscale{0.40}
\plotone{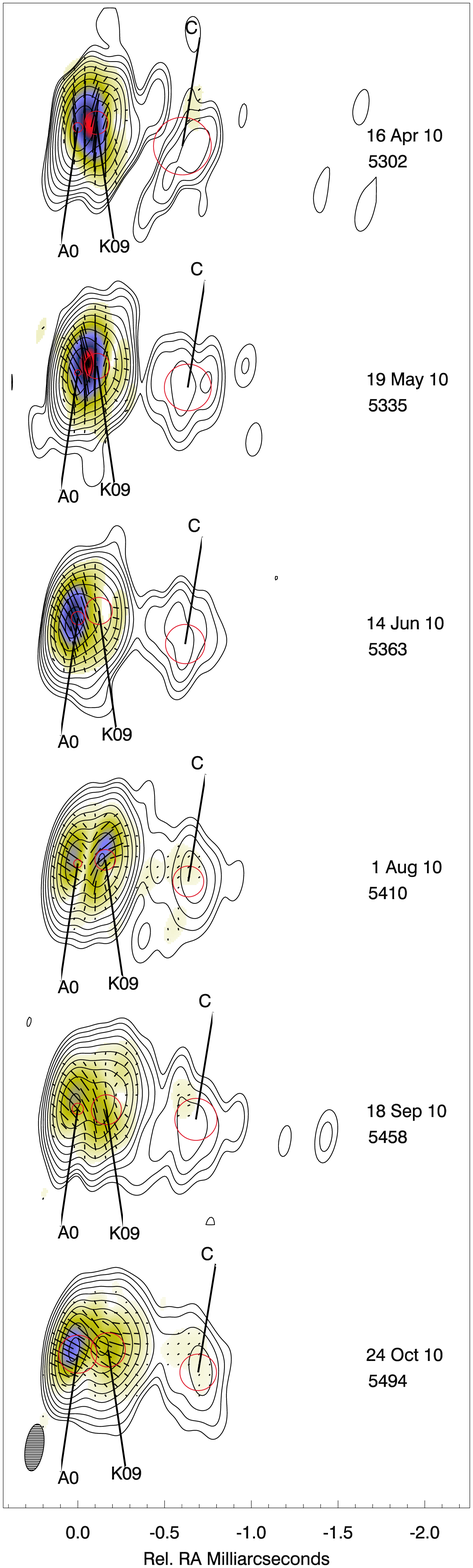}
\caption{43 GHz total (contours) and polarized (color scale) intensity images of 3C~454.3
with $I_{\rm peak}$=19.80~Jy/beam,  $I_{\rm peak}^{\rm pol}$=0.80~Jy/beam, and a Gaussian restoring beam=0.14$\times$0.33~mas$^2$ at $PA$=-10$^\circ$;
contours represent 0.1, 0.2,..., 51.2, 99.5\% of the peak intensity; line segments within the image show direction of linear polarization; red circles indicate position and size (FWHM) of components according to model fits.} \label{mapsK09}
\end{figure}

\begin{figure}
\epsscale{1.0}
\plotone{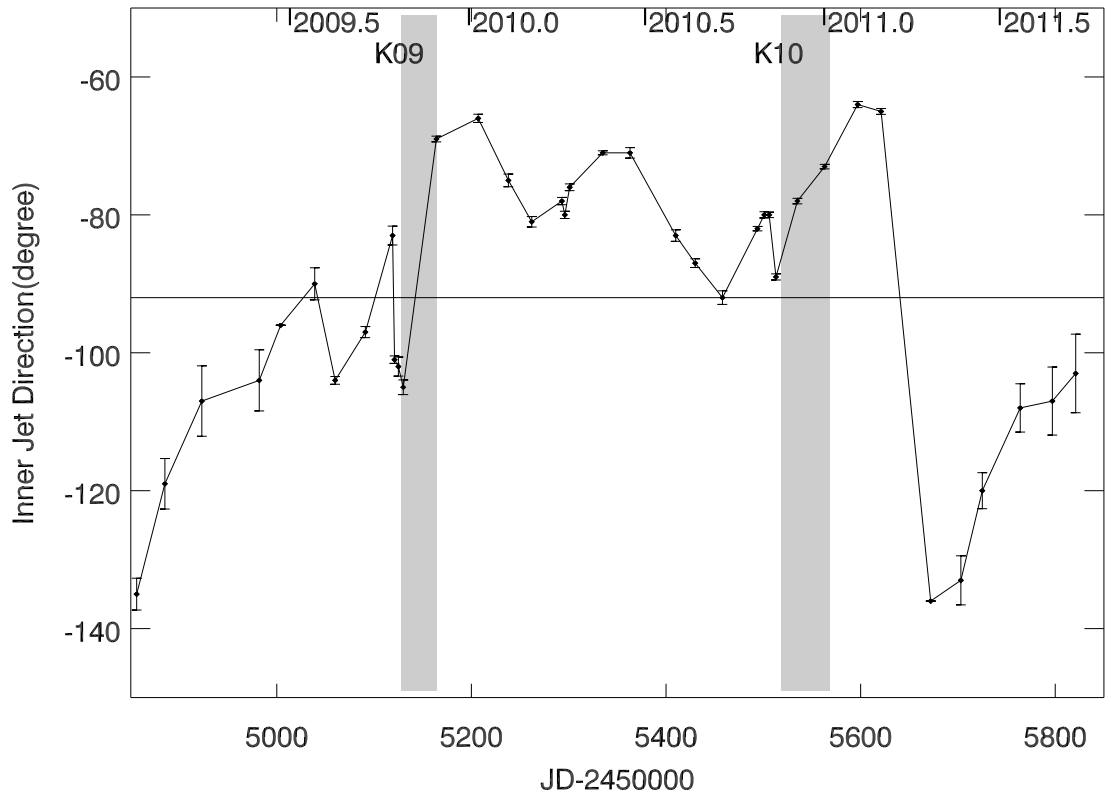}
\caption{The inner jet (within 0.2 mas of the core) direction vs. time. Horizontal solid line indicates the average direction of the jet, $-$92$\pm$20~deg. The vertical gray areas mark times of passage of knots $K09$ and $K10$ through the VLBI core, as derived from the kinematics of the knots.} \label{JetDir}
\end{figure}

 Figures~\ref{mapsK09} \& \ref{mapsK10} present sequences of 43 GHz total and polarized intensity VLBA images of 3C~454.3. We use these sequences to follow changes in the flux density and polarization of the core, as well as the appearance, motion, and evolution of new features in the jet. The total intensity images are modelled by components with circular Gaussian brightness distributions in the same manner as described in \citet{J05}. At each epoch we identify the core, $A0$, as the stationary feature located at the eastern end of the jet. The average angular size
of the core in the model fits is $<a_{\rm core}>=0.05\pm0.02$~mas ($\sim$0.4~pc).  
Parameters of components include flux density $S$ (in Jy), separation $r$ (in mas) and position angle $\Theta$ (in degrees) relative to the centroid of the core, and FWHM size $a$ (in mas).
We compute the degree $P$ and position angle $\chi$ of linear polarization of components
using an IDL program that calculates the mean values of the pixels at the position of each total intensity component and within an area equal to that of the size established by the modeling. All parameters of components are used to identify features across epochs in order to analyze the evolution of the jet. We define the inner direction of the jet $\Theta_{\rm jet}$ equal to the value of $\Theta$ of the brightest knot within 0.1-0.3~mas of the core. Figure~\ref{JetDir} plots $\Theta_{\rm jet}$, as well as the value of the average direction of the inner jet $<\Theta_{\rm jet}>$=-92$\pm$20~deg, vs.\ time. We find that the inner jet is oriented in the same direction as observed in 2004-2008, $<\Theta_{\rm jet}>$=-95$\pm$8~deg \citep{J10}, although the standard deviation of $<\Theta_{\rm jet}>$ indicates that the jet executed greater swings during 2009-2011 than reported previously. This is likely the result of a smaller viewing angle of the jet during 2009-2011, which amplified changes in the angle of the jet as projected on the sky plane.   
 
\subsection{Kinematics of the Parsec Scale Jet} 
\begin{figure}
\epsscale{1.0}
\plotone{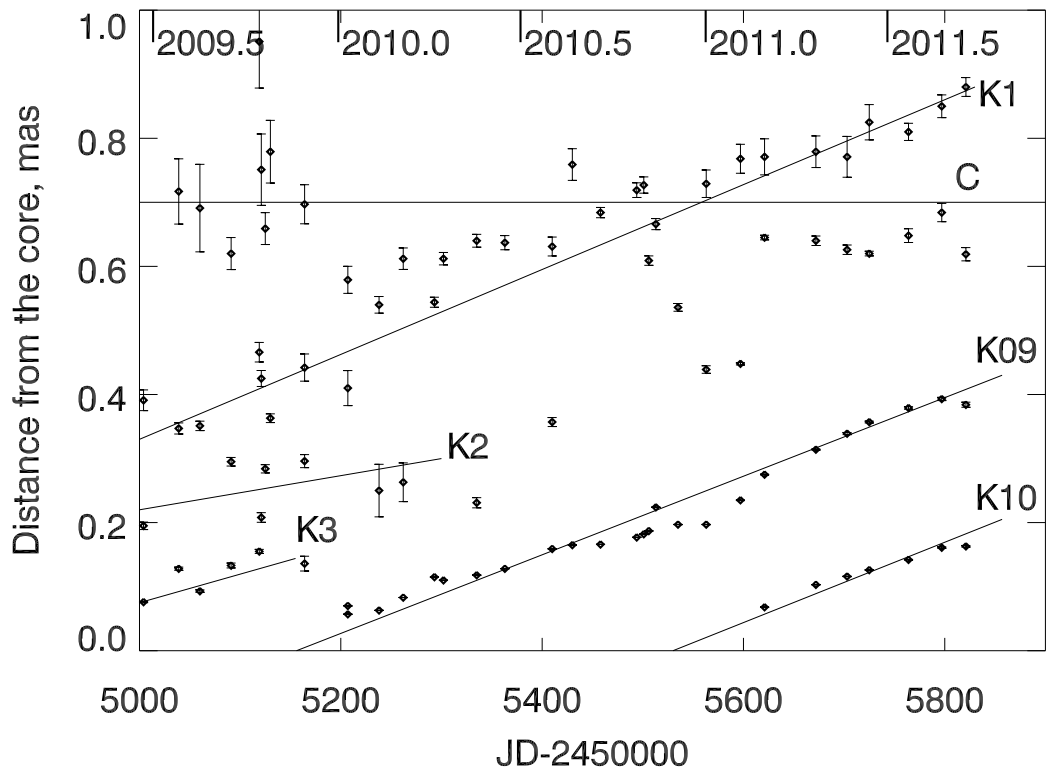}
\caption{Distance of components from the core within 1~mas of the core based on the model fitting. } \label{evol}
\end{figure}

 Figure~\ref{evol} shows
the results of modeling of the total intensity images within 1~mas of the core. We have identified features 
that can be associated with moving knots $K1$, $K2$, and $K3$, as well as quasi-stationary feature $C$ identified in \citet{J10}. 
In comparison with the results reported in \citet{J10}, knot $K1$ appears to have accelerated by a factor of $\sim$2, although at a number of epochs the knot is confused with either $K2$ or $C$.
Knot $K2$ moves with the same slow apparent speed, $\sim$3$c$, as seen during 
the later epochs analyzed in \citet{J10}. Although knot $K3$ has faded dramatically (S$_{\rm 7mm}\sim$0.3~Jy), it has the same speed $\sim$4~$c$ as reported previously. After the appearance of the new, very bright feature $K09$ at the end of 2009, 
knots $K2$ and $K3$ became too weak to be detected with a dynamic range of $\sim1500$:1. Knot $K09$ was as bright as the core in the beginning
of 2010  and even dominated the flux of the parsec scale jet in 2010 Summer. According to the modeling, knot $K10$ appeared to be ejected at the end of 2010.

\begin{figure}
\epsscale{1}
\plotone{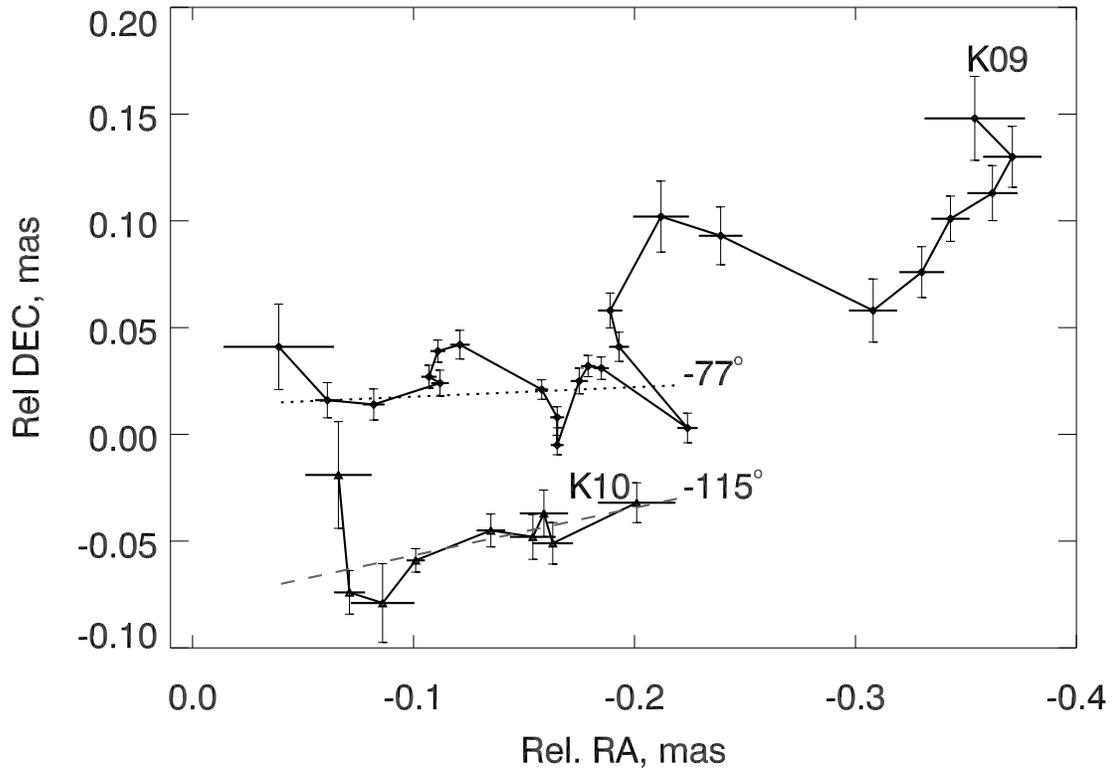}
\caption{Trajectories of knots $K09$ (diamonds) and $K10$ (triangles); the dotted and dashed lines show the average position angle of $K09$ and $K10$, respectively, within 0.2~mas of the core.} \label{Ktraj}
\end{figure}

The apparent motions of knots K09 and K10 are complex, as seen in Figure \ref{evol}. Knot K09 appears to decelerate at a distance of $\sim$0.15~mas from the core and then accelerate at a distance of $\sim$0.2~mas. However, this is almost surely an artifact of the blending of K10 with the core, which shifts the apparent centroid of the core downstream for some time, after which the separation of K09 from the core rejoins the line representing a ballistic trajectory.
Knot K10 appears in the jet at $\Theta\approx-136^\circ$, south of the usual jet direction (Fig.~\ref{mapsK10}), then its trajectory curves into the direction of the average projected jet axis, $\Theta_{\rm jet}\approx-92^\circ$ (Fig.~\ref{Ktraj}). Using the technique developed in \citet{J05},
we have calculated for $K09$ and $K10$ the apparent speed, $\beta_{\rm app}$, acceleration along and perpendicular
to the jet, $\dot{\mu}_\parallel$ and $\dot{\mu}_\perp$, time of ejection 
$T_\circ$\footnote{$T_\circ$ is the extrapolated time of coincidence of the centroid of a moving knot with the centroid of the core on the VLBA images}, timescale of flux variablity, $\tau_{\rm var}$, Doppler factor, $\delta$, 
Lorentz factor, $\Gamma$, and viewing angle, $\Theta_\circ$. The values of these parameters are given in Table~\ref{Kparm}.
According to Table~\ref{Kparm}, components $K09$ and $K10$ have similar apparent speeds, $\sim$9~$c$, a value that falls within the range of $\beta_{\rm app}$ observed previously in 3C~454.3 \citep{J01,J10,KL04,LIST09}. Both $K09$ and $K10$ execute an acceleration perpendicular to the jet, which can be related to the apparent change of position angle near the core. Figure~\ref{Ktraj} suggests that the knots were ejected along different position angles, $K09$ to the north and $K10$ to the south with respect to the average jet axis. $K09$ also increases its proper motion along the jet, which can be attributed to an intrinsic acceleration \citep{HOM09}. The values of the parameters $\delta$, $\Gamma_{\rm b}$, and $\Theta_\circ$ of $K09$ are very close to those derived by \citet{J05} from the kinematics of the jet in 1998-2001, while the Doppler factor of $K10$ is extreme, $\delta\sim$50. The latter yields a much smaller viewing angle for $K10$ with respect to $K09$, in agreement with the different projected trajectories of the knots, which differ by $\sim$38$^\circ$ (Fig.~\ref{Ktraj}). According to Table~\ref{Kparm}, the main difference in the derived values of $\delta$ results from the timescale of variability. $K10$ fades faster than $K09$ by a factor of 2.5. 

\subsection {Flux and Polarization Variability \label{FPV}} 
\begin{figure}
\epsscale{0.8}
\plotone{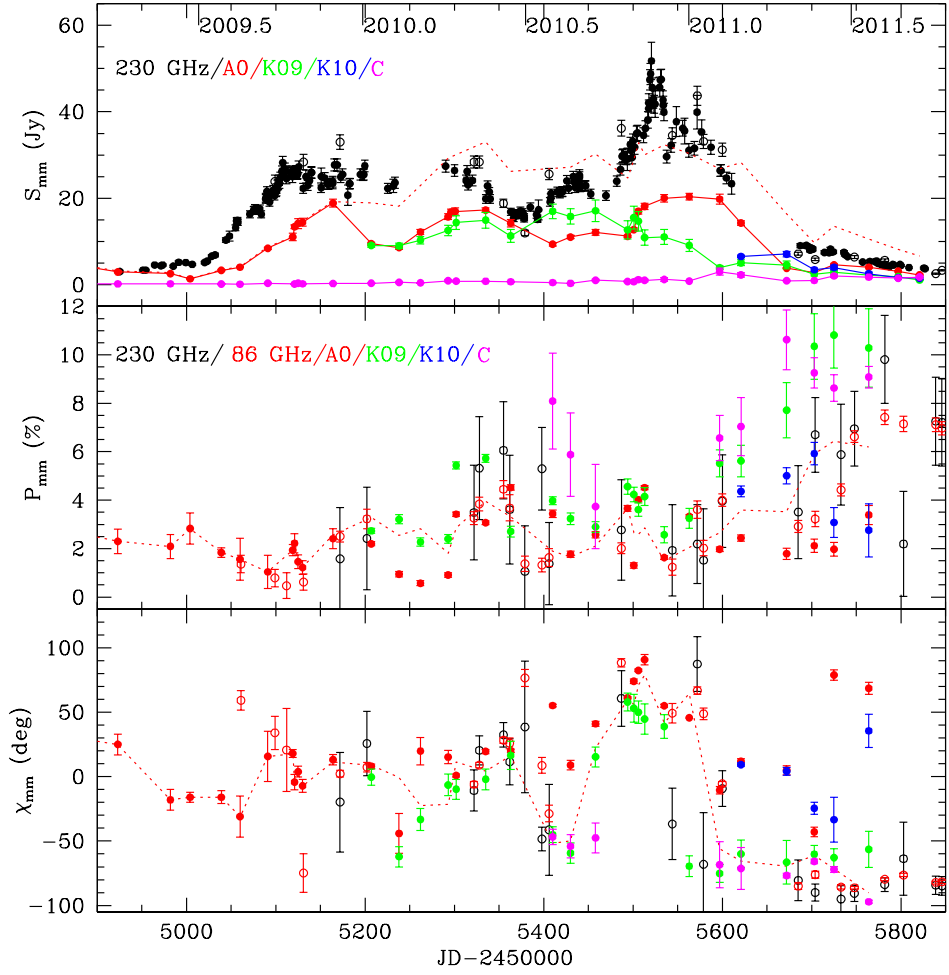}
\caption{Total flux, $S$, degree, $P$, and position angle, $\chi$, of linear polarization at mm wavelengths vs. time. {\it Top panel:} the light curve at 230~GHz obtained with the {\it SMA} (black filled cicles) and {\it IRAM} (black open circles) plus the light curves of the jet features: 
VLBI core, $A0$, - red circles connected by the red solid line, stationary knot $C$ - magenta circles and magenta line, knot $K09$ - green circles and green line, and knot $K10$ - blue circles and blue line; the red dotted line shows the summed flux of all 4 jet knots ($A0+C+K09+K10$). {\it Middle panel:} degree of polarization from the whole source at 230~GHz (black open circles) and 86 GHz (red open circles); $P$ of the jet features at 43~GHz: the core $A0$~(red filled circles), knot $C$ (magenta filled circles; $P$ of knot $C$ is divided by a factor of 3 to display alongside other features), $K09$ (green filled circles), and $K10$ (blue filled circles); the red dotted line shows the summed $P$ of all 4 jet knots. {\it Bottom panel:} position angle of polarization from the whole source at 230~GHz and 86~GHz, $\chi$ of the jet features at 43~GHz: $A0, C, K09$, and $K10$ (designations are the same as in the middle panel); the red dotted line shows the summed $\chi$ of all 4 jet knots.} \label{mmPol}
\end{figure}

Figure~\ref{mmPol} displays the overall 1~mm and 7~mm light curves of individual components in the inner
jet. The total 7~mm flux is calculated as the sum of $A0, K09, K10$, and $C$, depending on which feature is present at a given epoch according to the modeling of the images. The light curve of the core follows a smooth version of the variations at 1~mm, although the contribution of other
jet components to the 1~mm flux is significant, since in general the flux at 1~mm exceeds the core flux at 7~mm throughout the majority of epochs. Comparison of the 1~mm and inner jet light curves shows that during RJD: 5100-5200
and  RJD: 5500-5550 the flux at 1~mm is higher than that at 7~mm from the inner jet, and the opposite is observed
within RJD: 5300-5500 and after RJD: 5600. The bright 1~mm states, relative to 7~mm, are modeled to be    contemporaneous with the times when knots $K09$ and $K10$ were passing through the core. The lower 7~mm flux can be explained by a temporary suppression of the 7~mm flux outburst resulting from opacity increases in the 7~mm core as a superluminal knot moves through it. After a significant increase of the 1~mm flux in 2009 Autumn, the 1~mm flux remains at a high level for more than a year, a circumstance that we associate with the appearance in the jet of the very bright knot $K09$. Although the flux of the core decreased significantly after the ejection of $K09$, the core was still brighter than during quiescent states in 2009 Spring and 2011 Summer when there is  good agreement between the 1~mm and 7~mm core light curves. This implies that the contribution of the jet outside 1~mas to the emission at 1~mm is negligible.  
 
\begin{figure}
\epsscale{0.45}
\plotone{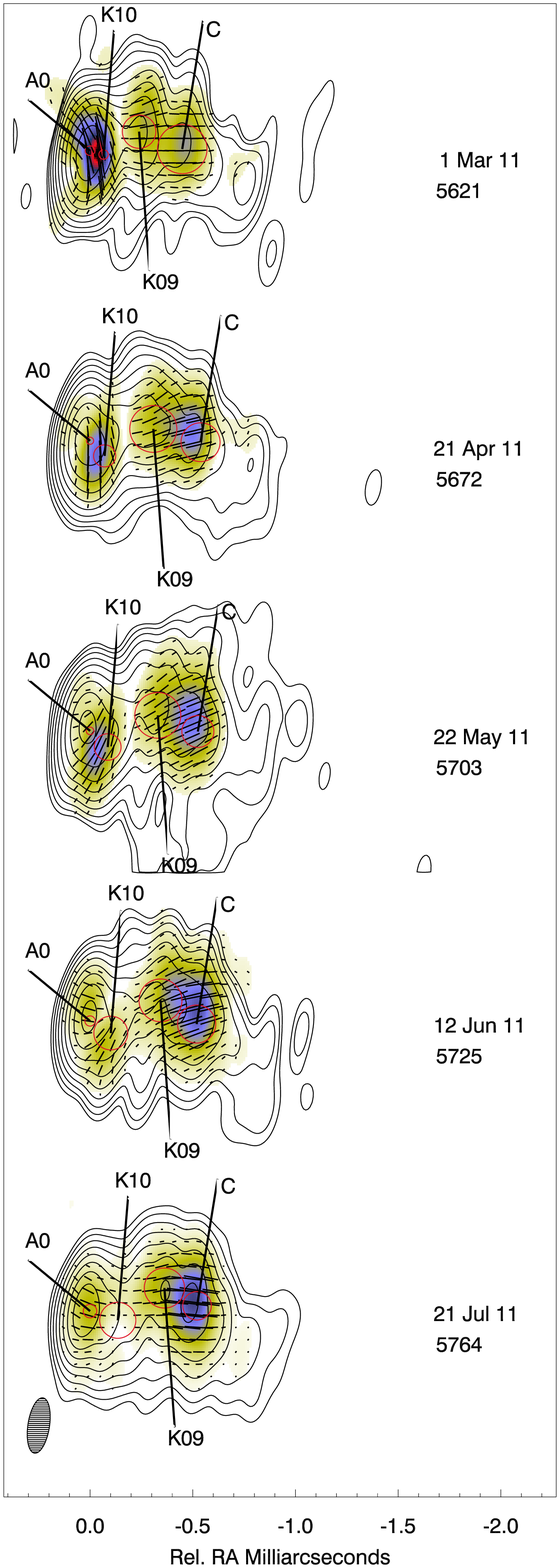}
\caption{43 GHz total (contours) and polarized (color scale) intensity images of 3C~454.3
with $I_{\rm peak}$=16.29~Jy/beam,  $I_{\rm peak}^{\rm pol}$=0.46~Jy/beam, and a Gaussian restoring beam=0.14$\times$0.33~mas$^2$ at $PA$=-10$^\circ$; contours represent 0.1, 0.2,..., 51.2, 99.5\% of the peak intensity; line segments within the image show direction of linear polarization; red circles indicate position and size (FWHM) of components according to model fits.} \label{mapsK10}
\end{figure}

Figure~\ref{mmPol} displays the degree of linear polarization $P$ vs. time at mm wavelengths. The  polarization from the whole source at 1 and 3~mm changes from 0 to $\sim$10\% and agrees very well with that of the inner jet (dashed line). This presents another argument in favor of the emission at 1-3~mm arising mostly from the inner jet that includes the core and components within 1~mas of the core.
The polarization of the core ranges from 1\% to $\sim$5\%, with a modest increase of $P$ during the outbursts. The polarization of moving knots $K09$ and $K10$ lies within 2-6\%; however, as $K09$ approaches stationary 
feature $C$, $P$ increases significantly for both knots and reaches 12\% and 30\% for $K09$ and $C$, respectively (see Fig.~\ref{mapsK10}). This supports the idea that the knots experience an interaction with the ambient jet, most likely in the form of a shock, since the position angle of polarization of both knots aligns with the jet direction. This is a primary signature of a transversely oriented shock, with the magnetic field compressed along the shock front \citep{HAA85,HAA89}. At this time, knots $K09$ and $C$ appear to contribute significantly to the polarized emission at 1 and 3~mm and $P_{\rm 1mm}$ rises up to 10\%. Although the flux of $C$ increases slightly, the total flux density continues to fade at mm wavelengths along with the VLBI core. 

\begin{figure}
\epsscale{1.0}
\plotone{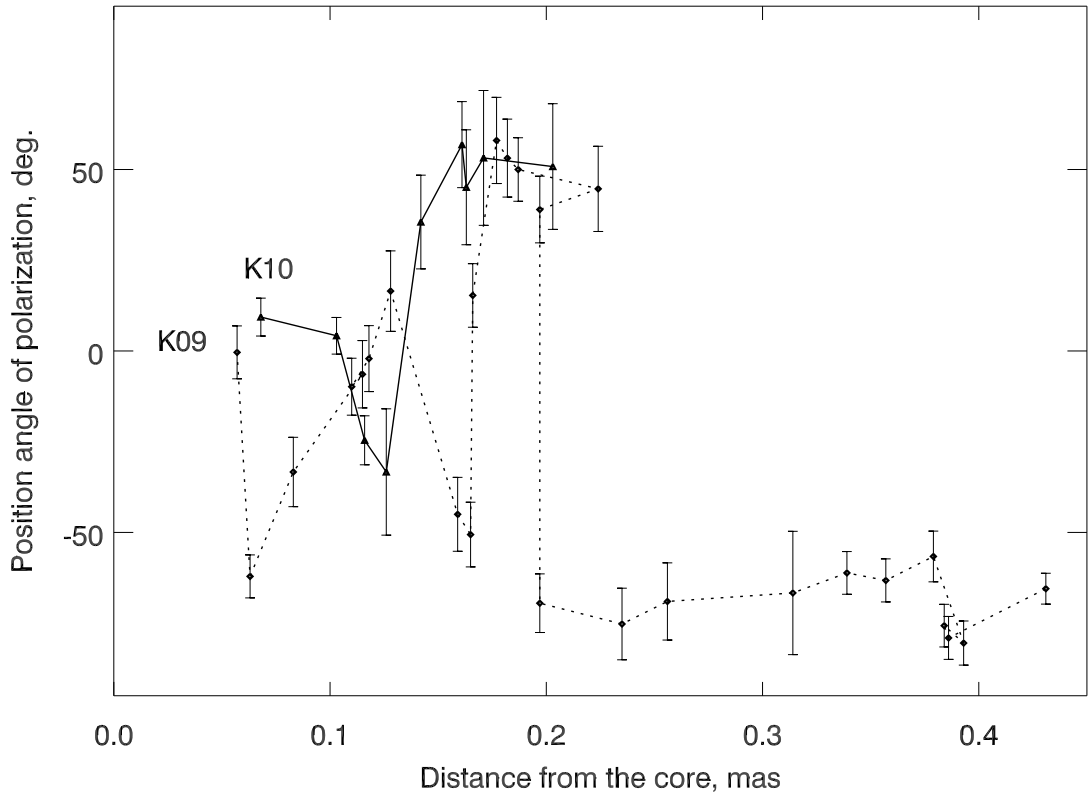}
\caption{Position angle of polarization of knots $K09$ (diamonds, dotted line) and $K10$ (triangles, solid line) vs. their distance from the core.} 
\label{Krevpa}
\end{figure}

\begin{figure}
\epsscale{1.0}
\plotone{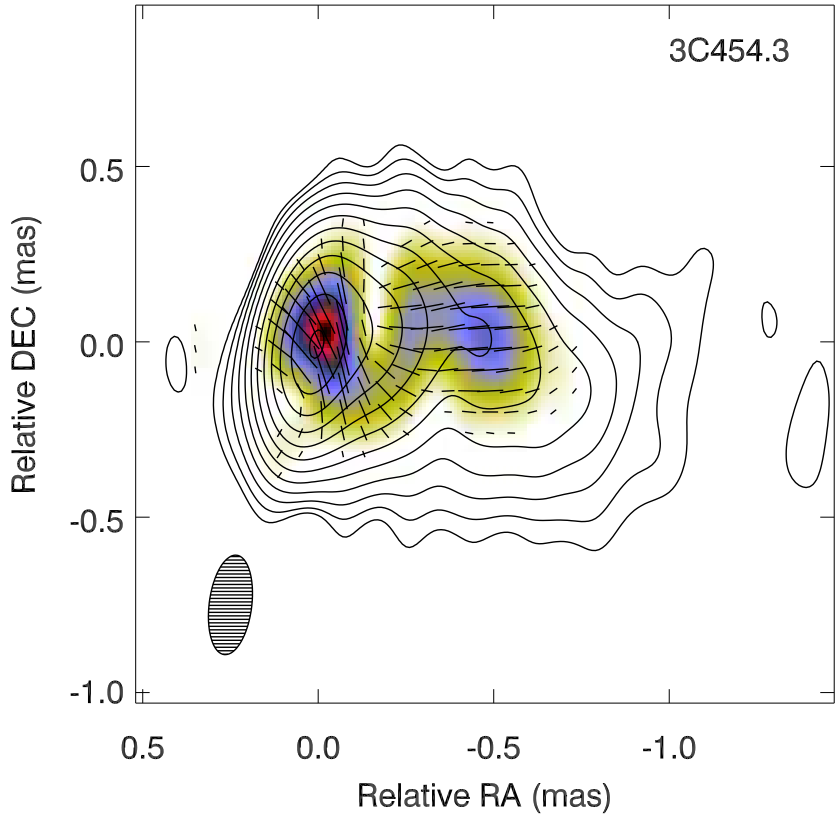}
\caption{Composite total (contours) and polarized (color) intensity image of 3C~454.3, which is the average of all 35 epochs of VLBA data at 43~GHz from April 2009 to August 2011, with $I_{\rm peak}$=11.51~Jy/beam,  $I_{\rm peak}^{\rm pol}$=0.11~Jy/beam, and beam=0.14$\times$0.33~mas$^2$ at $PA$=$-$10$^\circ$; contours represent 0.025, 0.05,..., 25.6, 51.2\% of the peak intensity; line segments within the image show direction of polarization.} 
\label{map_sum}
\end{figure}

During outbursts I and III the polarization position angle $\chi_{\rm core}$ of the core at 7~mm  rotates from +16$^\circ$ to $-$44$^\circ$  and from +91$^\circ$ to $-$11$^\circ$, respectively (Fig.~\ref{mmPol}). In addition, $\chi_{\rm 3mm}$ rotates in a similar manner, although the range of rotation is greater than for $\chi_{\rm core}$, especially during outburst I. 
Figures  \ref{mapsK09}, \ref{mapsK10}, \& \ref{mmPol} show that the polarization vectors of both $K09$ and $K10$ undergo rotations, as well. Figure~\ref{Krevpa} compares the evolution of $\chi_{K09}$  and $\chi_{K10}$ with distance from the core. Both knots appear in the jet with polarization vectors oriented perpendicular to the jet axis, but between separations of 0.05 and 0.18~mas the EVPAs of both knots swing, although the rotations are in the opposite direction at the same distance from the core. These rotations might be a signature of a large intrinsic Faraday rotation measure near the core caused by a toroidal structure of the magnetic field. In this case the different directions of rotation of $\chi_{K09}$ and $\chi_{K10}$ can be readily explained by a change in the sign of the magnetic field with respect to the line of sight,
assuming that the knots are propagating along different sides of the jet, as can be inferred from
the knot's trajectories (Fig.~\ref{Ktraj}). At $\sim$0.2~mas from the core, $\chi_{K09}\approx\chi_{K10}\approx$50$^\circ$, oblique to the jet direction, and at distances $>$0.2~mas $K09$ has a stable EVPA, aligned with the jet axis, as well as with the EVPA of stationary knot $C$. Such a behavior is consistent with the knots being transversely oriented shocks propagating down the jet, which has a turbulent magnetic field \citep{HAA85,HAA89}.
Figure~\ref{map_sum} shows the composite structure of the parsec-scale jet emission from 2009 April to 2011 August.
The image is obtained by summing Stokes $I, Q, U$ parameter maps over all 35 epochs obtained during this period (each map was convoled with the same restoring beam). The image represents an active state of the inner jet, since the $I_{\rm peak}$ is a factor of $\sim$10 brighter than during a quiescent state. A spiral-type structure is apparent in polarized intensity up to $\sim$0.4~mas from the core, with a different direction of the EVPAs on 
the southwest and northwest edges of the polarized emission region. Farther down the jet, the position angle of polarization aligns with the jet direction, as expected if a turbulent magnetic field becomes partially ordered along the front of a transverse shock. This region of the jet is dominated by the contribution from knots $K09$ and $C$. This picture is consistent with the scenario, proposed by \citet{J07}, that the mm-wave core is located at the end of the acceleration zone, where the jet energy density is dominated by the Poynting flux of a toroidal magnetic field.  Near the core the flow becomes kinetic energy dominated, with the magnetic field becoming turbulent.

\subsection{Connection between Jet Activity and the Gamma-Ray Outbursts}
\begin{figure}
\epsscale{1.0}
\plotone{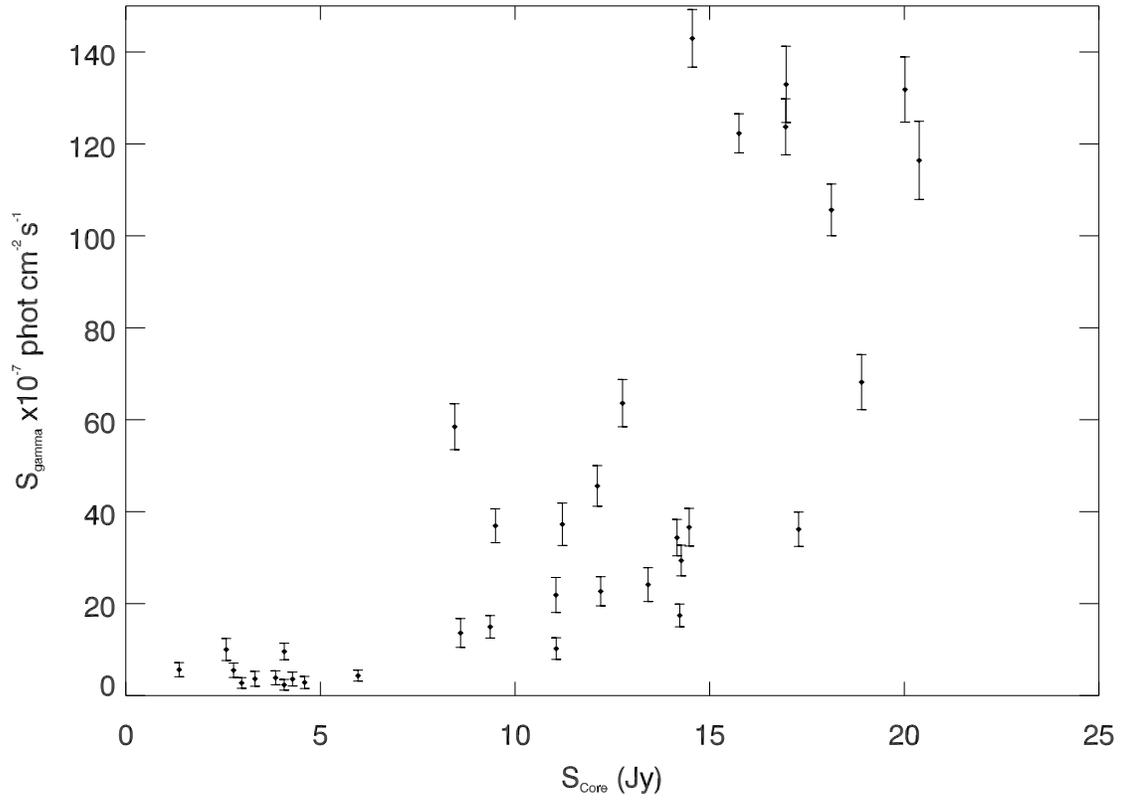}
\caption{Gamma-ray flux density vs. flux density at 43~GHz  in the VLBI core for simultaneous measurements.} 
\label{GA0flux}
\end{figure}

We analyze the relation between the $\gamma$-ray flux, $S_\gamma$, and the 43~GHz flux density of the VLBI core, $S_{\rm core}$. Figure~\ref{GA0flux} shows $S_\gamma$ vs. $S_{\rm core}$
for all VLBA epochs (35), with the $\gamma$-ray photons integrated over the 24 hours centered on the VLBA observation. There is a strong correlation between variations at $\gamma$-rays and in the core. The linear Pearson correlation coefficient, $r$=0.77, is statistically significant at the 99.9\% confidence level. The flux density of the core increases by a factor of 16 while the $\gamma$-ray flux rises by a factor of $\sim$65. This implies that the $\gamma$-ray flux is relativistically beamed in the same manner or more strongly than the radio flux, as suggested previously by \citet{J01} and \citet{KOV09}.
The relation between $S_\gamma$ and $S_{\rm core}$ does not fit a simple linear dependence, rather, two relationships can be inferred: (1) for $S_{\rm core} \lesssim$14~Jy the dependence is almost linear, and (2) for $S_{\rm core} \gtrsim$14~Jy the dependence is roughly quadratic. 
In addition, \citet{ANN12} have found a statistically significant correlation between the $\gamma$-ray and 1~mm light curves, with no lag.  
 
Enhanced flux density of the core region in a VLBI image of a blazar usually corresponds to the emergence of a new disturbance into the flow in the radio-emitting zone of the jet \citep[e.g.,][]{SAV02}. Comparison of Tables~\ref{Gparm} and \ref{Kparm} indicates that knots $K09$ and $K10$ passed through the mm-wave VLBI core close to the time of the $\gamma$-ray peaks of outbursts I and III (within the 1$\sigma$ uncertainty of the ejection time of 18 and 25~days, respectively). The duration of these $\gamma$-ray outbursts is comparable to the time needed for a knot to go through the core, 
90$\pm$15~days, for the average core size of 0.05~mas and average proper motion of knots of 0.20~mas~yr$^{-1}$.  In addition, during the $\gamma$-ray outbursts we observe an increase in the core opacity, as revealed by comparison of the 1~mm and 7~mm light curves, and rotation of the polarization vector in the core and at 3~mm. These trends argue in favor of the $\gamma$-ray outbursts coinciding with the passage of superluminal knots through the core. Although we did not detect a superluminal knot associated with $\gamma$-ray outburst II, Figure \ref{mmPol} shows that both the core and $K09$ underwent flares during outburst II (RJD: 5275-5349), with the flux reaching $S\sim$18~Jy. Moreover, we observed an increase of the degree of polarization in the core and at 1~mm and 3~mm, along with a rotation of $\chi$ at 1 and 3~mm. Based on the history of 3C~454.3, it is likely that this increase in the flux and polarization of the core during the $\gamma$-ray event was caused by the passage of a new superluminal knot through the core, with a flux of $\le$7~Jy. Summer - Autumn 2010 would have been the best period to detect this hypothetical new component in the jet. However, during this interval $K09$ was still close to the core, between 0.12~mas and 0.20~mas (Fig. ~\ref{evol}). Although $K09$ is distinct from the core at the high resolution of our observations (see Fig.~\ref{mapsK09}), the resolution is insufficient to identify a weaker knot situated between the core and an extremely bright knot within 0.20~mas. The increase in the flux of $K09$ along with the core during $\gamma$-ray outburst II might be an artifact of modeling with circular gaussians of such complex structure, but $K09$ continued to be brighter than the core as the core faded in mid-2010 (Fig.~\ref{mmPol}). This suggests that $K09$ might have been blended with the putative superluminal knot. The subsequent dramatic increase of the flux in the core in 2010 November and ejection of K10 further reduced the possibility of detecting any propagating disturbance associated with $\gamma$-ray outburst II. {\it Overall, comparison of the $\gamma$-ray and mm-wave behavior provides strong evidence of a tight connection between enhanced $\gamma$-ray emission and the passage of superluminal knots through the mm-wave core located at a distance $\sim$15-20~pc from the central engine 
\citep{J10,PUSH12}.}

\section{Spectral Behavior}
We analyze the spectral behavior of 3C~454.3 from $\gamma$-ray to mm wavelengths with emphasis on outbursts I, II, and III.
This includes studies of 1) the $\gamma$-ray spectral index in the energy range of 0.1-200~GeV;  2) the X-ray spectral index at 0.3-10~keV; 3) optical spectra in the range of 4000-7550~{\AA} (181 spectra); and 4) spectral indices based on fluxes measured simultaneously (within 1 day) at different bands from $UV$ to far-IR, and at wavelengths from 1.3 to 8~mm. Tables~\ref{OptVar} and \ref{IrVar} indicate the number of simultaneous flux measurements at a given pair of wavelengths. We define the spectral index $\alpha$ such that the flux density $S_\nu\propto\nu^{-\alpha}$. 

\subsection{Gamma-Ray, X-Ray, and Millimeter-wave Spectral Indices}
\begin{figure}
\epsscale{1.0}
\plotone{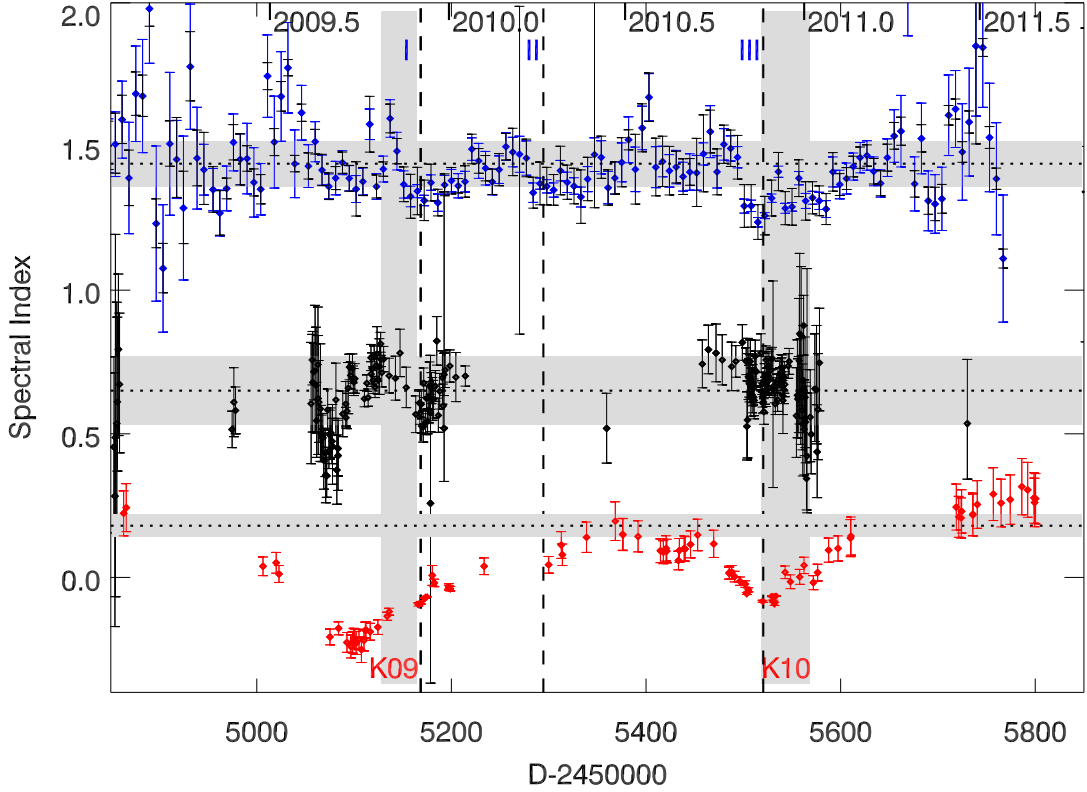}
\caption{Gamma-ray at 0.1-200~GeV (blue), X-ray at 0.3-10~keV (black), and mm-wave between 1 and 8~mm spectral indices vs. time; the dotted lines indicate the average spectral indices of $\alpha_\gamma$ and  $\alpha_{\rm X}$, and $\alpha_{\rm mm}$ of a quiescent state, corresponding 1$\sigma$ uncertainties are marked by the horizontal gray areas; the vertical gray areas mark times of passage of knots $K09$ and $K10$ through the VLBI core; the dashed vertical lines mark peaks of the $\gamma$-ray emission during flares I, II, and III.} 
\label{GXMInd}
\end{figure}

We have derived weekly $\gamma$-ray spectral indices $\alpha_\gamma$ between  0.1-200~GeV using a simple power law model
with variable photon index and normalization (``prefactor'') to represent the quasar emission. Frozen spectral parameters 
(values from the 2FGL catalog; \citealt{2FGL}) were used for other sources within 15$^\circ$ radius of 3C~454.3 to guide the maximum likelihood routine. Figure~\ref{GXMInd}
shows the derived values of $\alpha_\gamma$, which varies from 1.2 to 1.8 with an average $<\alpha_{\gamma}>=$1.44 and standard deviation $\sigma_\gamma$=0.08, which is less than the average uncertainty of an individual measurement $\sigma_\gamma^1=$0.13. 
However, the spectral indices averaged over a month-long interval centered on the peak of outburst are 1.33$\pm$0.03 (flare I), 1.37$\pm$0.03 (flare II), and 1.28$\pm$0.03 (flare III). This implies a harder $\gamma$-ray spectrum during the highest states of $\gamma$-ray emission, a trend studied in detail by \citet{ACKER10} and \citet{ABDO11}. 

We have derived 201 spectral indices, $\alpha_{\rm X}$, between 0.3 and 10~keV from 2009 April 25 to 2011 August 1. The X-ray spectral index varies from 0.3 to 0.8 with an average value  of $<\alpha_{\rm X}>=$0.65$\pm$0.12,
while the mean uncertainty of an individual measurement $\sigma_{\rm X}^1$=0.10 (Fig~\ref{GXMInd}).
This behavior is consistent with the conclusion of \citet{RAI11} that the spectral variations are dominated by the noise of the measurement uncertainties. Indeed, during both outbursts for which X-ray data are available (I and III), $\alpha_{\rm X}$ exhibits random fluctuations around $<\alpha_{\rm X}>$. However, $\sim$3~months before the peak of outburst I, the spectral index flattens over 19 consecutive measurements from RJD: 5065 to RJD: 5090, with an average value of 0.44$\pm$0.08.  

We use the data collected at the SMA, IRAM, and  Mets\"ahovi Radio Observatory facilities to derive the
mm-wave spectral index ($\alpha_{\rm mm}$) between 1.3 and 8~mm. We have obtained 92 spectral indices from
measurements simultaneous within 1~day. Figure~\ref{GXMInd} shows $\alpha_{\rm mm}$
versus epoch. The spectral index changes from $+$0.3 to $-$0.25, very similar to the range
observed in 2004-2008 \citep{J10}.
The figure also shows times of ejection of knots $K09$ and $K10$ from the core.
Although $\alpha_{\rm mm}$ is slightly steeper during quiescent states than $\alpha_{\rm mm}^{\rm q}=$0.18$\pm$0.04 
measured in 2004-2008, the main feature of the $\alpha_{\rm mm}$ behavior --- that the mm-wave spectral index reaches a local minimum just before 
the ejection of a superluminal knot --- holds for both knots $K09$ and $K10$. This pattern was also observed for knots $K1, K2$, and $K3$ \citep{J10}. Unfortunately, there is a gap in the radio data during $\gamma$-ray outburst II that prevents us from using changes in $\alpha_{\rm mm}$ to search for signs of the ejection of the hypothetical knot discussed in \S~4.3. During the events associated with knots $K09$ and $K10$, the spectrum becomes inverted with spectral indices of $-$0.25$\pm$0.05 ($\alpha_{\rm mm}^{\rm K09}$) and $-$0.08$\pm$0.01 ($\alpha_{\rm mm}^{\rm K10}$), implying an increase in opacity at mm wavelengths during the events.  The fact that $\alpha_{\rm mm}^{\rm K09}<\alpha_{\rm mm}^{\rm K10}$ suggests a significant contribution of optically thin emission from $K09$ to the total flux at mm-wavelengths during the $K10$ event.

Figure~\ref{GXMInd} reveals similarity in variations of $\alpha_{\gamma}$ and $\alpha_{\rm mm}$. Both spectral
indices flatten at/near the time of ejection of superluminal knots and steepen during quiescent periods (2009 Spring, 2010 Summer, and 2011 Spring). However, the similarity is not sufficient to claim a statisticially significant correlation between the indices. Although the measurements of $\alpha_{\rm X}$ are sparse and noisy, a flattening of $\alpha_{\rm X}$ at RJD: 5065-5090, which is significant at the 2$\sigma$ level, coincides with the period of the flattest $\alpha_{\rm mm}$ (RJD: 5075-5110) that precedes the ejection of $K09$.
    
\subsection{Broad Emission Lines and Optical Continuum}\label{BLROC}
\begin{figure}
\epsscale{1}
\plotone{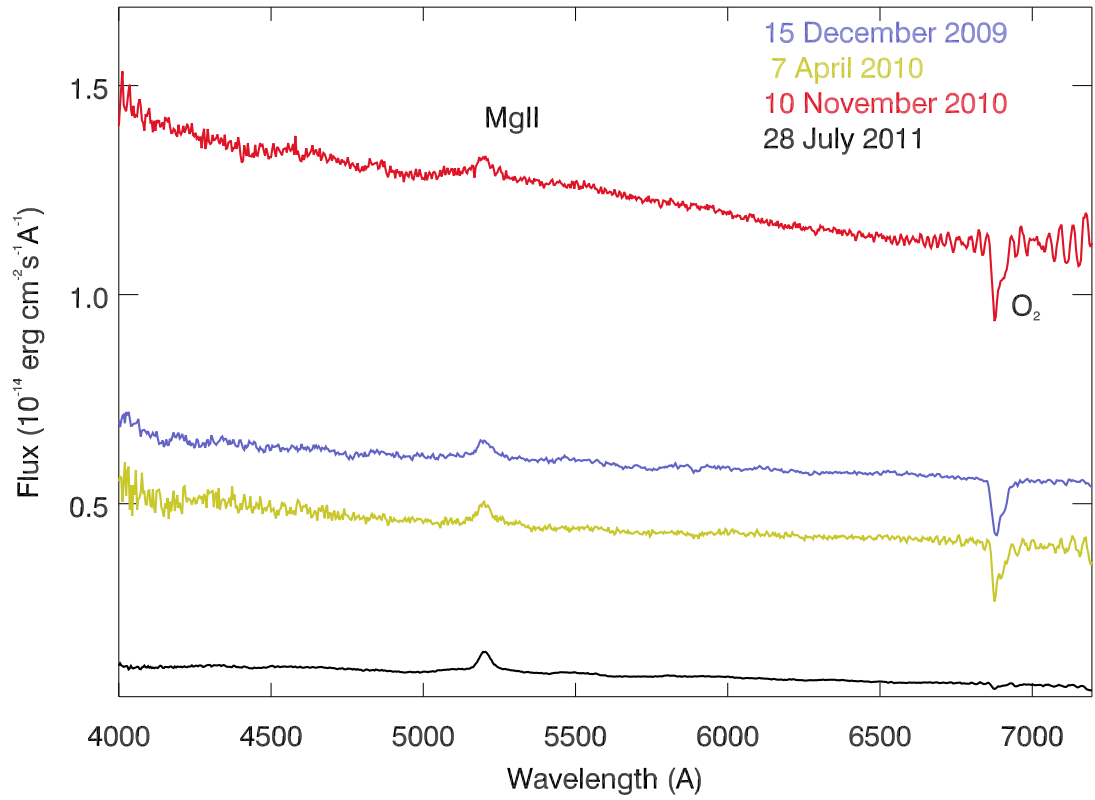}
\caption{Optical spectrum of 3C~454.3 in the observer's frame at four epochs.} \label{Mg2}
\end{figure}

Figure~\ref{Mg2} shows four spectra of the quasar at the epochs of the brightest (or next to brightest) $\gamma$-ray states during outbursts I, II and III, and at a quiescent state in 2011 Summer. The most prominent emission features in the Steward Observatory spectra (2150-4060~{\AA} in the rest frame) are broad MgII($\lambda=$2800~{\AA}) and blended FeII multiplets. A time series analysis between variations in the line and continuum emission is important for determining the size, geometry, and location of the broad emission line region (BLR). A number of studies have found that the size of BLR is proportional to the AGN luminosity, $R_{\rm BLR}\propto L_{\rm 5100}^{\kappa}$, where $\kappa$ varies from 0.5 to 0.7 \citep [e.g., ][] {KASPI00},
with $R_{\rm BLR}$ of low-luminosity AGN falling in the range of 10-100~lt-days. This implies that the size of BLR in luminous quasars $\ge$1~lt-yr. In general, analyses of emission line variations in blazars, including 3C~454.3, do not reveal a connection between line and continuum flux variations (3C~454.3: \citealt{RAI08}, 1222+216: \citealt{Smith_1222}, and 1633+384: \citealt{RAI12}). However, a recent analysis by \cite{LT13} of the optical spectra of 3C~454.3 obtained at Steward Observatory during 2008-2011 suggests the existence of significant variations in the MgII line, corresponding to a flare-like event, during outburst III that challenges the standard model of the BLR in blazars. We possess the same data as analyzed by \cite{LT13}. Unfortunately, the data miss the 2-week period from 2010 November 16 to 30 during the brightest $\gamma$-ray and optical state in 3C~454.3. Since the Yale University blazar monitoring group obtained optical spectra of the quasar during this period (C.M. Urry, private communication), we defer to an analysis of time variations of emission lines to the work by \citet{ISLER13}. 

\begin{figure}
\epsscale{1}
\plotone{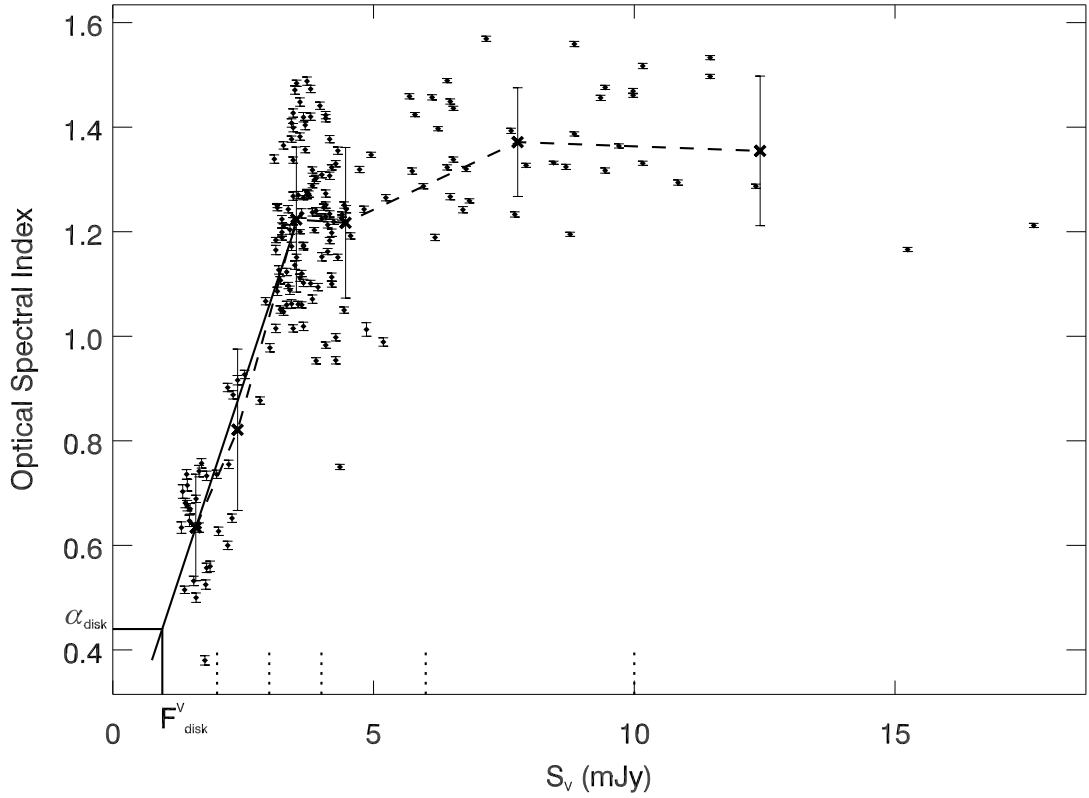}
\caption{Spectral index vs. brightness of the optical continuum in V band in the observer's frame;
crosses show average values of $\alpha_{\rm cont}$ within different intervals of flux, $S_{\rm V}$ (Table ~\ref{AcontT}),
connected by the dashed line; dotted vertical segments show intervals of the averaging; 
the solid line is a linear fit of the dependence for $S_{\rm V}< 4.0$~mJy.} 
\label{Acont}
\end{figure}

We have corrected the spectra of 3C~454.3 for Galactic extinction, performed Gaussian fits to the emission features, subtracted 
the fits from each of 181 spectra, and approximated the residual continuum by a power law, $S_\nu\propto\nu^{-\alpha_{\rm cont}}$, within the $\log_{10}\nu$ range from 14.65 to 14.85 ($\lambda$ from 4235 to 6712~{\AA}) to avoid the atmospheric $O_2$ absorption feature and noisy edges of the spectra (the wavelength range is from 2278 to 3611~{\AA} in the quasar rest frame). Figure~\ref{Acont} plots $\alpha_{\rm cont}$ versus flux in $V$ band from simultaneous measurements (within $\sim$1~hr).
Table~\ref{AcontT} lists average values of $\alpha_{\rm cont}$ with their standard deviations for different flux intervals. The spectral index of the continuum steepens 
by a factor of 2, from $\sim$0.6 to $\sim$1.2, as the flux increases from $\sim$1.5~mJy to 3.5~mJy, while the scatter of $\alpha_{\rm cont}$ values does not exceed 0.15 within each interval of averaging. A further increase of the flux up to $\sim$10~mJy does not change the average spectral index significantly, although a slight steepening, up to 1.35, toward higher flux
is observed. Such behavior of $\alpha_{\rm cont}$ is a direct indication that the continuum consists of at least two components, blue and red. The blue component dominates at low brightness states and can be attributed to big blue bump (BBB) emission produced by the accretion disk \citep{SMITH88,RAI07}. The red component is non-thermal optical emission that originates in the relativistic jet. Figure~\ref{Acont} can be interpreted within the assumption that the blue component is constant during our observations both in flux and in spectral index, while the red component varies significantly in flux and slightly (within $\pm$0.15) in spectral index. The latter can account for the scatter observed within different intervals of averaging. We
derive a linear approximation of $\alpha_{\rm cont}=(0.18\pm0.04) + (0.30\pm0.02)S_{\rm V}$ for $S_{\rm V}$ from 0 to 4.0~mJy, which is shown in Figure~\ref{Acont} by a solid line. According to the Pearson's $\chi^2$ test the linear fit corresponds to the data sufficiently well with $\chi^2$=1.83 for 103 measurements. \citet{VB01} have created a composite continuum spectrum of a quasar using a homogeneous data set consisting of over 2200 quasar spectra from the Sloan Digital Sky Survey (SDSS), which covers a rest-wavelength range from 800 to 8555~{\AA}. The quasars were selected from both optical and radio criteria. The authors have found that at wavelengths from $\sim$1300 to 5000~{\AA} the continuum is well fit by a power law with $\alpha_\nu$=0.44, which we denote as $\alpha_{\rm disk}$, assuming that this continuum should represent the emission from the accretion disk. Although it is not clear what fraction of quasars in the sample are radio loud,
the index is in good agreement with values found in optically selected quasar samples \citep[e.g., ][]{OSQ98}. Under the assumption that the accretion disk
in 3C~454.3 has similar properties as the disk of a generic quasar and, therefore, when $\alpha_{\rm cont}=\alpha_{\rm disk}$ the contribution of the nonthermal component to the continuum is negligible, we can use the
linear fit for $\alpha_{\rm cont}$ to estimate the flux of the accretion disk, $S^{\rm V}_{\rm disk}$=0.85$\pm$0.15~mJy, which corresponds to $S^{\rm R}_{\rm disk}$=0.91$\pm$0.16~mJy. As expected, these values are slightly less than the minimum flux of the quasar observed in $V$ and $R$ bands, respectively.  

\subsection{Spectral Properties of Optical and Near-IR Synchrotron Emission \label{HAG}}
A method proposed by \citet{HT97} allows one to subtract from the total near-IR to UV emission the contribution of components that are either constant or vary on long timescales. These are expected to include the accretion disk, BRL, and dusty torus. The result is a relative spectral energy distribution, $RSED$, of the component responsible for variable emission on timescales of hours, days, or weeks. Since it is synchrotron emission that varies on such short timescales in a blazar, we refer to this as the ``synchrotron component.'' The method is based on simultaneous multicolor observations and
assumes a linear dependence between variations at two different bands, $S_{\rm N} = A_{\rm N}+C_{\rm N}S_{\rm r}$,  where $N$ is the band at which the flux is measured, and $S_r$ is the flux at a reference band (see below). Figures~\ref{FFI} and \ref{FFIII} reveal approximately linear flux-flux relations at different bands that are statistically significant at a level $\ge99.9\%$ according to the $\chi^2$ criterion.  
We consider that two measurements at different wavelengths are simultaneous if they are performed within 2~hr of each other. Table~\ref{OptVar} shows the number of simultaneous observations at different wavelengths with respect to $R$ band for outbursts I and III, as well as the values of $\chi^2$ for
a linear fit to the flux-flux relations. Figures~\ref{FFI} and \ref{FFIII} show that the simultaneous
measurements cover a wide range of variability during both outbursts. Unfortunately,
in the case of outburst II there are no observations in the $UV$ region and only 5-6 observations
at near-IR wavelengths over a narrow range of flux levels. $R$ band is chosen 
as the reference band for the construction of the $RSED$, since the largest number of observations were obtained in this band. 
In the case of $UV$ observations, $B$ band serves as a primary reference band (see Figs.~\ref{FFI},~{\it left} and \ref{FFIII},~{\it left}) 
and the linear dependence between the $B$ and $R$ fluxes is used to derive the coefficients $C_{\rm U}$, $C_{\rm UVW1}$, $C_{\rm UVM2}$, and $C_{\rm UVW2}$. Similarly, $J$ band is the primary reference band for measurements in $H$ and $K$ bands. The $C$   coefficients are given in Table~\ref{OptVar}. The dependence of the coefficients on frequency represents a relative spectral energy distribution. Figure~\ref{RSED} shows the $RSED$ for outbursts I and III.  Both $RSED$s can be approximated by a power law, $S\propto\nu^{-\alpha_{\rm opt}^{\rm syn}}$, with similar spectral indices,
$\alpha_{\rm opt}^{\rm synI}=1.77\pm0.05$ and $\alpha_{\rm opt}^{\rm synIII}=1.71\pm0.04$.
This synchrotron emission represents a red spectral component that dominates
the optical continuum when the source is brighter than $S_{\rm V}\sim$3.5~mJy, as discussed
in Section~\ref{BLROC}. The spectral index of the red component is significantly steeper than 
the spectral index of the optical continuum, $\alpha_{\rm cont}\sim$1.30, during high optical states,
hence the disk emission is always significant for the quasar spectral energy distribution. 

\subsection{Spectral Properties of Far-IR Synchrotron Emission}
\begin{figure}
\plottwo{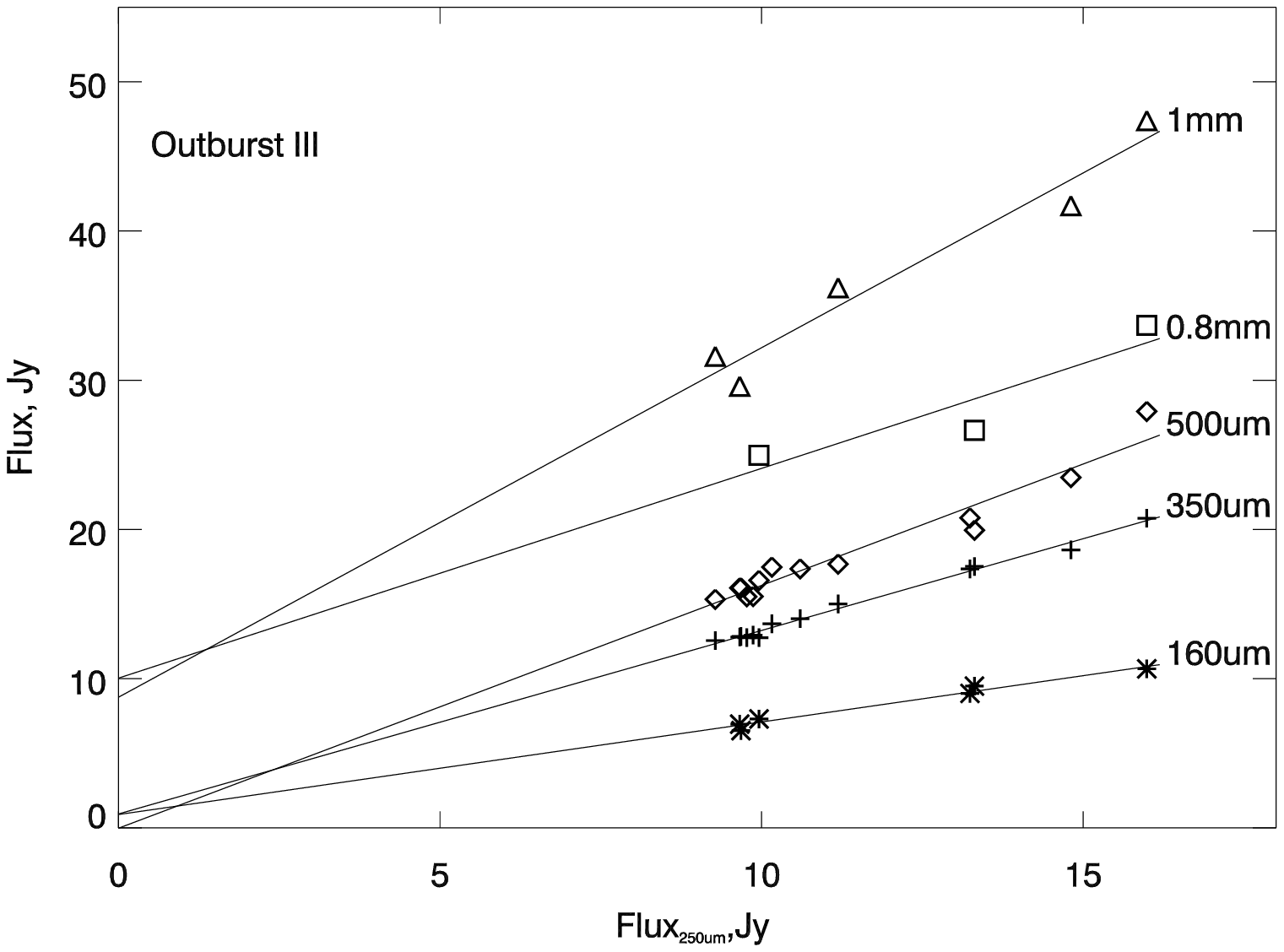}{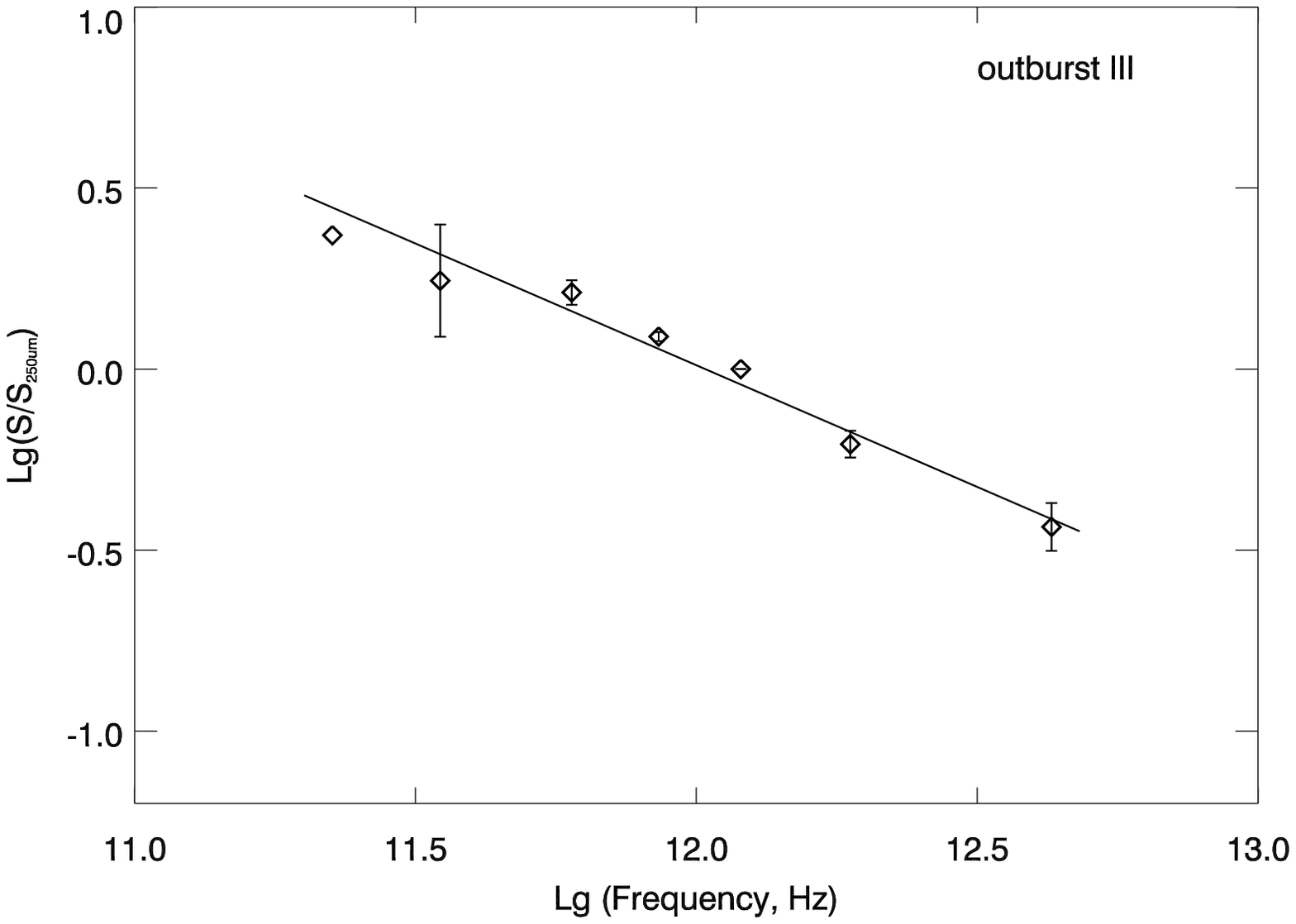}
\caption{{\it Left:} Flux-flux dependences for simultaneous measurements at different wavelengths;
flux densities at 160~$\mu$m (asterisks), 350~$\mu$m (crosses), 500~$\mu$m (diamonds), 0.85~mm (squares), and 1.3~mm (triangles)
vs. flux densities at 250~$\mu$m. {\it Right:} Relative spectral energy distribution of the synchrotron component responsible for the variability at far-IR wavelengths.} 
\label{FFIR}
\end{figure}

We have applied the same method as described in \S~\ref{HAG} for simultaneous far-IR data 
obtained during outburst III with the Herschel PACS and SPIRE photometers at 70, 160, 250, 350, and 500~$\mu$m,
along with 0.85 and 1~mm measurements obtained with the SMA and IRAM. In this case we treat two observations as simultaneous 
if they are performed within 24~hr of each other. We use 250~$\mu$m as a reference wavelength relative to which the $RSED$ is constructed,
except for measurements at 70~$\mu$m, for which 160~$\mu$m serves as a preliminary reference wavelength, with the linear dependence between 
fluxes at 160~$\mu$m and 250~$\mu$m employed to derive $C_{\rm 70}$. Figure~\ref{FFIR}~({\it left}) shows that the flux-flux relations follow linear dependences quite well.  Table~\ref{IrVar} lists the number of simultaneous observations at different wavelengths, values of $\chi^2$
for linear fits, and coefficients $C$ that represent the slopes of the linear fits.  The dependence of the coefficients on frequency is plotted in Figure~\ref{FFIR}~({\it right}), which shows a good correspondence ($\chi^2$=1.19) of far-IR points (from 70 to 850~$\mu$m) with a power law of spectral index $\alpha_{\rm IR}^{\rm syn}$=0.79$\pm$0.06. The 1~mm measurement deviates slightly from the fit, falling a factor of 0.84 below an extrapolation of the fit to the 70-850~$\mu$m $RSED$. This deviation most likely indicates that the variable component responsible for the synchrotron emission at 1~mm is partially optically thick. 
{\it The good correlation between far-IR and 1~mm flux variations (\citealt{ANN12} and Fig.~\ref{FFIR}, {\it left})
and between 1~mm and 7mm core flux and polarization behavior (Section~\ref{FPV}) suggests that
the mm-wave VLBI core is the region where the synchrotron far-IR emission orginates.}    

\subsection{Spectral Energy Distributions \label{SEDs}}
We construct spectral energy distributions, $SED$s, of the quasar from $UV$ to mm wavelengths for three epochs during outburst III: on 2010 November 3/4 (RJD: 5503), at the beginning of flare $a$; on November 19 (RJD: 5519), at the maximum of the outburst; and on December 7 (RJD:5537), during the fading branch of the outburst.
We model each spectral energy distribution of the quasar assuming that (1) the optical continuum from $UV$ to near-IR wavelengths consists of a blue component with a constant spectral index 
$\alpha_{\rm disk}$=0.44 and constant flux, $S^{\rm V}_{\rm disk}$=0.85~mJy plus a variable red (synchrotron) component with a constant spectral index $\alpha_{\rm opt}^{\rm syn}$=1.71, and (2) the mid and far-IR continuum is dominated by a variable (synchrotron) component with a constant spectral index $\alpha_{\rm IR}^{\rm syn}$=0.79 with a contribution from jet knot $K09$. We use {\it Herschel} observations to determine the minimum wavelength where the contribution from $K09$ is still significant, $\sim$80~$\mu$m.  We determine the flux density of the red variable component in $V$-band as $S_{\rm V}^{\rm syn}=S_{\rm V}^{\rm obs}-S^{\rm V}_{\rm disk}$, where $S_{\rm V}^{\rm obs}$ is the observed flux density at a given epoch. The flux density of the variable synchrotron component at 1.3~mm is calculated as 
$S_{\rm 1mm}^{\rm syn}=const\times[S_{\rm 1mm}^{\rm obs}-S_{K09}(\nu_{\rm 1mm}/\nu_{\rm 7mm})^{-\alpha_{K09}}]$, where $S_{K09}$ is the flux 
of knot $K09$ at 7~mm derived from the VLBA images. We adopt $\alpha_{K09}$=0.7, and set $const=$1.0 for November 3/4 and 1.19
after November 4 to correct for the higher opacity at mm wavelengths with respect to the far-IR measurements (see \S~5.5). 
Figure~\ref{SEDmod} shows the observed and modeled $SED$s, as well as $SED$s for different components: the variable synchrotron component, the accretion disk, and $K09$. Table~\ref{SEDpar} gives the parameters of the $SED$'s peak according to modeling.
Using the values of $S^{\rm V}_{\rm disk}$ and $\alpha_{\rm disk}$, we estimate the luminosity of the accretion, $L_{\rm disk} \approx 2.6\times10^{46}$ erg~s$^{-1}$, with integrating from 6500 to 2000~{\AA}.

Figure~\ref{SEDmod} shows good agreement between the observations and models for all three epochs. 
Especially promising is the close correspondence between the mid-IR measurements obtained with the IRTF on 2010 November at 4.9, 10.6, and 20.7~$\mu$m and the modeled $SED$, since these data-points were not involved in modeling. This supports the assumption that one synchrotron component with a constant spectral index and variable flux is responsible for the outburst at optical/IR wavelengths. The {\it IRTF} observations cover wavelengths affected by an IR excess if 3C~454.3 has a dust torus of similar properties as found in the $\gamma$-ray quasar 1222+216 \citep{MALM11}. Although neither the {\it IRTF}  nor  {\it Herschel} observations suggest the presence of an additional component between 5 and 160 $\mu$m, the measurement at 20.7~$\mu$m --- where the peak of a dust component (with temperature $\sim1200$~K) is expected in the observer's frame for 3C~454.3 --- is slightly higher than the model value. However, this measurement has substantial uncertainties. The maximum deviation from the model flux defines an upper limit to the luminosity of the dust torus of $L_{\rm dust}<5\times10^{46}$~erg~s$^{-1}$. This is quite high, allowing the dust torus to possess a higher luminosity than that of the accretion disk, while in the case of 1222+216 IR emission from hot dust has a luminosity of only 0.22 times that of the accretion disk \citep{MALM11}. 

The peak of the synchrotron {\it SED} corresponds to a break in the spectrum from a spectral index $\alpha_{\rm low} < 1$ to $\alpha_{\rm high} > 1$. The magnitude of the break, $\Delta\alpha \equiv \alpha_{\rm high} - \alpha_{\rm low}$. During Outburst III, $\alpha_{\rm IR}^{\rm syn}=$0.79 and $\alpha_{\rm opt}^{\rm syn}=$1.71, hence $\Delta\alpha = 0.92$. This is considerably greater than the value of 0.5 expected from radiative energy losses while relativistic electrons are constantly injected into the emission region. Interestingly, according to the data listed in Table~\ref{SEDpar}, the wavelength of peak flux, $\lambda_{\rm peak}$, remains essentially constant as the outburst proceeds to its global maximum on 2010 Nov 19.
If the value of $\lambda_{\rm peak}$ were determined by a balance between the timescale of radiative losses and the time for the energized electrons to cross the emission region, $\lambda_{\rm peak}$ would increase as the intensity of seed photons for IC scattering rises.
On the other hand, between 2010 Nov 19 and Dec 7, $\lambda_{\rm peak}$ increases by a factor of 2 while the optical flux drops by a factor of 2.75, consistent with the general trend expected from radiative losses as the rate of injection of relativistic electrons subsides.

\begin{figure}
\epsscale{1}
\plotone{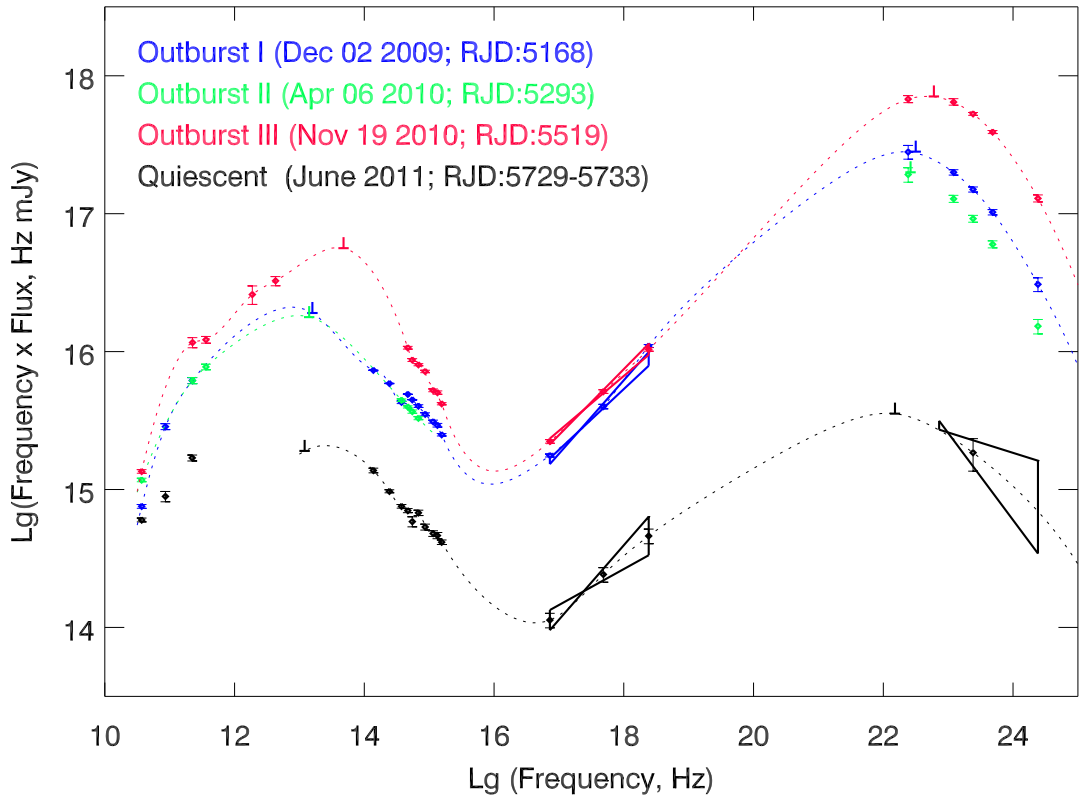}
\caption{Spectral energy distributions obtained during maxima of outbursts I (blue), II (green), III (red), and at a quiescent state (black);
the symbol ``$\perp$'' denotes low and high energy peaks of the {\it SEDs} according to modeling (see text \S~\ref{SEDs}), while dotted lines show spline approximations of the data points.} 
\label{SEDobs}
\end{figure}

Figure~\ref{SEDobs} presents {\it SEDs} of the quasar at the epochs of the $\gamma$-ray maximum, $T_\gamma^{\rm max}$, of each outburst 
as well as during a more quiescent state. The $\gamma$-ray fluxes during the outbursts are calculated using spectral parameters
given in Table~2 of \citet{ACKER10} and Table~1 of \citet{ABDO11} for a {\it LogParabola} spectrum.   
To analyze the {\it SEDs} we apply the Doppler factors derived for knots $K09$ and $K10$ (Table~\ref{Kparm}) for outburst I and III, respectively.
For outburst II we assume $\delta_{\rm II}$ to be slightly less than $\delta_{\rm I}$ (Table~\ref{SEDobsT}) because, on one hand, the outbursts have very similar {\it SEDs}, while on the other hand both the optical and $\gamma$-ray fluxes of outburst II are slightly less than those of outburst I.
For the quiescent state we use a minimum Doppler factor obtained by \citet{J05} in 1998-2001 when the flux of the VLBI core at 43~GHz 
did not exceed 5~Jy. All {\it SEDs} have a double peak shape typical of blazars with a low frequency peak  $S_{\rm peak}^{\rm LE}$ at frequency $\nu_{\rm peak}^{\rm LE}$, representing enhanced synchrotron emission, and a high frequency peak, $S_{\rm peak}^{\rm HE}$ at frequency $\nu_{\rm peak}^{\rm HE}$, which we attribute to inverse Compton scattering. The high energy peak dominates the {\it SEDs} during the active states. 

Since neither low nor high energy peaks of the {\it SEDs} are restricted by the measurements, we use the values of $S_{\rm peak}$ and $\nu_{\rm peak}$ for outburst III derived from the modeling (Table~\ref{SEDpar}) and adopt the values of $S_{\rm peak}^{\rm HE}$ and $\nu_{\rm peak}^{\rm HE}$ for outburst I from the {\it SED} presented by \citet{BON11} for 2009, December 2. Taking into account that  $\nu=\delta{\nu}'$, where ${\nu}'$ is the frequency of the emission emitted by the source and $\nu$ is the frequency of the emission received by the observer, we have estimated $\nu_{\rm peak}^{\rm LE}$ and $\nu_{\rm peak}^{\rm HE}$ for each {\it SED} assuming that a change in the Doppler factor is the main factor responsible for the outburst's increase in energy output. Using such peak frequencies, along with spline approximations of {\it SED's} data points, 
we estimate values of $S_{\rm peak}^{\rm LE}$ and $S_{\rm peak}^{\rm HE}$ for each {\it SED}. The values are given in Table~\ref{SEDobsT} and marked in Figure~\ref{SEDobs} by the symbol ``$\perp$''. Note that in 2011 June the mm-wave emission was still in some moderately active state with respect to the higher energy emission. We calculate the apparent luminosity of 3C~454.3 at the low and high energies for each activity state as $L\approx 4{\pi}D_{\rm L}^2S_{\rm peak}\nu_{\rm peak}$, where $D_{\rm L}$ is the luminosity distance, 5.489~Gpc. The values of $L^{\rm HE}$ for the outbursts listed in Table~\ref{SEDobsT} are by a factor of 2-5 lower than those presented by \citet{ACKER10} and \citet{ABDO11} because we use flux densities  at the peak frequency only to estimate the luminosity. Table~\ref{SEDobsT} shows that during the quiescent state the luminosity at low energies is comparable to the luminosity of the accretion disk. We estimate the luminosity of a synchrotron component of the emission at the quiescent state as $L_{\rm q}^{\rm syn}\approx L_{\rm q}^{\rm LE}-L_{\rm disk}$. Attributing an enhanced emission at low energies during the outbursts to a change of the Doppler factor, we derive the luminosity of the synchrotron component for each outburst as $L^{\rm syn}\approx L_{\rm q}^{\rm syn}(\delta/\delta_{\rm q})^4$. Making the same assumption that enhanced luminosity at high energies is caused by a change of the Doppler factor with respect to the quiescent state and adopting $L_{\rm q}^{\rm IC}=L_{\rm q}^{\rm HE}$, we calculate values of the luminosity at high energies expected within such assumptions,
$L^{\rm IC}\approx L_{\rm q}^{\rm IC}(\delta/\delta_{\rm q})^4$. Table~\ref{SEDobsT} shows reasonable agreement between $L^{\rm LE}$ and $L^{\rm syn}$, and $L^{\rm HE}$ and $L^{\rm IC}$. However, while for outburst III $L^{\rm HE}$ and $L^{\rm IC}$ are very consistent,
$L^{\rm syn}$ is larger than $L^{\rm LE}$  by a factor of 5. The opposite occurs for outbursts I and II: there is good agreement between
$L^{\rm syn}$ and $L^{\rm LE}$, while $L^{\rm IC}$ is underestimated by a factor of 5 with respect to $L^{\rm HE}$. Taking into account that
the mm-wave emission region is larger than the $\gamma$-ray emission region by a factor of $\sim$100, the discrepancies suggest slight variations
of the Doppler factor within the mm-wave emission region at a given epoch, with a higher $\delta$ than the mean value in the volume where the $\gamma$-ray emission originates. {\it Therefore, it appears that a change in Doppler factor can explain differences in the amplitudes of the outbursts, as was proposed previously by \citet{vil07} and \citet{RAI11}}. 

\section{Polarization Behavior}
Here we analyze our entire set of optical and VLBI polarization data, which were collected from 2008 June
to 2011 December. The data set includes 706 measurements of optical polarization 
along with simultaneous photometric measurements in $R$ band (390 cases), and 244 spectropolarimetric observations, as well as 52 measurements of polarization in the VLBI core at 43 GHz that coincide with optical observations within 2 days.
\subsection{Dependence of Optical Linear Polarization on Wavelength}
\begin{figure}
\epsscale{1}
\plotone{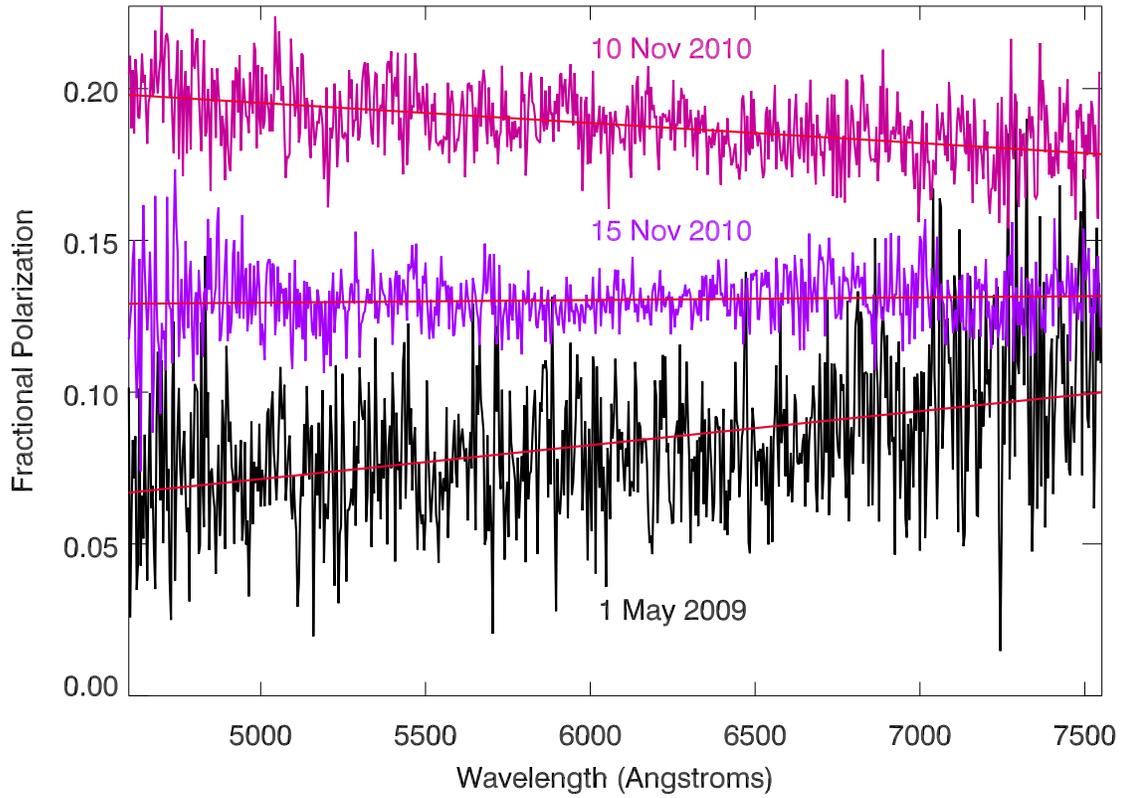}
\caption{Degree of polarization spectra of 3C~454.3 in the observer's frame at different brightness levels; the red solid lines represent a linear fit of the $P(\lambda)$ dependence.} 
\label{PolSpec}
\end{figure}

\begin{figure}
\epsscale{1}
\plotone{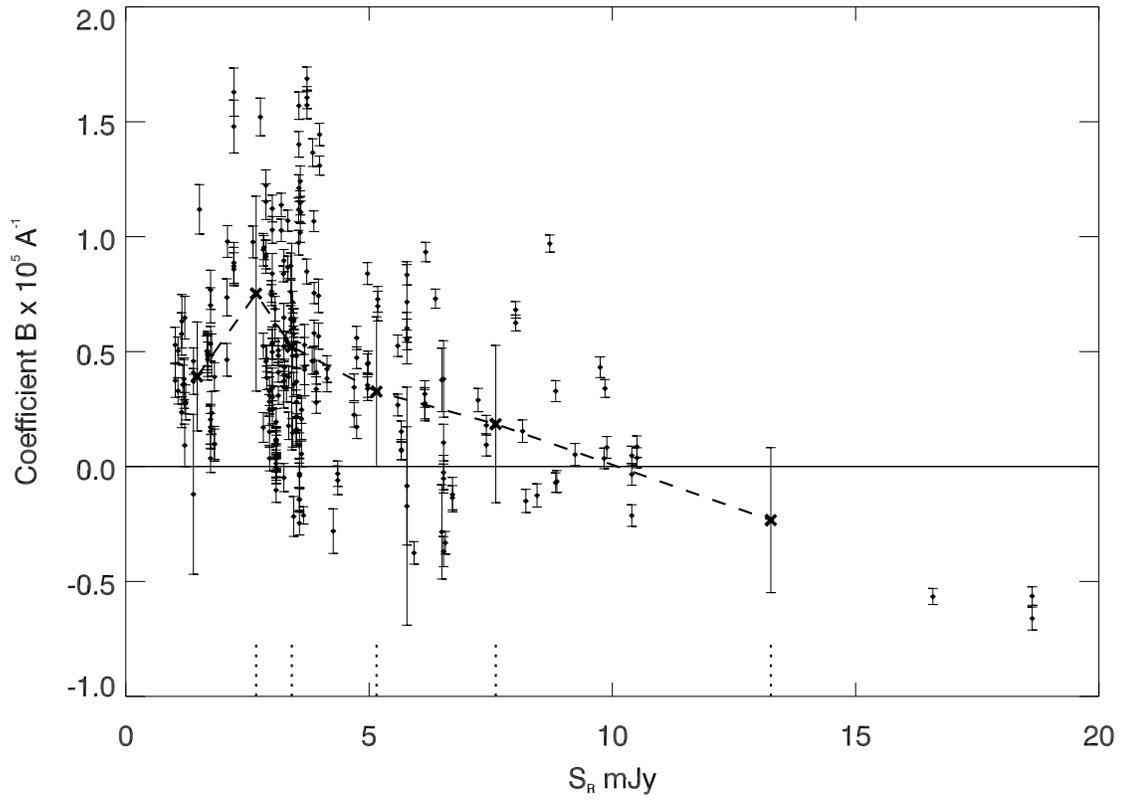}
\caption{Slope $B$ of $P(\lambda)$ dependence vs. flux density of 3C~454.3 in $R$ band; the dashed line connects the average values of $B$ (crosses); dotted vertical segments show intervals of the averaging; the solid line marks $B$=0.} 
\label{Plambda}
\end{figure}

 The spectropolarimetric observations performed at Steward Observatory provide spectra of the normalized linear polarization Stokes parameters $q$ and $u$ in the range of 4000--7550~{\AA} with a dispersion of 4~{\AA}. We use the Stokes spectra to calculate the degree of polarization as a function of wavelength, $P(\lambda)=\sqrt{q(\lambda)^2+u(\lambda)^2}$, within the range of 4500--7000~{\AA}, which avoids noisy edges of the $q$ and $u$ spectra. Since not all spectropolarimetric observations were accompanied by photometric measurements in $V$-band, 
we used the light curve in $R$ band to associate polarization and photometric measurements 
if they were performed within 3~hr of each other. Figure~\ref{PolSpec} shows three examples of the $P(\lambda)$ dependence that appear to be representative:
(1) the degree of polarization increases with wavelength when the optical emission is weak and $P_{\rm opt}$ is moderately high (e.g., on 2009 May 1, $S_{\rm R}=$1.51$\pm$0.02~mJy and $P_{\rm opt}=$7.94$\pm$0.09\%); (2) the degree of polarization does not depend on wavelength when the optical emission is sufficiently bright and $P_{\rm opt}$ is high (e.g., on 2010 Nov 15 $S_{\rm R}=$10.51$\pm$0.18~mJy and $P_{\rm opt}=$13.00$\pm$0.04\%); and (3) the degree of polarization decreases with wavelength when the optical emission is very bright and highly polarized (e.g., on 2010 Nov 10 $S_{\rm R}=$18.63$\pm$0.36~mJy and $P_{\rm opt}=$18.84$\pm$0.05\%). We use a linear fit, $P(\lambda)=A+B\lambda$, to approximate the dependence of the fractional polarization on wavelength for each spectrum obtained from 2009 April to 2011 August (244 spectra). Examples of the fits are shown in Figure~\ref{PolSpec} by red solid lines with slope $B$ equal to (0.112$\pm$0.011)$\times$10$^{-4}$~ {\AA}$^{-1}$, (0.38$\pm$0.44)$\times$10$^{-6}$~{\AA}$^{-1}$, and ($-$0.661$\pm$0.050)$\times$10$^{-5}$~{\AA}$^{-1}$ for 2009 May, 2010 Nov 15, and 2010 Nov 10, respectively. Figure~\ref{Plambda} plots derived values of slope $B$ vs. brightness in $R$ band for all observations, while Table~\ref{AcontT} lists the average values of slope $B$
for different brightness intervals. Figure~\ref{Plambda} and Table~\ref{AcontT} show that there is a change in the $P(\lambda)$ dependence with brightness: for $S_{\rm R}\le$4.5~mJy the
coefficient $B$ is positive-definite despite significant scatter, while for $S_{\rm R}>$4.5~mJy $B$ is close to zero. This agrees 
very well with the finding discussed in \S~\ref{BLROC} that the optical continuum consists of two components,
blue (BBB) and red (synchrotron). The contribution of the BBB to the optical emission is significant in the blue part of the spectrum, especially at low flux levels. Dilution by this component leads to a degree of polarization that increases at longer wavelengths (Fig. ~\ref{Plambda}),
as found previously by \citet{SMITH88}. The latter trend suggests that the emission of the BBB is unpolarized. 
The $P(\lambda)$ dependence disappears when the nonthermal component dominates the total optical emission, as expected for the synchrotron emission over a relatively narrow wavelength range.  Figure \ref{Plambda} and Table~\ref{AcontT} also show that at very bright flux levels ($S_{\rm R}>$10~mJy) $B$ is negative, which implies a higher degree of polarization at shorter wavelengths. Although the number of such observations is small, such a tendency supports models in which electrons are accelerated at a front and then lose energy to radiation. In this case, higher-energy electrons occupy a smaller volume beyond the front --- with a more uniform magnetic field --- than do lower-energy electrons that radiate at longer wavelengths \citep[e.g.,][]{MG85,MGT92,MJ10}. Note that we have investigated dependence of position angle of polarization on wavelength and found that $\chi_{\rm opt}$ does not depend on $\lambda$, either at high or moderate degrees of polarization, although at a low level
of polarization uncertainties of  $\chi_{\rm opt}(\lambda)$ increase significantly.

\subsection{Dependence of Optical Linear Polarization on Brightness}
\begin{figure}
\plottwo{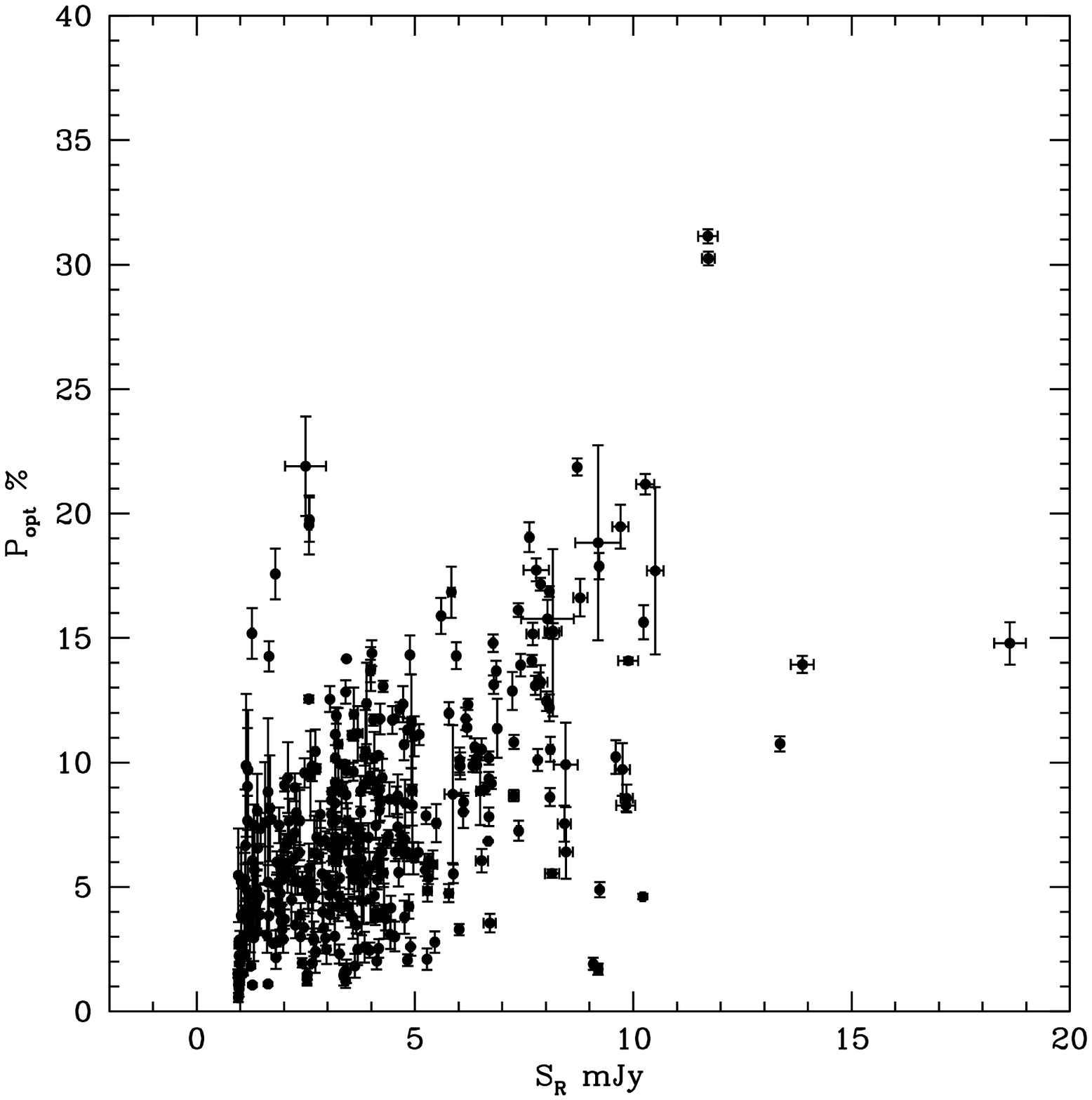}{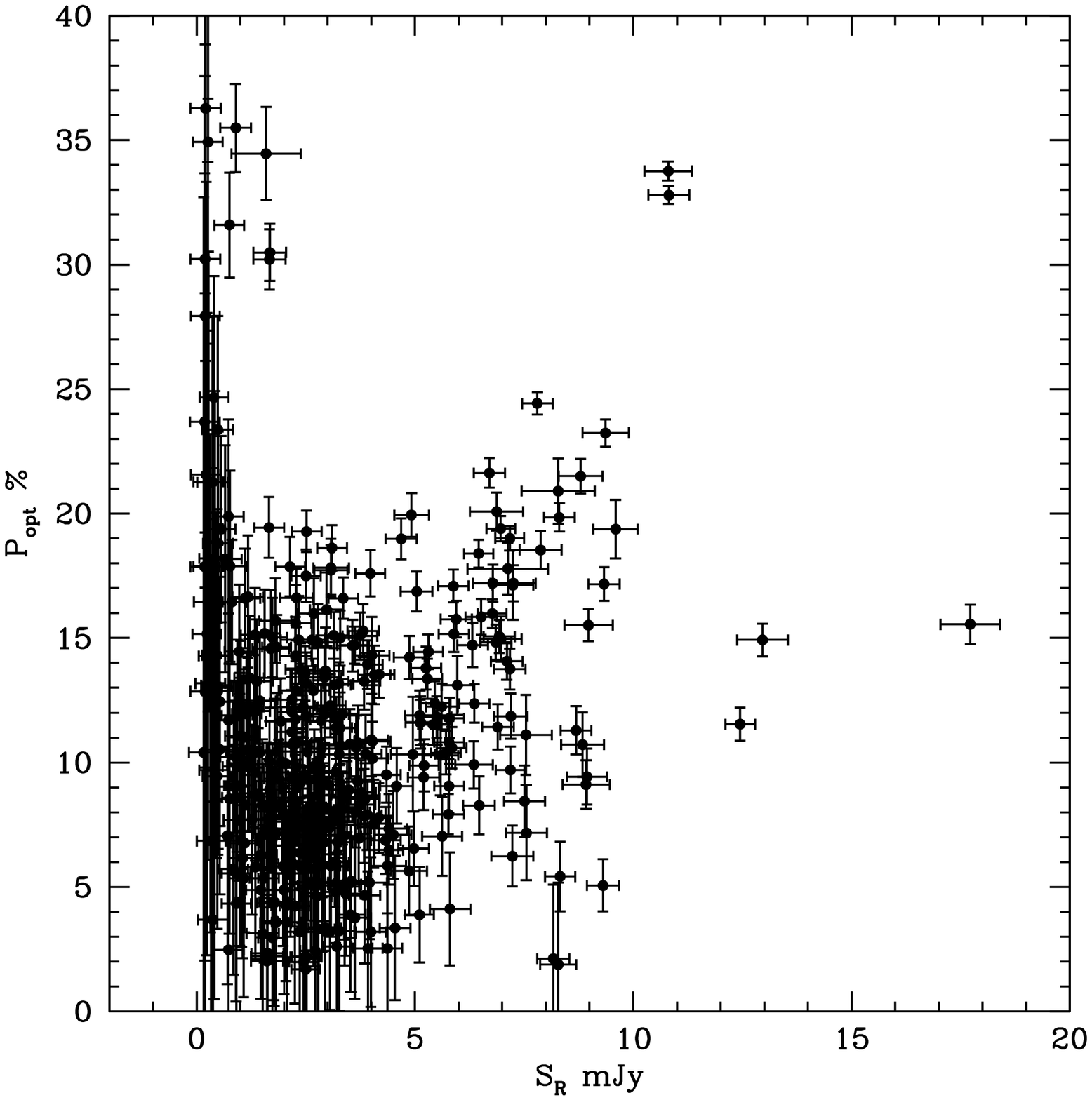}
\caption{{\it Left:} Dependence of the degree of observed optical polarization on brightness of the quasar in $R$ band. {\it Right:} Dependence of the degree of optical polarization of the synchrotron component  on its brightness in $R$ band.} 
\label{OPF}
\end{figure}

Figure~\ref{OPF} ({\it left}) shows the dependence between the degree of optical polarization and flux
of 3C~454.3 in $R$ band. The fractional polarization changes from $<$0.5\% to 30\%, while the flux 
varies from $\sim$1~mJy to 20~mJy.  The Spearman rank correlation coefficient, $\rho$=0.565, gives a statistical significance of 98.7\% that the values are related, with the degree of polarization rising along with the flux. We have shown above that the optical continuum consists of
two components: unpolarized thermal (BBB) and polarized synchrotron (jet) emission. According to \S\ref{BLROC} the contribution of the BBB in $R$ band is $S_{\rm R}^{\rm disk}\sim$0.91~mJy. 
If we assume that the thermal component was
constant during our observations and completely unpolarized, we can derive the flux and degree of polarization of the variable synchrotron component: $S_{\rm R}^{\rm var}=S_{\rm R}^{\rm obs}-S_{\rm R}^{\rm disk}$ and $P_{\rm opt}^{\rm var}= P_{\rm opt}\times S_{\rm R}^{\rm obs}/S_{\rm R}^{\rm var}$  with $S_{\rm R}^{\rm disk}$=0.91$\pm$0.16~mJy (\S~\ref{BLROC}).  Figure~\ref{OPF} ({\it right}) shows the dependence between the degree of optical polarization and flux for the variable component. In this case an increase of the degree of polarization is observed also at a very 
small flux level of the variable source, with $P_{\rm opt}^{\rm var}$ rising up to $\sim$40\% while 
$S_{\rm var}<$1~mJy, although the uncertainties in $P_{\rm opt}^{\rm var}$ are significant owning to an uncertainty in the derived BBB flux. 
An increase of the degree of polarization along with the flux level might be a signature 
of shock processes owing to ordering of the magnetic field in the shocked region if the quiescent 
jet has a chaotic magnetic field. However, Figure~\ref{OPF} ({\it right}) suggests that in a completely
quiescent state the synchrotron component originates in a region with a very well-ordered magnetic field.
This implies that the optical synchrotron emission during quiescent and active states arises from different 
regions in the jet. These two synchrotron components (quiescent and active) perhaps possess 
different polarization properties that can explain the minimum values of $P_{\rm opt}^{\rm var}$ at fluxes of 2-3~mJy when a quiescent synchrotron component
has a flux comparable to that of an active synchrotron component at a moderate stage of activity. A ``competition''
between the thermal and different synchrotron components at a moderate flux level might be responsible
also for the largest scatter in the slope of the $P(\lambda)$ dependence seen at flux level between 2 and 4~mJy (Table~\ref{AcontT}).

\begin{figure}
\plottwo{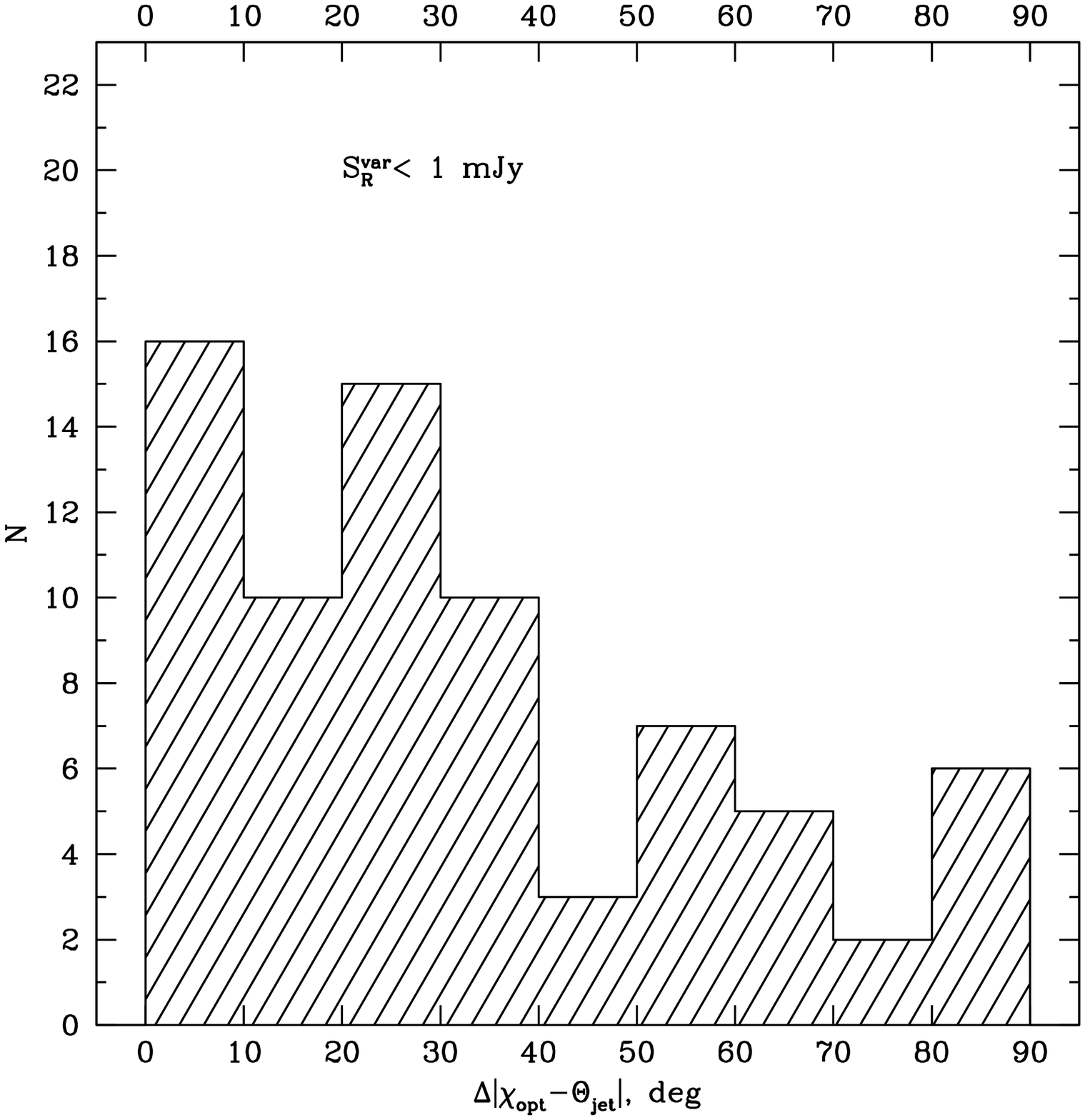}{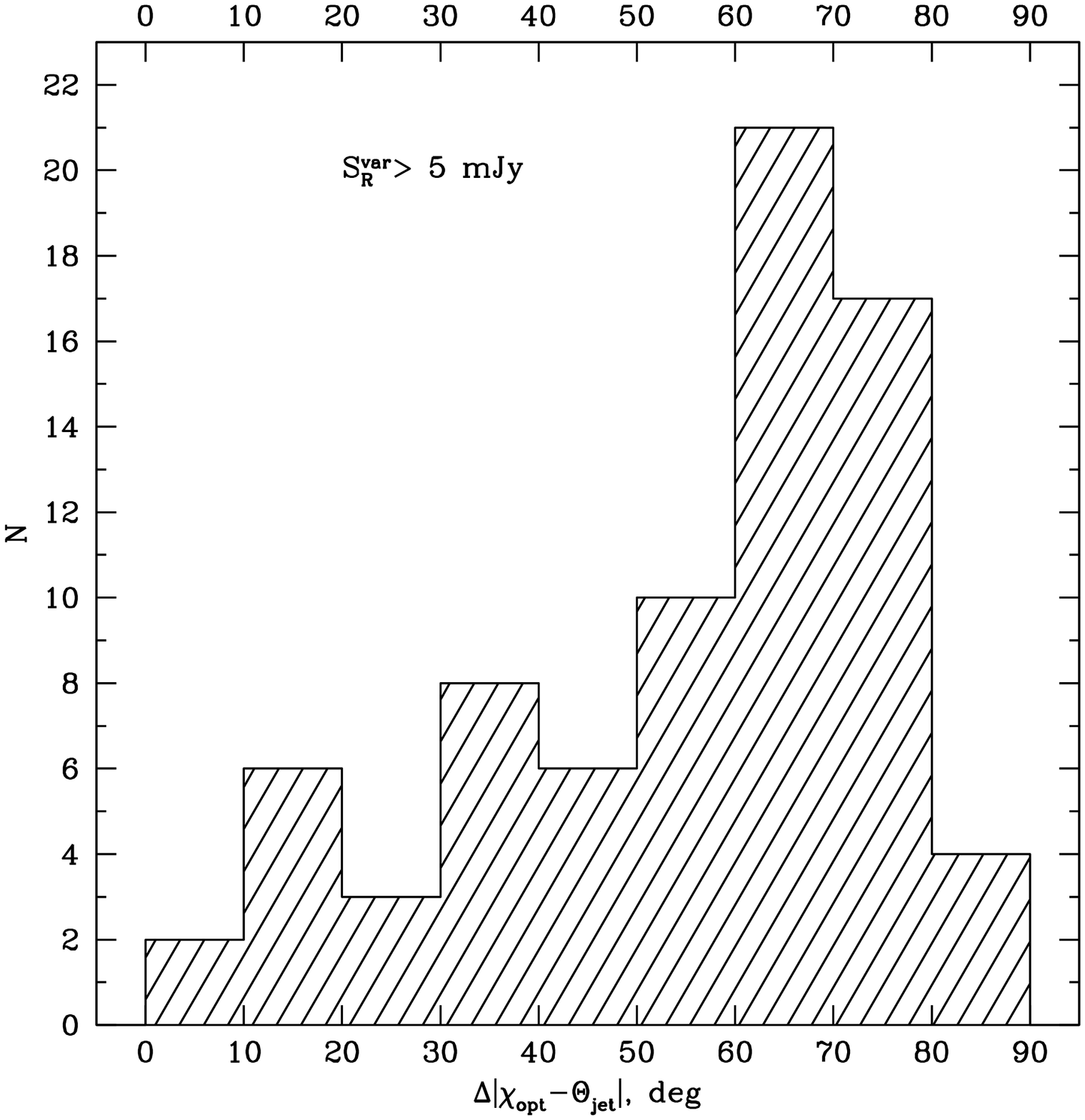}
\caption{Distribution of position angle of optical polarization with respect to jet axis
during low ({\it left}) and high ({\it right}) levels of optical synchrotron emission.} 
\label{EVPA}
\end{figure}

If the optical synchrotron emission originates in different regions during quiescent and active states,
we can expect that the magnetic field configuration of the regions is different.
Figures~\ref{EVPA} show the distributions of the position angle of optical polarization with
respect to the jet direction for $S_{\rm var}<$1~mJy ({\it left}) and $S_{\rm var}>$5~mJy
({\it right}). It is clear that the position angle of polarization differs during quiescent and active states, with $\chi_{\rm opt}$ tending to align nearly along the jet direction during quiescent states, while
being oblique, or nearly perpendicular to the jet direction, during active states. This suggests that the optical synchrotron emission during a quiescent state originates in a region with a well ordered toroidal magnetic field. Such a region is most likely located in the magnetically dominated part of the jet, relatively close to the black hole (within several thousand Schwarzshild radii; \citealt{MKU00,McK06}). On the other hand, the optical synchrotron emission during an active state originates in a region with a chaotic magnetic field, although a further increase of the flux leads to ordering the magnetic field
along the jet axis. The latter implies that the region of flaring synchrotron emission is located farther downstream the jet where either spiral loops of the toroidal magnetic field are very loose \citep{LPG05} or the effects of velocity shear align the magnetic field with the jet axis \citep{LANG80,FRANI09}. 
  
\subsection{Comparative Analysis of Optical and Millimeter-wave Polarization during Gamma-Ray Outbursts}
\begin{figure}
\plotone{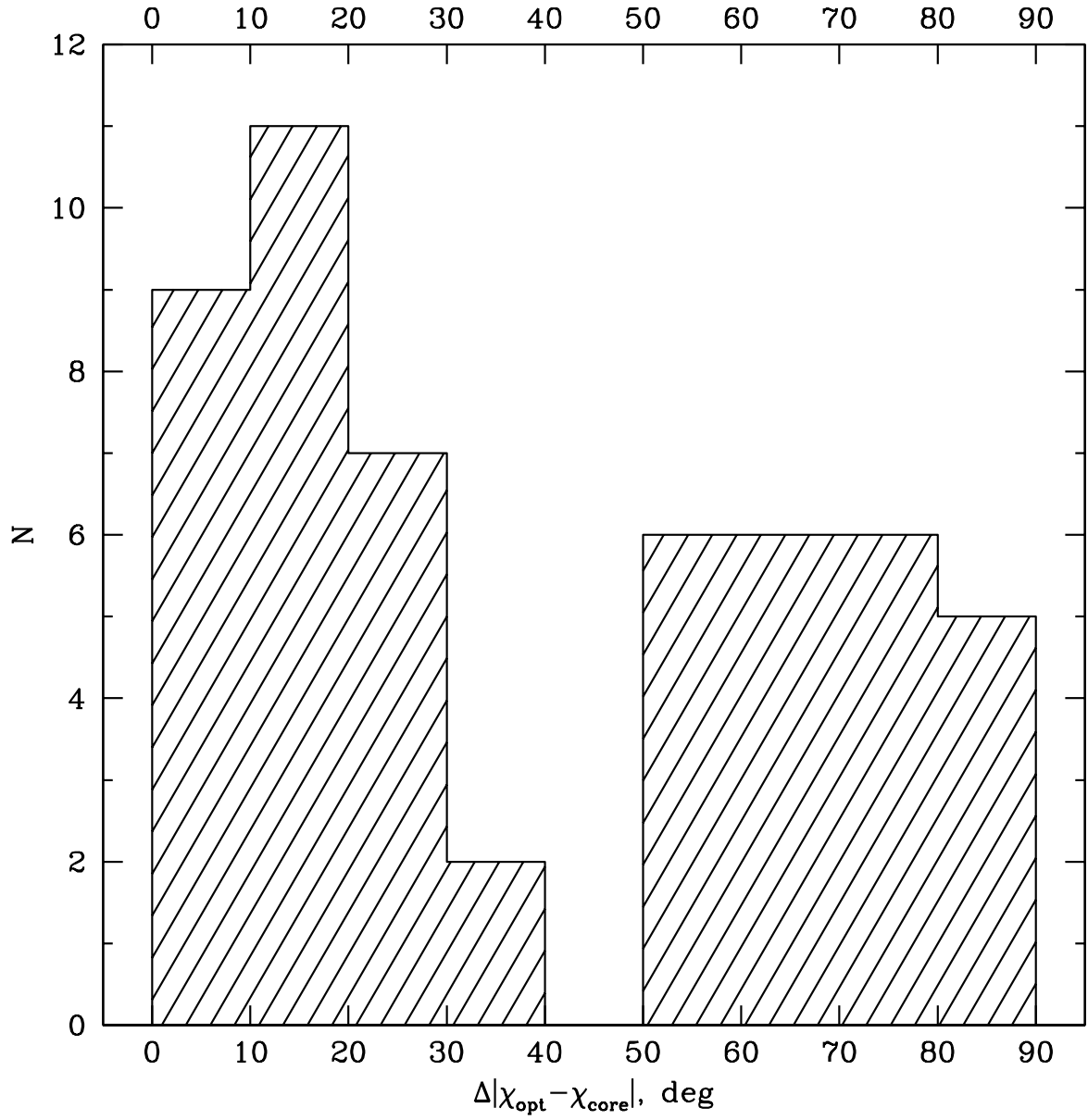}
\caption{Distribution of offsets between position angle of optical polarization and position angle 
of polarization in the VLBI core at 7~mm.} 
\label{dEVPA}
\end{figure}

We compare the position angle of the polarization (EVPA) at optical wavelengths and in the VLBI core at 43~GHz
for simultaneous observations. Figure~\ref{dEVPA} shows the distribution of differences between 
the optical EVPA, $\chi_{\rm opt}$, and EVPA in the VLBI core, $\chi_{\rm core}$. The distribution is bimodal, with  $\chi_{\rm opt}$ either similar 
to $\chi_{\rm core}$ or different by $>$50$^\circ$. Note that a
good agreement between the position angles, $|\chi_{\rm opt}-\chi_{\rm core}|<$20$^\circ$ (20 cases), is observed when the source is brighter than 2~mJy in $R$ band.  
 
The distribution therefore suggests that there may be a relationship between the properties of optical and VLBI core polarization when the source is in an active state. We find stronger evidence for such a connection in observed similarities between optical polarization parameters and those in the core during outbursts.
Figure~\ref{polG} shows the parameters of the optical and core polarization during outbursts I, II, and III (we also use the polarization data obtained in $V$-band by \citet{SAS12} during outburst I).  The data for each outburst are plotted relative to the corresponding time of $\gamma$-ray flux maximum, $T_\gamma^{\rm max}$, listed in Table~\ref{Gparm}. In general, there is an increase of $P_{\rm opt}$ during the outbursts. However, measurements at the peak of outbursts I and II reveal a significant drop of the degree of optical polarization (down to 2-3\%) over a period of $\sim 3-4$~days centered at $T_\gamma^{\rm max}$. The degree of polarization in the core reveals similar behavior: $P_{\rm core}$ increases during outbursts up to 4\%, but it drops below 1\% close to  $T_\gamma^{\rm max}$ for outburst II, for which there are observations within 2~days of $T_\gamma^{\rm max}$. 

The behavior of the position angle of polarization during outbursts is extremely interesting. Figure~\ref{polG} shows that: 
i) near the beginning of an outburst, $\chi_{\rm opt}$ is relatively stable at $\sim-$25$^\circ$, while  $\chi_{\rm core}$ differs from $\chi_{\rm opt}$
by $\sim$90$^\circ$ for outburst III, which indicates that the core is most likely optically thick at 43~GHz; ii) $\sim$10 days before 
$T_\gamma^{\rm max}$, $\chi_{\rm opt}$ starts to rotate, although it does so in opposite directions for outbursts I and III; iii) at the peak of a $\gamma$-ray outburst $\chi_{\rm opt}$
fluctuates on a timescale of several hours; iv) $\sim 5-10$~days after $T_\gamma^{\rm max}$,  $\chi_{\rm opt}$
starts a new cycle of rotation with the same counter-clockwise direction for both outbursts I and III, and at a similar rate of $\sim$9$^\circ$ per day over at least 10-15~days; v) despite similarities in the rotation of $\chi_{\rm opt}$, there is a constant offset of $\sim$40$^\circ$ between the rotation lines (the dashed lines in Fig.~\ref{polG}, {\it bottom panel}); this corresponds to a shift in the directions of the trajectories of knots $K09$ and $K10$ within 0.2~mas of the core (see Fig.~\ref{Ktraj});  vi) $\sim$35~days after $T_\gamma^{\rm max}$, the optical EVPA stabilizes at $\chi_{\rm opt}\sim$0$^\circ$, which is the EVPA of the core as well as that of $K09$ and $K10$ when they first appear in the jet (see Fig.~\ref{Krevpa}). {\it Therefore, although the optical polarization varies dramatically during $\gamma$-ray outbursts, the behavior of the optical polarization maintains a tight connection with the properties of the mm-wave core and superluminal knots $K09$ and $K10$.} 

\section{Discussion}
The multi-frequency outbursts of the quasar 3C~454.3 in 2009 and 2010 have been analyzed by different authors. \cite{BON11}
have modelled simultaneous SEDs at different stages of outburst I. These authors use a leptonic, one-zone synchrotron and inverse Compton model discussed in detail by \cite{GT09}. They successfully reproduce the large variations in the $\gamma$-ray flux by
varying the power injected in relativistic electrons (by a factor of 10 from the quiescent to the highest state), and the bulk 
Lorentz factor (from 15 to 20). They place the dissipation zone of the outburst within the BLR ($\sim$1000 Schwarzschild radii).
\cite{BON11} find it necessary to decrease the value of the magnetic field as the outburst progresses to fit the X-ray
spectrum. This increases with respect to the BH the location of the dissipation region during the highest state although by less than a factor of 1.4. 
A one-zone leptonic model has been used also by \cite{verc10}, \cite{pac10}, and \cite{verc11} to explain the dynamic behavior of the SEDs during outbursts I and III. The authors employ the synchrotron, SSC, and EC mechanisms, with seed photons for IC scattering provided by the accretion disk and BLR. The scattering is produced by a blob of relativistic plasma moving with $\Gamma\sim$20-25 at $\sim$0.05~pc from the BH. These authors reproduce the SEDs of the quasar rather well. However, our findings, summarized below, challenge one-zone leptonic models that place the dissipation zone of outbursts so close to the BH. Either the BLR of the quasar has a different geometry than assumed, as proposed by \cite{LT13} and \cite{ISLER13}, or more complicated models are neeeded to explain the multi-frequency behavior of 3C~454.3 during outbursts.

We find that the correlation analysis of the high-energy, optical, IR, and mm-wave variations observed in 3C~454.3 in 2009-2011
indicates that the events seen at different wavelengths were co-spatial. However, the size of the emission regions is different at different wavelengths, with the $\gamma$-ray radiation occupying the smallest volume. The behavior of the optical polarization along with the 43~GHz polarization in the parsec scale jet imply  that
i) the degree of optical polarization tends to increase when the flux in $R$ band is less than $\sim$2~mJy (a quiescent state) and 
more than  $\sim$4~mJy (an active state); ii) during a quiescent state the position angle of optical polarization tends to align 
with the jet direction, which suggests a toroidal configuration of the magnetic field; iii) there is better agreement between the optical and VLBI core polarization parameters during active states with both optical polarization angle and EVPA in the core having a preferred direction -- perpendicular to the jet axis; and iv) the optical synchrotron emission during quiescent states originates in a location where the magnetic field is well-ordered, perhaps in the acceleration and collimation zone (ACZ) upstream of the mm-wave core, while during active states the location of the optical synchrotron emission moves down the jet, closer to the VLBI core. 

We find that the VLBI core of the parsec scale jet was active during all three high energy events, and 
that the two events corresponding to the most dramatic $\gamma$-ray outbursts were accompanied by the ejection of 
superluminal knots with the highest Lorentz
factor of $\sim$30 corresponding to the most dramatic $\gamma$-ray outburst. These emission properties 
and connections cause us to place the event's site within the mm-wave VLBI core located $\sim$20~pc from the BH. The triple-peaked 
profile of the light curves during each outburst implies that the core with the angular size of 0.05$\pm$0.02~mas 
contains three locations where the
emission reaches a local maximum. In fact, \citet{J10} found evidence for such triple structure in
``super-resolved'' 43 GHz VLBA images of 3C~454.3. We discuss below three possible 
theoretical models that could be compatible with this general picture, although each requires future
detailed computations to verify how well they can reproduce the observed behavior of 3C~454.3.

1. {\it Recollimation Shocks and the Turbulent Extreme Multi-Zone Model (TEMZ).~~~}
In this model, we associate the triple struture of the core with a system of three alternating conical ``recollimation'' shocks and rarefactions, as suggested by \citet{dm88}, \citet{GOM97}, \citet{KF97}, \citet{MAR06}, and \citet{CAW06}. The outbursts occur as a disturbance --- an increase in the energy and/or velocity of the flow in the jet, presumably originating at the input site at the jet base --- crosses these standing shocks. The disturbance may correspond to a moving shock, but this is not a general requirement. Each standing shock increases the density, compresses the magnetic field component parallel to the shock front, and accelerates particles. The level of, and variations in, linear polarization
suggest that the magnetic field direction varies across the emitting region, as expected if the jet plasma is turbulent.
The TEMZ model \citep{MAR12,MAR13} calculates the emission expected from such a turbulent plasma flowing down the jet
after it crosses a single standing conical recollimation shock in the millimeter-wave core. The shock energizes electrons
and compresses the plasma, leading to strong emission downstream of the shock. The TEMZ code computes
the spectral energy distribution from synchrotron radiation and inverse
Compton scattering, as well as the linear polarization of the synchrotron
emission at various frequencies, as a function of both time and location
within the jet. The model for 3C~454.3 assumes a randomly
oriented magnetic field upstream of the shock, with each cell (after
compression by the shock) having its own field direction and maximum
electron energy. The energy density at the jet input varies with time
stochastically within a power spectrum of a power-law shape with a slope of $-$1.2,
similar to that observed for $\gamma$-ray flux variations \citep{LARS13}. The plasma in
each cell has a velocity that is the vector sum of the general flow and
local turbulent velocities. Seed photons for the scattering include
infrared emission by hot dust in a parsec-scale molecular torus, as well
as synchrotron and inverse Compton photons from a Mach disk on the jet axis.
The inclusion of a turbulent magnetic field and non-uniform maximum electron energy across the emission region 
reproduces in a general manner the fluctuations in polarization and flux observed in 3C~454.3 at different wavelengths \citep{MAR12,MAR13}. \cite{ANN12} use the TEMZ code to fit several SEDs of 3C~454.3 during outburst III. The derived SEDs match the millimeter to optical and $\gamma$-ray spectra quite well, although the observed X-ray spectrum is somewhat steeper than in the model calculations that challenges the model. In addition, the dust would need to have a luminosity $\sim$1$\times$10$^{46}$erg~s$^{−1}$, which is half the luminosity of the accretion disk. The distribution of dust would also need to be very patchy
in order to extend over a large enough volume to provide a high density of seed photons at distances $\sim$15-20~pc from
the BH. However, in the TEMZ model, the  emission at different frequencies occupies a volume whose size is inversely related to the frequency of observation. This dictates that the average degree of linear polarization, as well as the level of variability of both the flux and polarization, should increase with frequency \citep{MAR13}. Indeed, we see such a behavior during the outbursts (see Tables~1-4 and Fig.~\ref{PolSpec}).

2. {\it Mini-Jet Model.~~~} 
In the ``mini-jet'' scenario \citep{BFR08,gian09,gian13} compact emetting regions (blobs) move relativistically
with a Lorentz factor $\sim$100 within a jet with a bulk Lorentz factor $\sim$10. Such extremely fast 
motions are possible in a magnetically dominated flow where magnetohydrodynamical waves approach the speed of light and a substantional fraction of the jet luminosity is dissipated in reconnection events. In addition, the beaming can be supplemented by an anistropic electron distribution, such that the electrons stream toward the line of sight in some of the blobs \citep{cer12}. Although the model has been proposed to explain dramatic TeV energy variability on timescale of $<$1~hr in some BL~Lac objects, it can be adapted to 3C~454.3 with less severe constraints on the Lorentz factors.
The model does not depend significantly on the location of the dissipation zone. The main constraint
is connected with the timescale for the reconnection to occur, which must be shorter than the observed 
timescale of the variability, $<$3~hr in the case of 3C~454.3. The triple structure of the outburst light curves
and core would require three different physical locations where magnetic reconnections occur.

3. {\it Current-Driven Instability (CDI).~~~}
According to analytical studies \citep [e.g., ][]{VK04} and numerical simulations \citep [e.g., ][]{McK06, TCH08}, relativistic jets are accelerated by magnetic stresses in an extended region dominated by the Poynting flux that do not operate uniformly across the jet radius. This creates a gradient in the bulk Lorentz factor with distance from the jet axis. Magnetically dominated plasma with a toroidal magnetic field is known to be subject to CDI. \citet{NT09} find that the sign of the poloidal
velocity shear is important for stability of the jet: jets with positive velocity shear (Lorentz factor increasing with radius) are stable, while jets with velocity shear changing sign are unstable.
\citet{NB12} identify two types of unstable modes, {\it exponential} and {\it overstable}, and show
that the growth rates of exponential modes decrease with increasing velocity shear. These authors note
that their results are most suitable at distance scales beyond the main
acceleration/collimation zone (ACZ), where the effects of velocity shear are expected to be most prominent. In 
3C~454.3 the position angle of optical polarization and polarization in the VLBI core during active states
is predominantly perpendicular to the jet, so that the mean magnetic field is parallel to the jet.
Such a configuration of the magnetic field is expected if velocity shear stretches and orders the fields 
lines along the flow \citep{LANG80,FRANI09}. \citet{NB12} propose
that CDI can provide an important energy dissipation mechanism and play a crucial role in converting 
a magnetically dominated jet into a matter-dominated flow that produces the observed emission from blazars.
The polarization behavior of 3C~454.3 suggests that the mm-wave core is located at the end of 
the ACZ. A similar conclusion was drawn from polarization studies
of a number of blazars  at optical and mm wavelengths by \citet{J07}. They suggested that the ACZ ends
between the VLBI cores at 3 and 7~mm, which might also be the radiative dissipation zone.
 
These three models are not necessarily mutually exclusive, i.e., more than one physical mechanism might be
operating within the core. For example, CDI could cause the plasma to become turbulent downstream of the jet
ACZ, after which the plasma crosses recollimation shocks; or the turbulence
instigated by CDI could create the conditions under which magnetic reconnection events are common.
The models are all potentially capable of explaining how the timescale of
variability at optical
and $\gamma$-ray frequencies can be as short as hours for emission arising
in the parsec-scale
jet. One factor is that the jet is very narrow, with an opening half-angle
of 0.014 radians \citep{J05}.
The width of the parsec-scale jet is therefore of order $10^{17}$~cm. Taking into
account polarization properties of the quasar, 
the size of a turbulent cell or reconnection region could be $\sim 0.1$ times
this width. Turbulent motion could enhance
the Doppler factor $\delta_{\rm c}$ of a cell of plasma \citep{NP12} above the mean value ($\sim 30$),
as could fast streams originating in magnetic reconnections \citep{BFR08,gian09,gian13}. The
timescale of flux variability resulting from these factors can be as short as
$\tau_{\rm var}<10^{16}(1+z)/(c\delta_{\rm c})\lesssim$5.7~hrs, compatible with the minimum observed value.

\section{Conclusions}

\begin{figure}
\epsscale{1}
\plotone{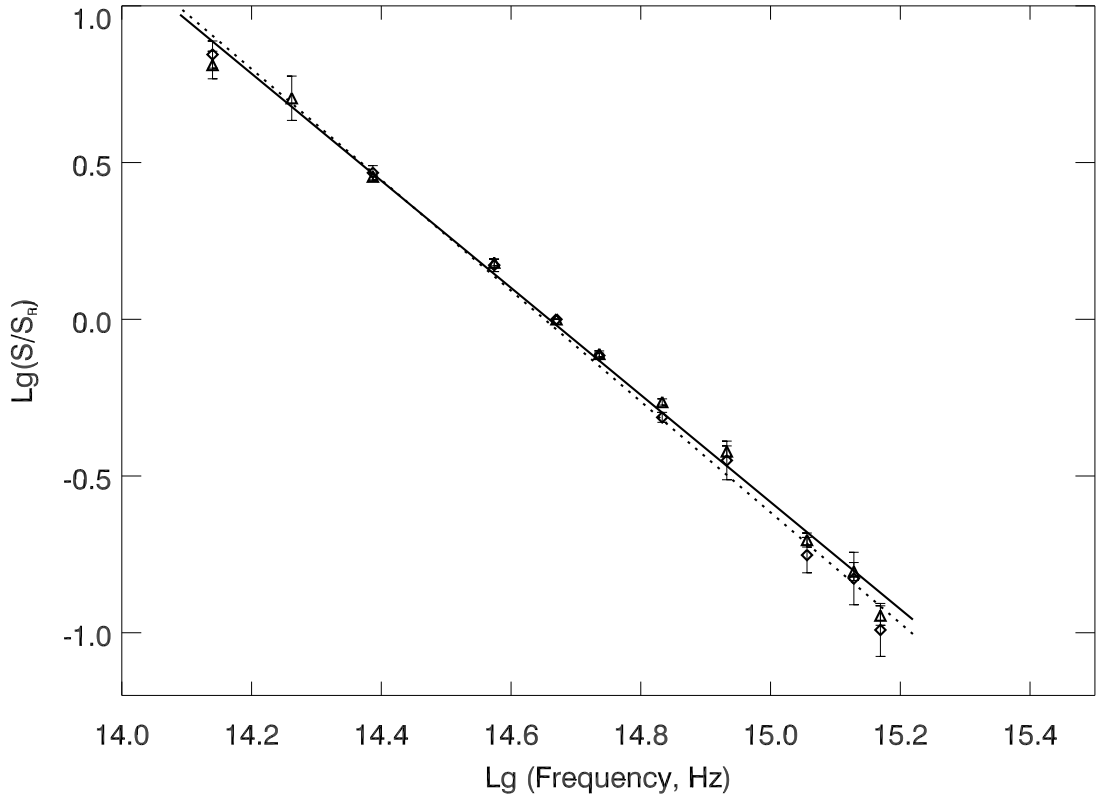}
\caption{Relative spectral energy distribution of the synchrotron component responsible for the variability at optical and near-IR
wavelengths during outbursts I (diamonds and dotted line) and III (triangles and solid line).} 
\label{RSED}
\end{figure}

\begin{figure}
\epsscale{1}
\plotone{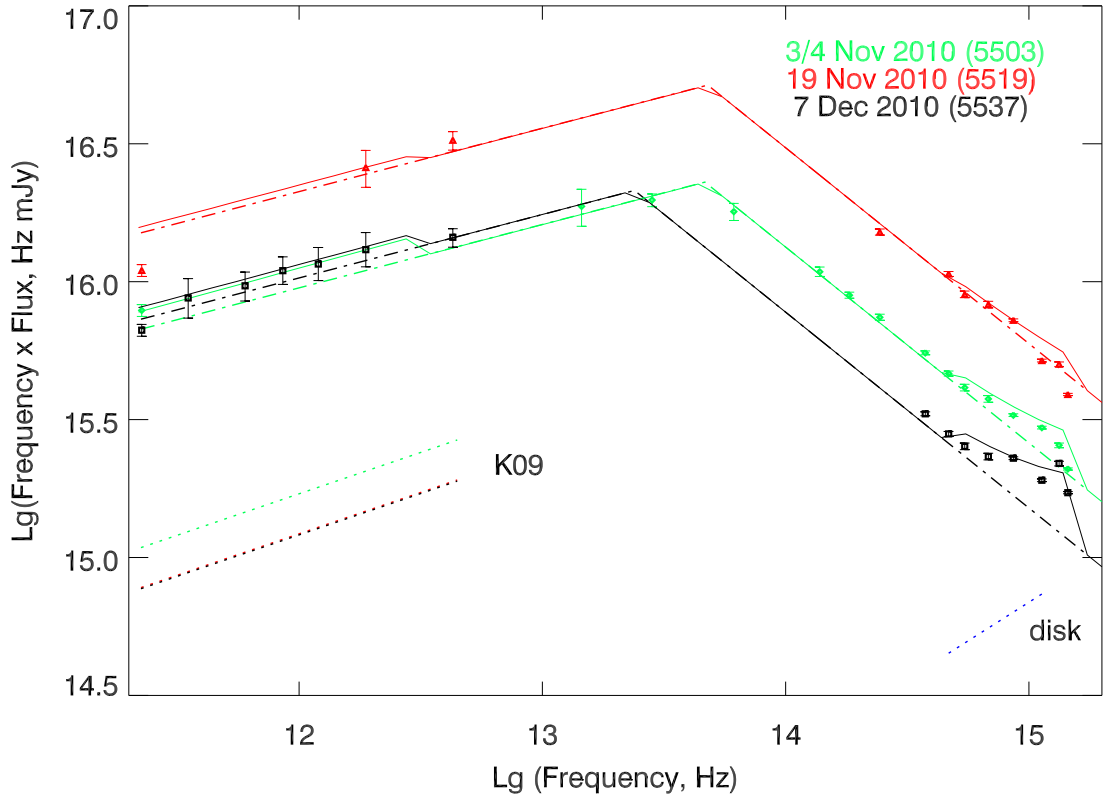}
\caption{Spectral energy distributions obtained on 2010 November 3/4 (red triangles), November 19 (green diamonds), and December 7 (black squares) and modeled by the sum (solid line) of the emission from a synchrotron component (dash-dotted line), accretion disk (blue dotted line), and knot $K09$ (black and green dotted lines).} 
\label{SEDmod}
\end{figure}

\begin{figure}
\plotone{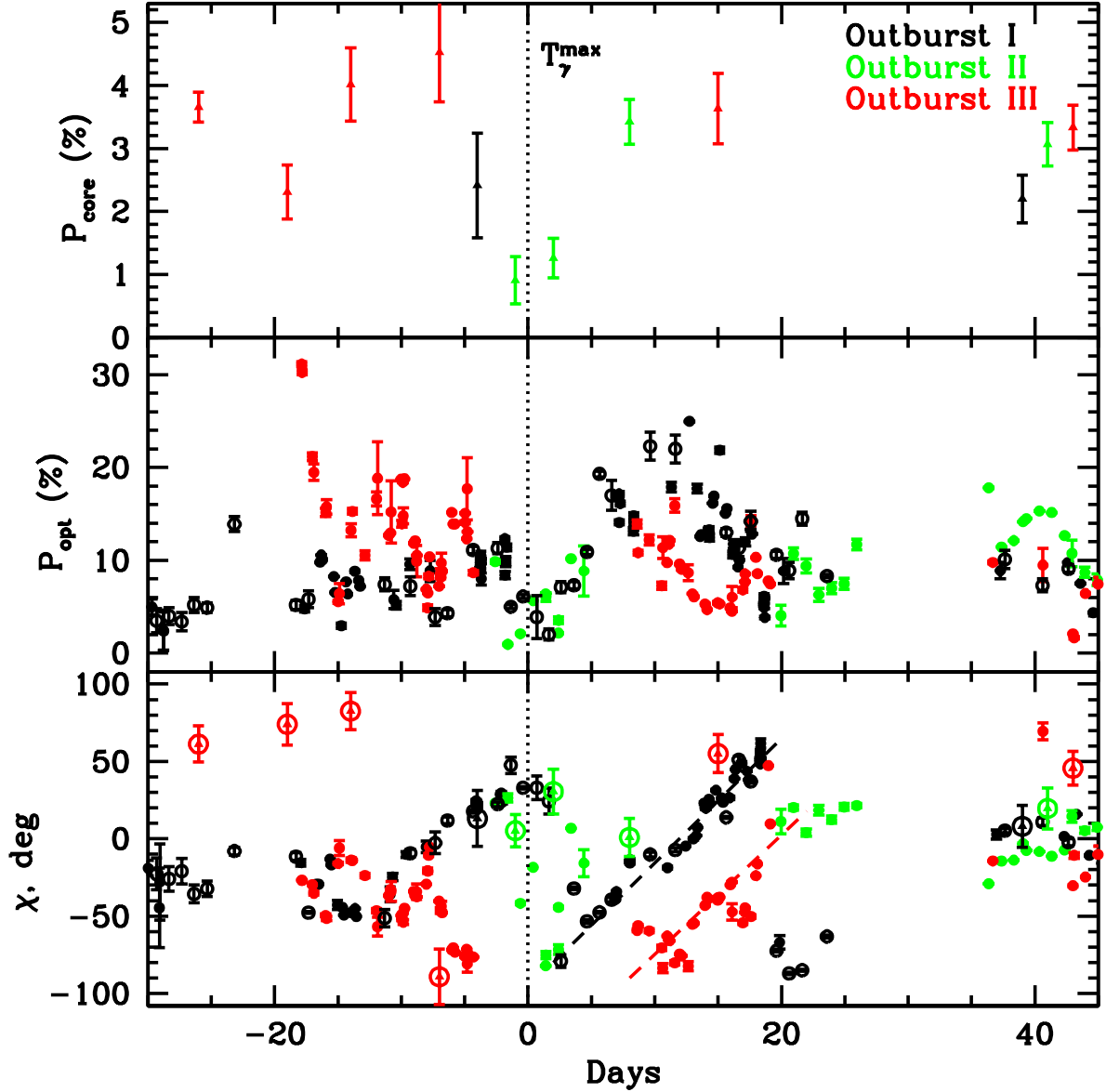}
\caption{Millimeter-wave core and optical polarization parameters during outburst I (black, open circles are measurements from \citet{SAS12}), II (green), and III (red) vs. time
relative to $T_\gamma^{\rm max}$ of each outburst; ({\it from the top}): degree of polarization in the core, degree of optical polarization, and position angle of optical polarization (circles) and in the core (triangles inside of circles), the dashed lines show a rotation of $\chi_{\rm opt}$
during outbursts I (black) and III (red).} 
\label{polG}
\end{figure}

We have discovered a repeating pattern in the properties of the three major $\gamma$-ray plus lower frequency outbursts observed in 3C~454.3 from 2009 to 2011. The duration, shape, and timescale of variability are similar, although the amplitudes of $\gamma$-ray outbursts are different (Table~\ref{Gparm}). This strongly suggests that the mechanism(s) and location of the high energy events are the same for all three outbursts. 
The $\gamma$-ray
variations are strongly correlated with those at optical, far-IR, and mm-wavelengths with a delay $<$1~day,  \citep[see also][]{ANN12}, although
the timescale of $\gamma$-ray variability is significantly shorter than at longer wavelengths. We have determined that a single synchrotron component is responsible for the variability from UV to IR wavelengths during an outburst (Fig.~\ref{SEDmod}), and that the properties of this component --- spectral index (Fig.~\ref{RSED}), timescale of variability (Table~\ref{Optparm}), and  polariazation parameters (Fig. \ref{polG}) --- are similar for the different outbursts. We have found interesting optical polarization
behavior during the outbursts that has not been noted previously: despite a general increase in the degree of polarization during an outburst, the degree of polarization drops significantly at the peak of the $\gamma$-ray event. In addition, the position angle of polarization varies on a timescale comparable to that of the $\gamma$-ray flux variations. This argues in favor of turbulence playing a significant role in the variations near the peak of a $\gamma$-ray event. 

We have detected apparent superluminal disturbances (knots) in the parcec-scale jet that we associate with the outbursts based on an analysis of the motions of the knots. We have found that the duration of the outbursts matches the time needed for a knot to pass through the mm-wave VLBI core.
We have derived the Doppler factors of the knots and shown that the differences in the Doppler factors can explain differences in the amplitudes of the outbursts. We have also shown that the polarization properties of the core and knots, as well as the trajectories of the knots, are connected with the optical polarization properties during the outbursts (Fig. \ref{polG}). 

Our multi-frequency analysis shows that, in the absence of relativistic boosting, the luminosity of the quasar 3C~454.3 would be dominated by accretion disk emission, in accordance with the unified scheme of AGN. The dramatic outbursts from radio wavelengths to $\gamma$-rays are certainly connected with the relativistic jet. The multi-frequency variability along with analysis of the parsec scale jet behavior 
favor in the localization of the outbursts in the mm-wave VLBI core, which  
is most likely located at the end of the acceleration zone where the magnetically dominated jet is converted into a matter dominated jet \citep{MAR08,J07}.

\acknowledgments
The research at Boston University (BU) was funded in part by NASA Fermi Guest Investigator grants NNX08AV65G, NNX10AO59G, NNX10AU15G, NNX11AO37G,
and NNX11AQ03G and Swift Guest Investigator grant NNX10AF88G. 
The VLBA is operated by the National Radio Astronomy Observatory. The
National Radio Astronomy Observatory is a facility of the National Science
Foundation operated under cooperative agreement by Associated Universities, Inc.
The PRISM camera at
Lowell Observatory was developed by K.\ Janes et al. at BU and Lowell Observatory, with funding from
the NSF, BU, and Lowell Observatory. The Liverpool Telescope is operated on the island of La Palma by Liverpool John Moores University in the
Spanish Observatorio del Roque de los Muchachos of the Instituto de Astrofisica de Canarias, with funding
from the UK Science and Technology Facilities Council. The effort at Steward Observatory
was funded in part by NASA through Fermi Guest Investigator grant NNX08AW56G and NNX09AU93G. The St. Petersburg State University team acknowledges support from RFBR grants 12-02-00452 and 12-02-31193. 
The research at the IAA-CSIC is supported by the Spanish Ministry of Economy and Competitiveness and the Regional Government of Andaluc\'{i}a (Spain) through grants AYA2010-14844 and P09-FQM-4784, respectively.
A. Wehrle acknowledges Guest Investigator support from NASA via Herschel RSA 1427799 and Fermi Guest Investigator grant NNX11AAO85G.
The Submillimeter Array is a joint project 
between the Smithsonian Astrophysical Observatory and the Academia Sinica Institute of Astronomy and Astrophysics, and is funded by the Smithsonian Institution and the Academia Sinica. The Abastumani team thanks the Georgian National Science Foundation for
support through grant GNSF/ST09/4-521. The {\it Swift} effort at PSU is supported by NASA contract NAS5-00136.
The Calar Alto Observatory is jointly operated by the Max-Planck-Institut f\"ur Astronomie and the Instituto de 
Astrof\'{\i}sica de Andaluc\'{\i}a-CSIC. 
The IRAM 30\,m telescope is supported by INSU/CNRS (France), MPG (Germany), and IGN (Spain).
The Mets\"ahovi team acknowledges support from the Academy of Finland. This study is partly based on data taken and assembled by the WEBT collaboration 
and stored in the WEBT archive at the Osservatorio Astronomico di Torino - INAF (http://www.oato.inaf.it/blazars/webt/).
This paper has made use of up-to-date SMARTS optical/near-infrared light curves that are available at www.astro.yale.edu/smarts/glast/home.php.

{\it Facilities:} Fermi, VLBA, SMA, Swift, Herschel, IRTF, IRAM, Lowell Obs., Steward Obs., Calar Alto Obs., Mets\"ahovi Obs, SMARTS

\begin{deluxetable}{crrr}
\singlespace
\tablecolumns{4}
\tablecaption{\bf Parameters of Gamma-Ray Outbursts \label{Gparm}}
\tabletypesize{\footnotesize}
\tablehead{
\colhead{Parameter}&\colhead{Outburst I}&\colhead{Outburst II}&\colhead{Outburst III} }
\startdata
$\Delta T_{\gamma}$, day&82&66&113\\
$<S_\gamma>$, 10$^{-6}$~ph~cm$^{-2}$s$^{-1}$ &5.84$\pm$3.79&5.94$\pm$3.34&12.36$\pm$11.66 \\
$<\sigma_\gamma>$, 10$^{-6}$~ph~cm$^{-2}$s$^{-1}$ &1.04&1.05&1.87 \\
$\Delta T_\gamma^{\rm a}$, day&11&20&5 \\
$T_\gamma^{\rm max}$&2009/12/02&2010/04/08&2010/11/20 \\
$T_\gamma^{\rm max}$, RJD&5168.343&5294.593&5520.593 \\
$S_\gamma^{\rm max}$, 10$^{-5}$~ph~cm$^{-2}$s$^{-1}$ &2.41$\pm$0.23&1.60$\pm$0.15&8.38$\pm$0.40 \\
$\alpha_\gamma^{\rm max}$&1.32$\pm$0.04&1.34$\pm$0.05&1.20$\pm$0.02 \\
$\tau_\gamma^{\rm a}$, hr&5.2&4.3&5.4 \\
$f_\gamma^{\rm a}$&1.79&2.02&1.74 \\
$\Delta T_\gamma^{\rm pre}$, day&8&7&13 \\
$S_\gamma^{\rm pre}$, 10$^{-6}$~ph~cm$^{-2}$s$^{-1}$&6.98$\pm$1.05&5.54$\pm$0.83&11.03$\pm$1.02 \\
$\alpha_\gamma^{\rm pre}$&1.34$\pm$0.10&1.36$\pm$0.11&1.30$\pm$0.06 \\
$\tau_\gamma^{\rm pre}$, hr&3.9&3.1&4.4 \\
$f_\gamma^{\rm pre}$&2.18&2.64&1.97 \\
$\Delta T_\gamma^{\rm post}$, day&9&6&13 \\
$S_\gamma^{\rm post}$, 10$^{-6}$~ph~cm$^{-2}$s$^{-1}$&6.51$\pm$0.92&5.34$\pm$0.89&20.91$\pm$4.20 \\
$\alpha_\gamma^{\rm post}$&1.31$\pm$0.08&1.41$\pm$0.07&1.35$\pm$0.06 \\
$\tau_\gamma^{\rm post}$, hr&5.6&3.5&7.3 \\
$f_\gamma^{\rm post}$&1.71&2.35&1.51 \\
$T_\gamma^{\rm b}$&2009/12/29&2010/05/05&2010/12/20 \\
$T_\gamma^{\rm b}$, RJD&5195.47&5322.22&5550.84 \\
$S_\gamma^{\rm b}$, 10$^{-5}$~ph~cm$^{-2}$s$^{-1}$ &1.07$\pm$0.16&1.14$\pm$0.16&2.53$\pm$0.22 \\
$\alpha_\gamma^{\rm b}$&1.36$\pm$0.04&1.36$\pm$0.03&1.29$\pm$0.02 \\
$T_\gamma^{\rm c}$&2010/01/18&2010/05/18&2011/01/06 \\
$T_\gamma^{\rm c}$, RJD&5214.59&5340.59&5568.22 \\
$S_\gamma^{\rm c}$, 10$^{-6}$~ph~cm$^{-2}$s$^{-1}$ &8.1$\pm$1.4&6.5$\pm$1.1&22.3$\pm$4.4 \\
$\alpha_\gamma^{\rm c}$&1.38$\pm$0.04&1.32$\pm$0.04&1.26$\pm$0.02 \\
$\tau_\gamma^{\rm min}$, hr&3.9&3.1&4.4 \\
$T_\gamma^{\tau_{\rm min}}$, RJD&5161.343&5281.593&5506.093 \\
$<\tau_{\gamma,2}>$, hr&19&21&22 \\
\enddata
\tablecomments{$\Delta T_{\gamma}$ - duration of $\gamma$-ray outburst (see text \S{3.1}); $<S_\gamma>$ - the average flux density during
the outburst and its standard deviation; $<\sigma_\gamma>$ - the average 1$\sigma$ uncertainty of an individual measurement during the outburst; $\Delta T_\gamma^{\rm a}$ - duration of the main sub-flare in flare $a$ (FWHM); $S_\gamma^{\rm max}$ - $\gamma$-ray flux at the peak of flare $a$ at 0.1-200~GeV calculated with a 3~hr integration interval; $\alpha_\gamma^{\rm max}$ - spectral energy index at 0.1-200~GeV calculated for a simple power law model for a 1~day integration interval centered at $T_\gamma^{\rm max}$; $\tau_\gamma^{\rm a}$ - minimum timescale of variability of $\gamma$-ray flux during the main flare; $f_\gamma^{\rm a}$ - factor of the $\gamma$-ray flux change over $\tau_\gamma^{\rm a}$;  $\Delta T_\gamma^{\rm pre}$ - duration of the pre-flare plateau during an $a$ flare; $S_\gamma^{\rm pre}$ - the average $\gamma$-ray flux and its standard deviation over period of $\Delta T_\gamma^{\rm pre}$; $\alpha_\gamma^{\rm pre}$ - spectral index at 0.1-200~GeV averaged over $\Delta T_\gamma^{\rm pre}$; $\tau_\gamma^{\rm pre}$ - minimum timescale of variability of $\gamma$-ray flux during $\Delta T_\gamma^{\rm pre}$; $f_\gamma^{\rm pre}$ - factor of the $\gamma$-ray flux change over $\tau_\gamma^{\rm pre}$;
$\Delta T_\gamma^{\rm post}$, $S_\gamma^{\rm post}$, $\alpha_\gamma^{\rm post}$, $\tau_\gamma^{\rm post}$, and 
$f_\gamma^{\rm post}$ - parameters for the post-flare plateau obtained in the same manner as for the pre-flare plateau; $T_\gamma^{\rm b},S_\gamma^{\rm b}$, and $\alpha_\gamma^{\rm b}$ - epoch, maximum flux, and spectral index, respectively, for flare $b$, calculated in the same manner as for flare $a$; $T_\gamma^{\rm c}, S_\gamma^{\rm c}$, and $\alpha_\gamma^{\rm c}$ are epoch, maximum flux, and spectral index, respectively, for flare $c$; $\tau_\gamma^{\rm min}$ - minimum timescale of variability of $\gamma$-ray flux during an outburst; $T_\gamma^{\rm\tau}$ - epoch of the start of an event with minimum timescale of variability; $<\tau_{\gamma,2}>$ - typical timescale of flux doubling (see text).}
\end{deluxetable}

\begin{deluxetable}{crr}
\singlespace
\tablecolumns{4}
\tablecaption{\bf Parameters of X-ray Outbursts \label{Xparm}}
\tabletypesize{\footnotesize}
\tablehead{
\colhead{Parameter}&\colhead{Outburst I}&\colhead{Outburst III} }
\startdata
$M$&33&96\\
$<S_{\rm X}>$,10$^{-11}$,~erg~cm$^{-2}$~s$^{-1}$&8.46$\pm$3.54&7.50$\pm$3.26 \\
$<\sigma_{\rm X}>$,10$^{-11}$,~erg~cm$^{-2}$~s$^{-1}$&0.45&0.40 \\
$\Delta T_{\rm X}^{\rm a}$, day&13.5&5 \\
$T_{\rm X}^{\rm max}$&2009/12/04&2010/11/19 \\
$T_{\rm X}^{\rm max}$, RJD&5169.531&5519.594 \\
$S_{\rm X}^{\rm max}$,10$^{-11}$,~erg~cm$^{-2}$~s$^{-1}$&16.73$\pm$0.40&17.92$\pm$0.64 \\
$\Delta T_{\rm X}^{\rm pre}$, day&\nodata&24 \\
$S_{\rm X}^{\rm pre}$,10$^{-11}$,~erg~cm$^{-2}$~s$^{-1}$&\nodata&6.48$\pm$0.64 \\
$\Delta T_{\rm X}^{\rm post}$, day&7 &12 \\
$S_{\rm X}^{\rm post}$,10$^{-11}$,~erg~cm$^{-2}$~s$^{-1}$&6.61$\pm$0.68&7.85$\pm$0.71 \\
$\tau_{\rm X}^{\rm min}$, hr&33.1&26.2 \\
$f_{\rm X}$, hr&1.71&1.80 \\
$T_{\rm X}^{\tau_{\rm min}}$, RJD&5176.89&5518.28 \\
$<\tau_{\rm X,2}>$, hr&48&57 \\
\enddata
\tablecomments{$M$ - number of X-ray measurements at 0.3-10~keV obtained during the ouburst;  $<S_{\rm X}>$ - the average flux density during
the outburst and its standard deviation; $<\sigma_{\rm X}>$ - the average 1$\sigma$ uncertainty of an individual measurement during the outburst; $\Delta T_{\rm X}^{\rm a}$ - duration of the main sub-flare in flare $a$ (FWHM); $T_{\rm X}^{\rm max}$ - epoch of the global maximum; $S_{\rm X}^{\rm max}$ - the flux density at the peak of flare $a$; $\Delta T_{\rm X}^{\rm pre}$ - duration of the pre-flare plateau during an $a$ flare, the value in parenthesises
indicate the average uncertainty of an individual measurement of the flux density in 10$^{-11}$,~erg~cm$^{-2}$~s$^{-1}$; $S_{\rm X}^{\rm pre}$ - the average flux density and its standard deviation over period of  $\Delta T_{\rm X}^{\rm pre}$;
$\Delta T_{\rm X}^{\rm post}$ and $S_{\rm X}^{\rm post}$ - similar parameters for the post-flare plateau; 
$\tau_{\rm X}^{\rm min}$ - minimum timescale of variability of X-ray flux during an outburst;  $f_{\rm X}$ - factor of the flux change over $\tau_{\rm X}^{\rm min}$;
$T_{\rm X}^{\tau_{min}}$ - epoch of the start of an event exhibiting the minimum timescale of variability;
$<\tau_{\rm X,2}>$ - typical timescale of flux doubling.}
\end{deluxetable}

\begin{deluxetable}{crrr}
\singlespace
\tablecolumns{4}
\tablecaption{\bf Parameters of Optical Outbursts \label{Optparm}}
\tabletypesize{\footnotesize}
\tablehead{
\colhead{Parameter}&\colhead{Outburst I}&\colhead{Outburst II}&\colhead{Outburst III} }
\startdata
$M$&306&103&2767\\
$<S_{\rm opt}>$, mJy &6.49$\pm$3.08&3.68$\pm$1.43&10.93$\pm$5.72 \\
$<\sigma_{\rm opt}>$, mJy &0.14&0.07&0.19 \\
$\Delta T_{\rm opt}^{\rm a}$, day&10&\nodata&8 \\
$T_{\rm opt}^{\rm max}$&2009/12/06&2010/04/10&2010/11/20 \\
$T_{\rm opt}^{\rm max}$, RJD&5172.273&5297.010&5520.673 \\
$S_{\rm opt}^{\rm max}$, mJy &12.71$\pm$0.19&6.71$\pm$0.14&24.40$\pm$1.10 \\
$\Delta T_{\rm opt}^{\rm pre}$, day&6&\nodata&7 \\
$S_{\rm opt}^{\rm pre}$, mJy&6.29$\pm$0.59&\nodata&8.30$\pm$0.72 \\
$\Delta T_{\rm opt}^{\rm post}$, day&10&\nodata&6 \\
$S_{\rm opt}^{\rm post}$, mJy&7.57$\pm$0.99&\nodata&7.42$\pm$0.34 \\
$T_{\rm opt}^{\rm b}$, RJD&5193.153&5324.521&5541.180 \\
$S_{\rm opt}^{\rm b}$, mJy &6.62$\pm$0.15&6.60$\pm$0.10&7.12$\pm$0.15 \\
$T_{\rm opt}^{\rm c}$, RJD&5215.144&5338.522&\nodata \\
$S_{\rm opt}^{\rm }$, mJy &4.97$\pm$0.05&3.46$\pm$0.06&\nodata \\
$\tau_{\rm opt}^{\rm min}$, hr&18.3&\nodata&23.7 \\
$f_{\rm opt}$, hr&1.44&\nodata&2.57 \\
$T_{\rm opt}^{\tau_{\rm min}}$, RJD&5168.26&\nodata&5509.55 \\
$<\tau_{\rm X,2}>$, hr&58&\nodata&40 \\
\enddata
\tablecomments{$M$ - number of optical measurements in $R$ band obtained during the ouburst; $<S_{\rm opt}>$ - the average flux density during
the outburst and its standard deviation; $<\sigma_{\rm opt}>$ - the average 1$\sigma$ uncertainty of an individual measurement during the outburst; $\Delta T_{\rm opt}^{\rm a}$ - duration of the main sub-flare in flare $a$ (FWHM); $S_{\rm opt}^{\rm max}$ - the flux density in $R$ band at the peak of flare $a$;
$\Delta T_{\rm opt}^{\rm pre}$ - duration of the pre-flare plateau during an $a$ flare; $S_{\rm opt}^{\rm pre}$ - the average flux in $R$-band and its standard deviation over period of  $\Delta T_{\rm opt}^{\rm pre}$;
$\Delta T_{\rm opt}^{\rm post}$ and $S_{\rm opt}^{\rm post}$ - parameters for the post-flare plateau obtained in the same manner as for the pre-flare plateau; $T_{\rm opt}^{\rm b}$ and $S_{\rm opt}^{\rm b}$ - epoch and maximum flux, respectively, for flare $b$; $T_{\rm opt}^{\rm c}$ and $S_{\rm opt}^{\rm c}$ - epoch and maximum flux, respectively, for flare $c$;
$\tau_{\rm opt}^{\rm min}$ - minimum timescale of variability of optical flux during an outburst; 
$f_{\rm opt}$ - factor of the flux change over $\tau_{\rm opt}^{\rm min}$;
 $T_{\rm opt}^{\tau_{min}}$ - epoch of the start of an event with minimum timescale of variability; $<\tau_{\rm opt,2}>$ - typical timescale of flux doubling (see text).}
\end{deluxetable}

\begin{deluxetable}{crr}
\singlespace
\tablecolumns{4}
\tablecaption{\bf Parameters of Outbursts at 1~mm\label{mmparm}}
\tabletypesize{\footnotesize}
\tablehead{
\colhead{Parameter}&\colhead{Outburst I}&\colhead{Outburst III} }
\startdata
$M$&25&44\\
$<S_{\rm mm}>$, Jy&22.32$\pm$4.27&36.27$\pm$7.04 \\
$<\sigma_{\rm mm}>$, Jy&1.17&1.63 \\
$\Delta T_{\rm mm}^{\rm a}$, day&14&6 \\
$T_{\rm mm}^{\rm max}$&2009/12/03&2010/11/20 \\
$T_{\rm mm}^{\rm max}$, RJD&5168.726&5520.657 \\
$S_{\rm mm}^{\rm max}$, Jy&27.70$\pm$1.39&51.77$\pm$4.31 \\
$T_{\rm mm}^{\rm b}$, RJD&5199.668&5548.587 \\
$S_{\rm mm}^{\rm b}$, mJy &27.49$\pm$1.38&37.73$\pm$3.46 \\
$T_{\rm mm}^{\rm c}$, RJD&\nodata&5571.764 \\
$S_{\rm mm}^{\rm c}$, mJy &\nodata&39.91$\pm$3.88 \\
$\tau_{\rm mm}^{\rm min}$, day&41&10 \\
$f_{\rm mm}$, day&1.34&1.35 \\
$T_{\rm mm}^{\tau_{\rm min}}$, RJD&5168.726&5534.64 \\
\enddata
\tablecomments{$M$ - number of measurements at 1~mm obtained during the outburst; $<S_{\rm mm}>$ - the average flux density during
the outburst and its standard deviation; $<\sigma_{\rm mm}>$ - the average 1$\sigma$ uncertainty of an individual measurement during the outburst;
$\Delta T_{\rm mm}^{\rm a}$ - duration of flare $a$ (FWHM);
$T_{\rm mm}^{\rm max}$ - epoch of the global maximum; $S_{\rm mm}^{\rm max}$ - the flux density at the peak of flare $a$; $T_{\rm mm}^{\rm b}$ and $S_{\rm mm}^{\rm b}$ - epoch and maximum flux, respectively, for flare $b$; $T_{\rm mm}^{\rm c}$ and $S_{\rm mm}^{\rm c}$ - epoch and maximum flux, respectively, for flare $c$;
$\tau_{\rm mm}^{\rm min}$ - minimum timescale of flux variability during an outburst; 
$f_{\rm mm}$ - factor of the flux change during $\tau_{\rm mm}^{\rm min}$;  $T_{\rm mm}^{\tau_{min}}$ - epoch of the start of an event exhibiting the minimum timescale of variability.}
\end{deluxetable}
\begin{deluxetable}{lrr}
\singlespace
\tablecolumns{3}
\tablecaption{\bf Parameters of Knots K09 and K10 \label{Kparm}}
\tabletypesize{\footnotesize}
\tablehead{
\colhead{Parameter}&\colhead{$K09$}&\colhead{$K10$} }
\startdata
$\mu$,mas~yr$^{-1}$&0.21$\pm$0.02&0.19$\pm$0.03 \\
$\dot{\mu}_\parallel$, mas~yr$^{-2}$&0.10$\pm$0.01&\nodata \\
$\dot{\mu}_\perp$, mas~yr$^{-2}$&0.13$\pm$0.02&1.10$\pm$0.22 \\
$\beta_{\rm app}$, $c$&9.6$\pm$0.6&8.9$\pm$1.7 \\
$T_\circ$, yr &2009.86$\pm$0.05&2010.95$\pm$0.07 \\
$T_\circ$, RJD &5146$\pm$18&5543$\pm$25 \\
$S_{\rm max}$, Jy&17.00$\pm$0.45&7.10$\pm$0.16 \\
$\tau_{\rm var}$, yr&0.67$\pm$0.06&0.24$\pm$0.02 \\
$a$, mas&0.12$\pm$0.02&0.08$\pm$0.01 \\
$\delta$&27$\pm$3&51$\pm$4 \\
$\Gamma_{\rm b}$&15$\pm$2&26$\pm$3 \\
$\Theta_\circ$, deg&1.35$\pm$0.2&0.4$\pm$0.1 \\
$N$&24& 7 \\
\enddata
\tablecomments{$\mu$ - proper motion; $\dot{\mu}_\parallel$ - angular acceleration along the jet;
$\dot{\mu}_\perp$ - angular acceleration perpendicular to the jet; $\beta_{\rm app}$ - apparent speed;
$T_\circ$ - time of ejection; $S_{\rm max}$ - maximum flux; $\tau_{\rm var}$ - timescale of flux variability;
$a$ - angular size of component at epoch of maximum flux;  $\delta$ - Doppler factor, 
$\Gamma_{\rm b}$ - Lorentz factor;
$\Theta_\circ$ - angle between velocity of component and line of sight; $N$ - number of epochs at which component was detected.}
\end{deluxetable}

\begin{deluxetable}{lrrrr}
\singlespace
\tablecolumns{5}
\tablecaption{\bf Spectral Index of Optical Continuum and Slope of $P(\lambda)$ Dependence \label{AcontT}}
\tabletypesize{\footnotesize}
\tablehead{
\colhead{Interval of $S^a$, mJy}&\colhead{$\alpha_{\rm cont}$}&\colhead{N}&\colhead{$B\times 10^{-5}, {\AA}^{-1}$}&\colhead{N$_{\rm B}$} }
\startdata
0$-$2.0&0.63$\pm$0.10&21 & 0.39$\pm$0.24&35\\
2.0$-$3.0&0.82$\pm$0.15&10& 0.75$\pm$0.42&28 \\
3.0$-$4.0&1.22$\pm$0.14&72& 0.52$\pm$0.45&109\\
4.0$-$6.0&1.22$\pm$0.14&43& 0.33$\pm$0.33&30 \\
6.0$-$10.0&1.37$\pm$0.10&27&0.18$\pm$0.34&32 \\
$>$10.0&1.35$\pm$0.14&8&$-$0.23$\pm$0.31&8   \\
\enddata
\tablecomments{$^a$ $S$ is the flux in $V$ band for $\alpha_{\rm cont}$ measurements and  in $R$ band for slope $B$ measurements; 
$N$,$N_{\rm B}$ is the number of $\alpha_{\rm cont}$ and slope $B$ measurements in each flux interval, respectively.}
\end{deluxetable}
\clearpage
\begin{deluxetable}{llrrrrrr}
\singlespace
\tablecolumns{8}
\tablecaption{\bf Relative Spectral Energy Distribution of Optical/near-IR Synchrotron Components \label{OptVar}}
\tabletypesize{\footnotesize}
\tablehead{
\colhead{Band}&\colhead{$log_{10}(\nu)$}&\multicolumn{3}{c}{Outburst I}&\multicolumn{3}{c}{Outburst III} \\
\colhead{}&\colhead{Hz}&\colhead{$N$}&\colhead{$log_{10}(S_{\rm band}/S_{\rm R})$}&\colhead{$\chi^2$}&\colhead{$N$}&\colhead{$log_{10} (S_{\rm band}/S_{\rm R})$}&\colhead{$\chi^2$} \\
\colhead{(1)}&\colhead{(2)}&\colhead{(3)}&\colhead{(4)}&\colhead{(5)}&\colhead{(6)}&\colhead{(7)}&\colhead{(8)}
}
\startdata
UVW2&15.169&6&$-$0.991$\pm$0.085&0.35&37&$-$0.945$\pm$0.031&0.59 \\
UVM2&15.128&6&$-$0.827$\pm$0.084&1.36&44&$-$0.805$\pm$0.029&4.05 \\
UVW1&15.056&6&$-$0.752$\pm$0.057&0.42&36&$-$0.704$\pm$0.022&0.70 \\
U&14.933&6&$-$0.450$\pm$0.062&2.08&41&$-$0.422$\pm$0.018&2.52 \\
B&14.833&33&$-$0.313$\pm$0.016&0.73&125&$-$0.264$\pm$0.010&0.54 \\
V&14.736&38&$-$0.116$\pm$0.008&0.47&147&$-$0.110$\pm$0.009&0.71 \\
R&14.760&\nodata&0.0&\nodata&\nodata&0.0&\nodata \\
I&14.574&21&0.173$\pm$0.021&0.57&67&0.181$\pm$0.010&1.26 \\
J&14.387&25&0.467$\pm$0.023&3.93&44&0.455$\pm$0.012&4.03 \\
H&14.262&\nodata&\nodata&\nodata&9&0.705$\pm$0.071&3.30 \\
K&14.140&25&0.844$\pm$0.043&9.25&44&0.811$\pm$0.044&4.21 \\
\enddata
\tablecomments{Columns: 1 - band of observations; 2 - logarithm of effective frequency of band; 
3 - number of simultaneous observations during outburst I in given band and $R$ band
(in $B$ band for $UV$ filters, in $J$ band for $H$ and $K$ filters); 4 - logarithm of the
slope of the linear dependence between the flux in a given band and $R$-band and its 1$\sigma$ uncertainty 
for outburst I; 5 - the $\chi^2$ error statistic for a linear approximation of the flux-flux dependence 
during outburst I; 6,7,8 - the same as 3,4,5, respectively, for outburst III.} 
\end{deluxetable}

\begin{deluxetable}{llrrr}
\singlespace
\tablecolumns{5}
\tablecaption{\bf Relative Spectral Energy Distribution of far-IR Synchrotron Component during Outburst III \label{IrVar}}
\tabletypesize{\footnotesize}
\tablehead{
\colhead{$\lambda$}&\colhead{$log_{10}(\nu)$}&\colhead{$N$}&\colhead{$log_{10} (S_{\lambda}/S_{250})$}&\colhead{$\chi^2$} \\
\colhead{(1)}&\colhead{(2)}&\colhead{(3)}&\colhead{(4)}&\colhead{(5)}
}
\startdata
70$\mu$m&12.633&15&$-$0.436$\pm$0.066&1.40 \\
160$\mu$m&12.274&6&$-$0.208$\pm$0.037&1.36 \\
250$\mu$m&12.079&\nodata&0.0&\nodata \\
350$\mu$m&11.933&13&0.090$\pm$0.013&0.62 \\
500$\mu$m&11.778&13&0.211$\pm$0.033&0.73 \\
850$\mu$m&11.544&3&0.244$\pm$0.155&2.22 \\
1.3mm&11.352&5&0.370$\pm$0.061&2.07 \\
\enddata
\tablecomments{Columns: 1 - wavelength of observations; 2 - logarithm of frequency of observations; 
3 - number of simultaneous observations at a given wavelength and 250~$\mu$m (at 160~$\mu$m for 70~$\mu$m); 4 - logarithm of
slope of linear dependence between the flux at a given wavelength and 250~$\mu$m  and its 1$\sigma$ uncertainty; 
5 - $\chi^2$ error statistic for a linear approximation of the flux-flux dependence.}
\end{deluxetable}

\begin{deluxetable}{lrrr}
\singlespace
\tablecolumns{4}
\tablecaption{\bf Parameters of SEDs during Outburst III \label{SEDpar}}
\tabletypesize{\footnotesize}
\tablehead{
\colhead{Parameter}&\colhead{3/4 Nov}&\colhead{19 Nov}&\colhead{7 Dec}
}
\startdata
$S_{\rm V}^{\rm obs}, mJy$&7.59$\pm$0.28&16.52$\pm$0.15&6.00$\pm$0.12 \\
$S_{\rm 1mm}^{\rm obs}, Jy$&34.9$\pm$1.8&48.8$\pm$2.40&29.6$\pm$1.5 \\
$S_{K09}, Jy$&15.5$\pm$0.3&11.1$\pm$0.2&11.0$\pm$0.2\\
$\nu_{\rm peak}$, Hz&4.66E+13&4.78E+13&2.39E+13 \\
$\lambda_{\rm peak}$, $\mu$m&6.4&6.3&12.5 \\
$S_{\rm peak}$, mJy&493&1079&895 \\
\enddata
\tablecomments{Parameters of $SED$s in the observer's frame:
$S_{\rm V}^{\rm obs}$ - flux density observed in $V$ band; $S_{\rm 1mm}^{\rm obs}$ - flux density observed at 1.3~mm,
$S_{K09}$ - flux density of knot $K09$ at 7~mm;
$\nu_{\rm peak}$ - frequency of peak of $SED$, $\lambda_{\rm peak}$ - wavelength of peak of $SED$,
$S_{\rm peak}$ - flux density at peak of $SED$.
$L_{\rm disk}$ - luminosity of accretion disk integrated from 6500 to 2000~{\AA}}
\end{deluxetable}

\begin{deluxetable}{lrrrr}
\singlespace
\tablecolumns{5}
\tablecaption{\bf Parameters of SEDs during Outbursts and Quiescent State \label{SEDobsT}}
\tabletypesize{\footnotesize}
\tablehead{
\colhead{Parameter}&\colhead{Outburst I}&\colhead{Outburst II}&\colhead{Outburst III}&\colhead{Quiescent}
}
\startdata
$\delta$&27&25&51&13 \\
$\nu_{\rm peak}^{\rm LE}$, Hz&2.53E+13&2.34E+13&4.78E+13&1.22E+13 \\
$\lambda_{\rm peak}^{\rm LE}$, $\mu$m&11.8&12.8&6.3&24.6 \\
$S_{\rm peak}^{\rm LE}$, mJy&760&750&1100&160 \\
$L_{\rm LE}$, erg~s$^{-1}$&6.9E+47&6.8E+47&1.9E+48&7.0E+46\\
$L_{\rm syn}$, erg~s$^{-1}$&8.2E+47&6.0E+47&1.0E+49&4.4E+46\\
$\nu_{\rm peak}^{\rm HE}$, Hz&3.16E+22&2.63E+22&6.02E+22&1.51E+22 \\
$E_{\rm peak}^{\rm HE}$, GeV&0.13&0.11&0.25&0.06 \\
$S_{\rm peak}^{\rm HE}$, mJy&9.0E-6&7.6E-6&1.2E-5&2.3E-7 \\
$L_{\rm HE}$, erg~s$^{-1}$&1.0E+49&7.2E+48&2.6E+49&1.2E+47\\
$L_{\rm IC}$, erg~s$^{-1}$&2.2E+48&1.6E+48&2.8E+49&1.2E+47\\
\enddata
\end{deluxetable}
\clearpage

\clearpage

\begin{figure}
\plottwo{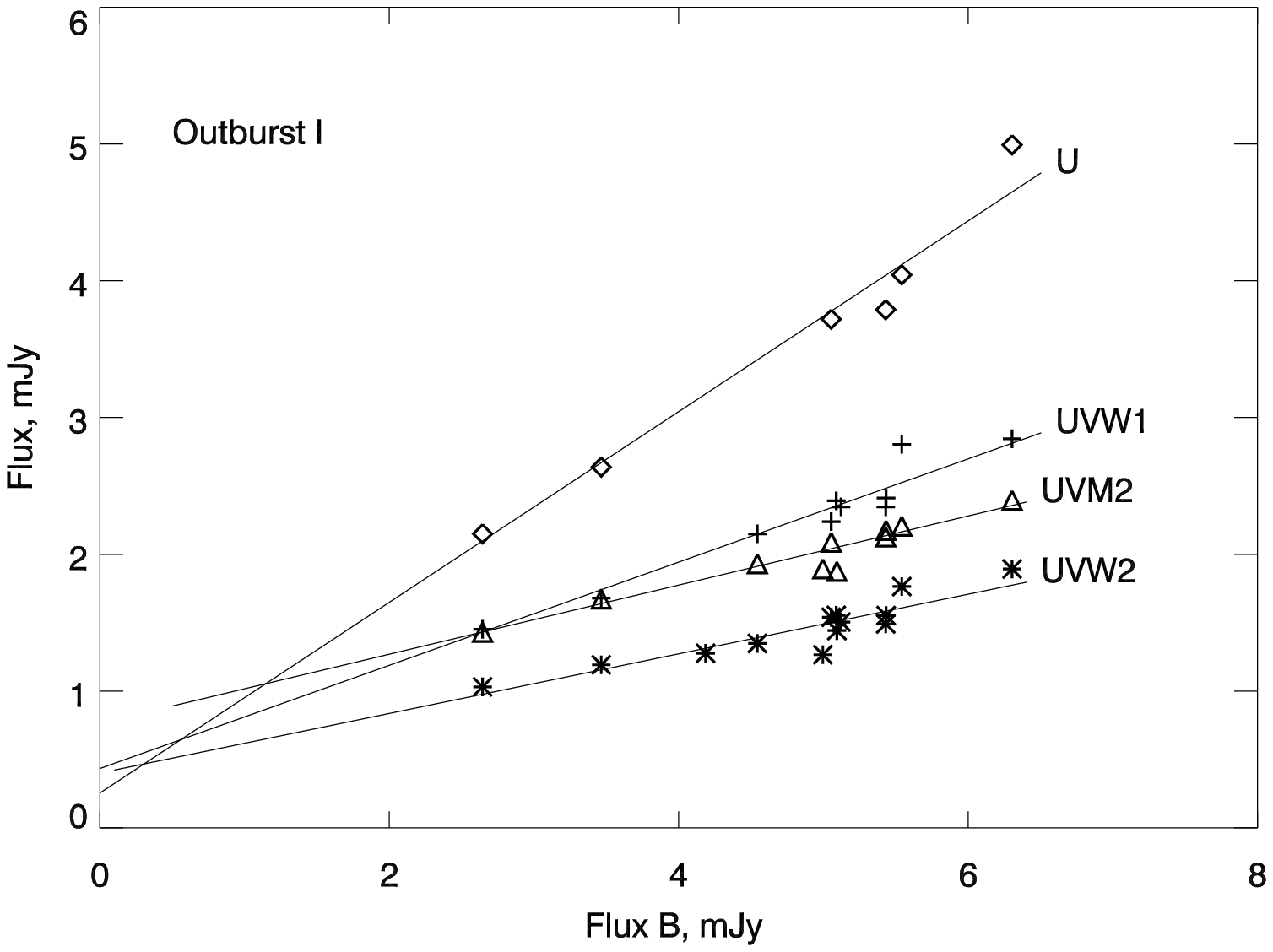}{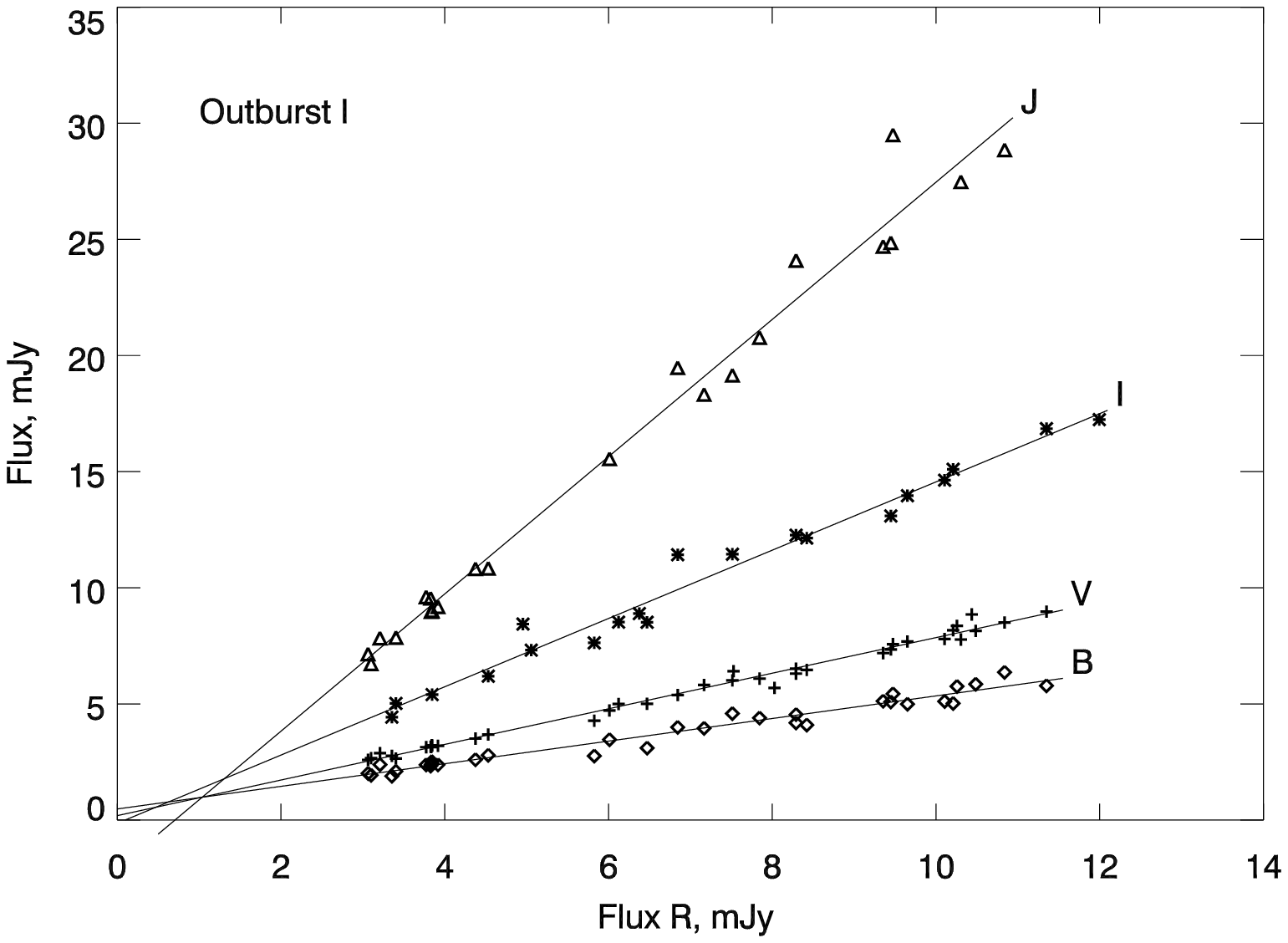}
\caption{Flux-flux dependences during outburst I for simultaneous measurements at different wavelengths. 
{\it Left:} Flux densities at $UV$ bands ($U$ - diamonds, $UVW1$ - crosses, $UVM2$ - triangles,
and $UVW2$ - asterisks) vs. flux densities in $B$ band. {\it Right:} Flux densities in optical and near-IR bands
($B$ - diamonds, $V$ - crosses, $I$ - asterisks, and $J$ - triangles) vs. flux densities in $R$ band. } 
\label{FFI}
\end{figure}

\begin{figure}
\plottwo{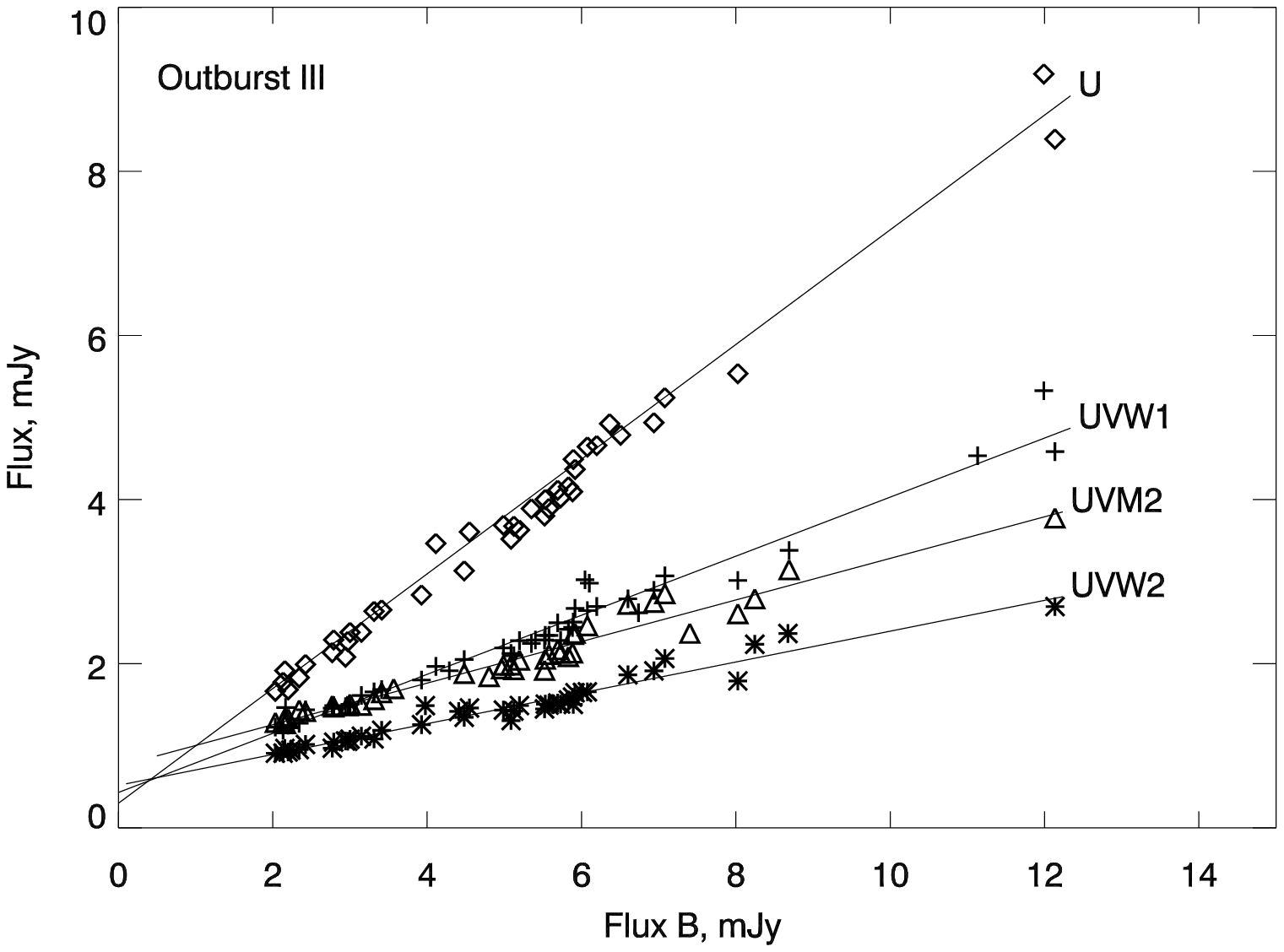}{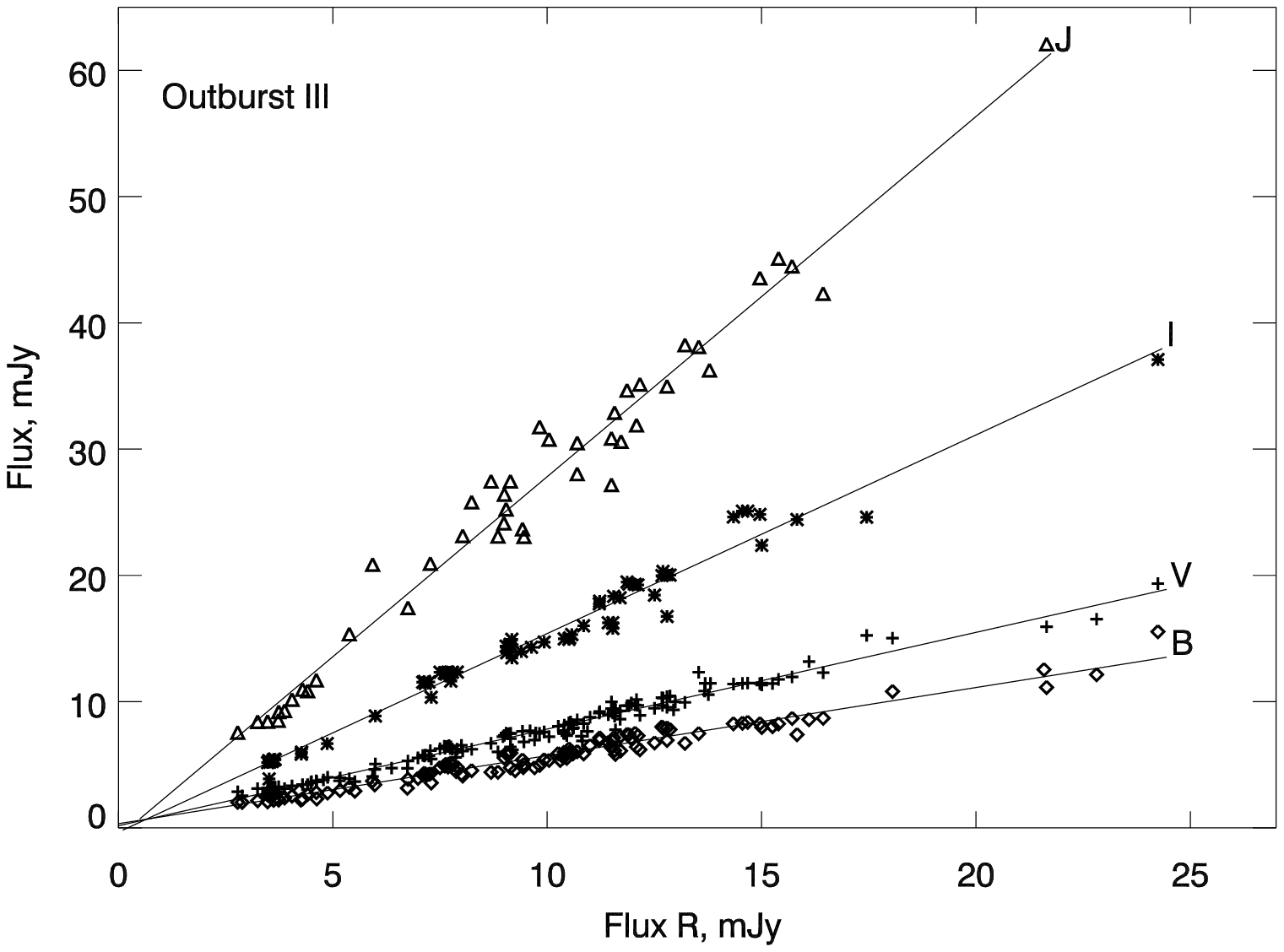}
\caption{Flux-flux dependences during outburst III for simultaneous measurements at different wavelengths. 
{\it Left:} Flux densities at $UV$ bands ($U$ - diamonds, $UVW1$ - crosses, $UVM2$ - triangles,
and $UVW2$ - asterisks) vs. flux densities in $B$ band. {\it Right:} Flux densities in optical and near-IR bands
($B$ - diamonds, $V$ - crosses, $I$ - asterisks, and $J$ - triangles) vs. flux densities in $R$ band. } 
\label{FFIII}
\end{figure}

\end{document}